\documentclass[aps,prd,reprint,twocolumn,superscriptaddress,longbibliography,nofootinbib,floatfix,showpacs]{revtex4-2}
\usepackage{epsfig}
\usepackage{amsmath,amssymb,amsfonts}
\usepackage{hyperref}
\usepackage{mathrsfs}
\usepackage{bbm}
\usepackage{slashed}
\usepackage{graphicx}
\usepackage{verbatim}

\usepackage{bm}

\usepackage[usenames]{color}

\definecolor{darkgreen}{rgb}{0.2,0.6,0}

\newcommand{\be}{\begin{equation}}
\newcommand{\ee}{\end{equation}}
\newcommand{\bw}{\begin{widetext}}
\newcommand{\ew}{\end{widetext}}
\newcommand{\bi}{\begin{itemize}}
\newcommand{\ei}{\end{itemize}}
\newcommand{\ud}{\mathrm{d}}

\newcommand{\LCm}{{\scriptscriptstyle -}} \newcommand{\LCp}{{\scriptscriptstyle +}}
\newcommand{\LCpm}{{\scriptscriptstyle \pm}}

\newcommand{\LCperp}{{\scriptscriptstyle \perp}}

\newcommand{\LCpara}{{\scriptscriptstyle \parallel}}

\usepackage[T1]{fontenc} \usepackage[latin1]{inputenc}

\newcommand{\pa}{\partial}
\let\Re\relax
\let\Im\relax
\DeclareMathOperator{\Re}{Re}
\DeclareMathOperator{\Im}{Im}

\begin{document}

\title{Nonlinear trident using WKB and worldline instantons}

\author{Gianluca Degli Esposti}
\email{g.degli-esposti@hzdr.de}
\affiliation{Helmholtz-Zentrum Dresden-Rossendorf, Bautzner Landstra{\ss}e 400, 01328 Dresden, Germany}

\affiliation{Institut f\"ur Theoretische Physik, 
Technische Universit\"at Dresden, 01062 Dresden, Germany}

\author{Greger Torgrimsson}
\email{greger.torgrimsson@umu.se}
\affiliation{Department of Physics, Ume{\aa} University, SE-901 87 Ume{\aa}, Sweden}

\begin{abstract}

We consider nonlinear trident, $e^\LCm\to e^\LCm e^\LCm e^\LCp$, in various electric background fields. This process has so far been studied for plane-wave backgrounds, using Volkov solutions. Here we first use WKB for trident in time-dependent electric fields, and then for fields which vary slowly in space. Then we show how to use worldline instantons for more general fields which depend on both time and space. For time-dependent fields the WKB approach is at least as simple to use as the worldline approach, but already the relatively modest step of including a slow spatial dependence makes the worldline approach much more efficient.   

\end{abstract}
\maketitle

\section{Introduction}

There continues to be growing progress in strong-field-QED processes, especially those occurring when electrons or photons collide with high-intensity laser fields~\cite{DiPiazza:2011tq,Gonoskov:2021hwf,Fedotov:2022ely}. One process is the nonlinear trident process, $e^\LCm\to e^\LCm e^\LCm e^\LCp$. Many particles can be produced by a single electron, because factors of $\alpha=1/137$, which would otherwise suppress higher-order processes, are compensated for by large factors due to the high intensity or extended pulse length of the background field. To theoretically model such higher-order processes, one needs to employ various approximations. One approximation, in particular, is to treat higher-order processes as incoherent products of first-order processes. One reason that we are interested in studying trident is that it is the first nontrivial step in the production of more particles while still being sufficiently ``simple'' so that one can calculate and check the corrections to the incoherent-product approximation. The calculations of these corrections are actually quite nontrivial. In fact, while the simplest part of the correction (the direct part) was calculated already in the 70's by~\cite{Baier,Ritus:1972nf}, it was not until~\cite{Dinu:2017uoj,King:2018ibi} that the rest of the correction (the exchange part) was calculated. It turned out that this previously omitted exchange term is in fact on the same order of magnitude as the simpler direct term. Here we should mention that there have also been fully numerical computations of the trident spectrum which did not worry about separating the probability into an incoherent product and corrections to it. A not complete list of papers studying trident is~\cite{Baier,Ritus:1972nf,Hu:2010ye,Ilderton:2010wr,King:2013osa,Hu:2014ooa,Krajewska15,Dinu:2017uoj,King:2018ibi,Mackenroth:2018smh,Acosta:2019bvh,Dinu:2019pau,Dinu:2019wdw,Torgrimsson:2020wlz,Torgrimsson:2022ndq}.

All these previous studies of trident have in common that they considered the background field to be a plane wave, i.e. with a gauge field given by\footnote{We absorb the charge into the definition of the background field, $eF^{\mu\nu}\to F^{\mu\nu}$, and set $c=\hbar=m=1$, where $m$ is the electron mass.} $A_\mu(kx)$, where $k_\mu=\omega(1,0,0,1)$ is the wave vector and $kA=0$. There are good reasons for this. One is the fact that an electron with sufficiently high energy tends to see a quite general background as if it were a plane wave. However, if the electron's momentum is (approximately) parallel to an electric field, then the electron will not see that field as a plane wave. And if the field is sufficiently strong to significantly bend the electron trajectory, then one can also expect differences from the plane-wave approximation. Then there are of course cases where the electron energy is not high. 

There is thus motivation for trying to go beyond plane waves. However, doing so is challenging. For plane waves the solution to the Dirac equation is the Volkov solution, which is particularly simple. There are a couple of 1D fields for which one can solve the Dirac equation exactly, e.g. a Sauter pulse, $A_3(t)=(E/\omega)\tanh(\omega t)$, but the solutions for those fields are already much more complicated than the Volkov solutions for plane waves. One can therefore not expect to be able to find exact solutions for realistic multidimensional fields, let alone use them to calculate the probability exactly. 

In this paper we will use weak-field approximations to calculate the trident probability for non-plane-wave backgrounds. We will start with a WKB treatment of trident in purely time-dependent electric fields, $E(t)=A_3'(t)$. One might argue about whether this can be seen as going ``beyond'' plane waves. We are not saying that fields of the form $E(t)$ actually give a more accurate model of realistic laser fields. For one thing, $E(t)$ does not solve Maxwell's equation, while plane waves do. However, from an analytical point of view, there is a sense in which considering $E(t)$ is going beyond plane waves. The fact that the two field invariants, ${\bf E}^2-{\bf B}^2$ and ${\bf E}\cdot{\bf B}$, are both zero for plane waves means that the probabilities have less structure. For $E(t)$, on the other hand, one of the field invariants is nonzero, so even before doing any calculations, one can expect the result to have more structure. In fact, results for a general plane wave can be obtained from the results for $E(t)$ by taking the limit where the transverse momentum of the electron is very high. In other words, the plane-wave results are contained in the results for $E(t)$, so in this sense we are going beyond plane waves when studying $E(t)$. Considering $E(t)$ therefore allows us to get a sense of what is neglected when approximating the background as a plane wave, and when it is not possible to do so. We show, in particular, that if we instead take the limit where the momentum of the electron is large but parallel to the electric field, then one does not recover the plane-wave result.    

Then we show how to use worldline instantons to study trident in general space-time dependent fields. Both the WKB and the instanton approach calculate the same leading order in a weak-field approximation. In the WKB approach one first has to solve the Dirac equation, i.e. one has to find an approximation of a field, $\psi(x^\mu)$, which in general depends nontrivially on $t$, $x$, $y$ and $z$ in addition to the field and momentum parameters, and then one has to plug that solution into the Feynman rules and perform the integrals (which one can do with the saddle-point approximation). Note that $\psi(x^\mu)$ contains more information than what is actually needed for any particular process. In contrast, in the instanton approach one goes directly for the amplitude of the particular process one considers, i.e. without first calculating some intermediate result. Moreover, the instantons are 1D trajectories which permit a more intuitive understanding of these processes. With this in mind, it is perhaps not so surprising that we can calculate things with the instanton approach which, although in principle possible, would be difficult to do with a WKB approach. We will show how to apply the instanton approach for fields which depend on both time and space separately (i.e. not just on e.g. $t-z$ as in the plane-wave case). This can be seen as going beyond plane waves both in the analytical sense described above as well as in a more direct sense of more accurately modeling realistic lasers.

Of course, both the WKB and the instanton approaches are dealing with the leading-order in a weak-field approximation, while for plane waves it is possible to calculate the probability exactly. In the plane-wave case the relevant expansion parameter is the quantum nonlinearity parameter $\chi=\sqrt{-(F^{\mu\nu}p_\nu)^2}$. However, as demonstrated in~\cite{Dinu:2019pau}, the saddle-point approximation does in fact offer a good qualitative estimate even when $\chi~\sim0.5$ is not very small.

Our main motivation and goals can be summarized as:

\bi

\item We want to understand how and under what conditions probabilities of higher-order processes ($\mathcal{O}(\alpha^2)$ in the case of trident) can be approximated by incoherent products of $\mathcal{O}(\alpha)$ processes. This problem is by now well-studied for the case of plane waves. However, while plane waves are well motivated, they are also special, e.g. because the field invariants vanish. So even by considering a relatively simple class of fields such as $E(t)$ one can get insights or ideas for how this factorization generalizes beyond plane waves.

\item We want to develop methods to be able to calculate the probability of trident and similar processes for fields which depend on more than one space-time coordinate. Our open-worldline approach seems promising for that.

\ei

\section{Trident in $E(t)$ using WKB}

We first consider time-dependent fields of the form $A_3(-t)=-A_3(t)$, $A_3(t)=f(\omega t)/\gamma$, where $f$ is a dimensionless function and $\gamma=\omega/E$ is the Keldysh parameter. In the high-energy limit we will often express things in terms of $a_0=1/\gamma$ (the classical nonlinearity parameter) instead of $\gamma$, because $a_0$ is more common when dealing with plane-wave backgrounds.
We use the following notation for the momentum, $\pi_\LCperp=q_\LCperp=\{q_1,q_2\}$, $\pi_3(t)=q_3-A_3(t)$, $\pi_0=\sqrt{m_\LCperp^2+\pi_3^2(t)}$, where $m_\LCperp=\sqrt{1+q_\LCperp^2}$. A typical example is a Sauter pulse, 
\be
f(v)=\tanh(v) \;.
\ee
The electron and positron wave functions are given by
\be\label{UandV}
\begin{split}
	{\sf U}_r(t,{\bf q})&=(\gamma^0\pi_0+\gamma^i\pi_i+1){\sf G}^+(t,{\bf q})R_r \\
	{\sf V}_r(t,-{\bf q})&=(-\gamma^0\pi_0+\gamma^i\pi_i+1){\sf G}^-(t,{\bf q})R_r \;,
\end{split}
\ee
where the spinors are chosen as $\gamma^0\gamma^3 R_s=R_s$ for $r=1,2$, and
\be\label{Gpmdef}
{\sf G}^\pm(t,{\bf q})=[2\pi_0(\pi_0\pm\pi_3)]^{-\frac{1}{2}}\exp\bigg[\mp i\int_{t_r}^t\!\ud t'\,\pi_0(t')\bigg] \;,
\ee
where $t_r\in\mathbb{R}$ is arbitrary, but, since we are interested in fields which are symmetric around $t=0$, it is natural to choose $t_r=0$.

We use ${\bf p}$, ${\bf p}_1$, ${\bf p}_2$ and ${\bf p}_3$ for the momenta of the initial electron, the two electrons in the final state, and the positron, respectively.
The amplitude for trident is given by $M=\langle0|a_{r_1}({\bf p}_1)a_{r_2}({\bf p}_2)a_{r_3}({\bf p}_3)U a_r^\dagger({\bf p})|0\rangle=M^{12}-M^{21}$ ($U$ is the time evolution operator), where
\be
\begin{split}
&(2\pi)^3\delta^3({\bf p}_1+{\bf p}_2+{\bf p}_3-{\bf p})M^{12}\\
&=e^2\int\ud^4x\ud^4y\bar{\psi}(p_2,y)\gamma^\mu\psi^{(\LCm)}(p_3,y)D_{\mu\nu}(y-x)\\
&\hspace{1cm}\times\bar{\psi}(p_1,x)\gamma^\nu\psi(p,x)
\end{split}
\ee
and $M^{21}$ is obtained from $M^{12}$ by swapping place of the two electrons in the final state. We use the Feynman gauge for the photon propagator
\be\label{photonPropagator}
D_{\mu\nu}(y-x)=-i\int\frac{\ud^4l}{(2\pi)^4}g_{\mu\nu}\frac{e^{-il(y-x)}}{l^2+i\epsilon} \;.
\ee
The fermion wave functions $\psi$ and $\psi^{(\LCm)}$ are approximated by ${\sf U}(t)$ and ${\sf V}(t)$ in~\eqref{UandV} times the trivial spatial parts $e^{-ip_jx^j}$.

We start by performing the trivial integrals in $M^{12}$. We first have
\be
\int\ud^3{\bf x}\to(2\pi)^3\delta^3({\bf l}-[{\bf p}-{\bf p}_1]) \;,
\ee
which is used to perform the photon momentum ${\bf l}$ integrals. Then the other spatial integrals give the overall momentum conservation, 
\be
\int\ud^3{\bf y}\to(2\pi)^3\delta^3({\bf p}_1+{\bf p}_2+{\bf p}_3-{\bf p}) \;.
\ee
We use this delta function to perform the integral over the positron momentum, i.e. ${\bf p}_3={\bf p}-{\bf p}_1-{\bf p}_2$. 

The initial state is described by a wave packet
\be
|\text{in}\rangle=\int\frac{\ud^3{\bf p}}{(2\pi)^3}f({\bf p})a_r^\dagger({\bf p})|0\rangle \;.
\ee
With the mode operators normalized as
\be
\{a_r({\bf p}),a_{r'}^\dagger({\bf p}')\}=\delta_{rr'}(2\pi)^3\delta^3({\bf p}-{\bf p}') \;,
\ee
$\langle\text{in}|\text{in}\rangle=1$ implies
\be
\int\frac{\ud^3{\bf p}}{(2\pi)^3}|f({\bf p})|^2=1 \;.
\ee
The (spin-polarized) probability is given by
\be\label{Pint123}
\begin{split}
P=&\int\frac{\ud^3{\bf p}_1}{(2\pi)^3}\frac{\ud^3{\bf p}_2}{(2\pi)^3}\frac{\ud^3{\bf p}_3}{(2\pi)^3}\\
&\times\left|\int\frac{\ud^3{\bf p}}{(2\pi)^3}f({\bf p})(2\pi)^3\delta^3({\bf p}_1+{\bf p}_2+{\bf p}_3-{\bf p})M\right|^2\\
=&\int\frac{\ud^3{\bf p}_1}{(2\pi)^3}\frac{\ud^3{\bf p}_2}{(2\pi)^3}\frac{\ud^3{\bf p}}{(2\pi)^3}|f({\bf p})|^2|M|^2\\
\to&\int\frac{\ud^3{\bf p}_1}{(2\pi)^3}\frac{\ud^3{\bf p}_2}{(2\pi)^3}|M|^2 \;,
\end{split}
\ee
where in the last step we have assumed that the wave packet is sharply peaked (and from now on we use ${\bf p}$ for the position of the peak). We will also average and sum over the initial- and final-state spins.

The remaining integrals are nontrivial. The exponential part of the integrand is now given by
\be\label{expGen}
e^{i[l_0(t-t')-I({\bf p},t)+I({\bf p}_1,t)+I({\bf p}_2,t')+I(-{\bf p}_3,t')]} \;,
\ee
where $t=x^0$ and $t'=y^0$, and
\be\label{Iint}
I({\bf p},t)=\int^t\ud t''\sqrt{m_\LCperp^2+\left(p_3-A_3(t'')\right)^2} \;.
\ee
We change variable from the (off-shell) photon energy $l_0=X+|{\bf l}|$ to $X$, where $|{\bf l}|=\sqrt{({\bf p}-{\bf p}_1)\cdot({\bf p}-{\bf p}_1)}$ is the on-shell energy.

The expansion parameter is $E\ll1$, with $\gamma=\mathcal{O}(E^0)$. By rescaling $t\to T/E$ and $t'\to T'/E$, the exponential part of the integrand goes like $e^{\dots/E}$ with no factors of $E$ in the ellipses. This justifies using the saddle-point method for weak fields $E\ll1$. We will show that all the remaining integrals can be performed by expanding the exponent around the saddle point. This usually results in Gaussian integrals, but for the photon-energy integral $X$ we find something more nontrivial due to the fact that $1/(l^2+i\epsilon)$ in the photon propagator has a pole (when $\epsilon\to0$) at the same point where the exponent has a saddle point.

We begin by performing the time-integrals on the amplitude level. We have a saddle point at $t=t'=\tilde{t}$, where $\tilde{t}=i\tilde{u}/\omega$ and $\tilde{u}$ is given by 
\be\label{ftildeutilde}
\tilde{f}(\tilde{u})=\sqrt{1+\frac{p_\LCpara^2}{p_\LCperp^2}}\gamma \;,
\ee
where $f(iu)=i\tilde{f}(u)$.
\eqref{ftildeutilde} is the same equation as in the Breit-Wheeler case (see Eq.~(47) in~\cite{Brodin:2022dkd}).
So the time saddle point is imaginary. Note that it cannot be real because the exponential part of the integrand is a pure phase for real integration variables but we know that the probability we are calculating has an exponential scaling, $P\sim e^{-(\text{something real})/E}$.  
We change variables from $t=\tilde{t}+\delta t$ and $t'=\tilde{t}+\delta t'$ to the fluctuations around the saddle points, $\delta t$ and $\delta t'$, and then expand the exponent in~\eqref{expGen} to $\mathcal{O}(\delta^2)$ using
\be
\begin{split}
I({\bf p},t)\approx&I({\bf p},\tilde{t})+\sqrt{m_\LCperp^2+(p_3-A(\tilde{t}))^2}\delta t\\
&+\frac{A'(\tilde{t})(A(\tilde{t})-p_3)}{\sqrt{m_\LCperp^2+(p_3-A(\tilde{t}))^2}}\frac{\delta t^2}{2} 
\end{split}
\ee
and similarly for the other terms, which gives
\be
\begin{split}
	&i[l_0(t-t')-I({\bf p},t)+I({\bf p}_1,t)+I({\bf p}_2,t')+I(-{\bf p}_3,t')]\\
	&\approx F_0({\bf p}_1,{\bf p}_2)+F_1({\bf p}_1,l_0)\delta t+F_2({\bf p}_1)\delta t^2\\
	&+F'_1({\bf p}_1,{\bf p}_2,l_0)\delta t'+F'_2({\bf p}_1,{\bf p}_2)\delta t^{\prime2} \;.
\end{split}
\ee
This equation defines $F$ and $F'$ and we have only written out how they depend on the integration variables. We refrain from writing out explicit expressions for the intermediate steps in the calculation, because they are long and not illuminating (the calculations have been done with Mathematica).
$F_1$ and $F'_1$ are nonzero because $\tilde{t}$ is only the saddle point for $t$ and $t'$ when the other integration variables are replaced by their saddle-point values. 

By an educated guess or otherwise, we find a saddle point for the momentum variables at
\be\label{eq:equalOutMomenta}
{\bf p}_1={\bf p}_2={\bf p}_3=\frac{{\bf p}}{3}=:{\bf p}' \;,
\ee
so all three final-state particles have approximately the same momentum, which is equal to one third of the initial momentum. For the photon-momentum we have a saddle point at $X=0$, so the intermediate photon is approximately on shell\footnote{Peaks in the probability associated with intermediate particles which are on or almost on shell have been studied in e.g~\cite{Oleinik,Seipt:2012tn,Sizykh:2021ywt}.}. We change variables from ${\bf p}_1={\bf p}'+\delta{\bf p}_1$, ${\bf p}_2={\bf p}'+\delta{\bf p}_2$ and $X=\delta X$ to $\delta{\bf p}_1$, $\delta{\bf p}_2$ and $\delta X$. All the integration variables with names starting with $\delta$ are $\mathcal{O}(\sqrt{E})$, because $\exp(-h[x_s+\delta x]/E)\approx\exp(-[h(x_s)/E]-[\delta x^2/(2E)])$. To simplify we will introduce the average and difference of the electron momenta,
\be\label{aveDiff12}
\delta{\bf p}_1=\delta{\bf P}-\frac{\delta{\bf p}}{2}
\qquad
\delta{\bf p}_2=\delta{\bf P}+\frac{\delta{\bf p}}{2} \;.
\ee
For the terms which are linear in $\delta t$ and $\delta t'$ we have 
\be
\begin{split}
&F_1({\bf p}_1,l_0)\delta t\approx[{\bf c}\cdot\delta{\bf p}_1+i\delta X]\delta t \\
&F'_1({\bf p}_1,{\bf p}_2,l_0)\approx[{\bf c}_1'\cdot\delta{\bf p}_1+{\bf c}_2'\cdot\delta{\bf p}_2-i\delta X]\delta t' \;,
\end{split}
\ee
where the coefficients ${\bf c}$, ${\bf c}'_1$ and ${\bf c}'_2$ are functions of ${\bf p}$.
We also have
\be
\begin{split}
&F_2({\bf p}_1)\delta t^2\approx F_2({\bf p}')\delta t^2 \\
&F_2'({\bf p}_1,{\bf p}_2)\delta t^{\prime2}\approx F_2({\bf p}',{\bf p}')\delta t^{\prime2} \;.
\end{split}
\ee  
In the pre-exponential part of the integrand we can set $\delta t\to0$ and $\delta t'\to0$.
The $\delta t$ and $\delta t'$ integrals are simple Gaussians,
\be
\int\ud\delta t\exp[F_1\delta t+F_2\delta t^2]=\sqrt{\frac{\pi}{-F_2}}\exp\left(-\frac{F_1^2}{4F_2}\right) 
\ee 
and similarly for $\delta t'$.

From the denominator of the photon propagator we have
\be
l^2+i\epsilon=X(X+2|{\bf l}|)+i\epsilon\approx2|{\bf l}|(X+i\epsilon) \;.
\ee
Since $F_1$ and $F_1'$ contain terms proportional to $X$, the photon-momentum integral now takes the form
\be\label{Xerfc}
\int\frac{\ud X}{2\pi}\frac{\exp[-aX^2+ibX]}{X+i\epsilon}=-\frac{i}{2}\text{erfc}\left(\frac{b}{2\sqrt{a}}\right) \;,
\ee
where $\text{erfc}(x)=1-\text{erf}(x)$ is the complementary error function, and
\be\label{ab}
a=-\frac{A}{p_1' A'}(p_1^{\prime2}A+p_3')
\qquad
b=\frac{1}{p_1' A'}(p_3'\delta p_1-p_1'\delta p_3) \;,
\ee
where $A$ and $A'$ are evaluated at $\tilde{t}$, so from~\eqref{ftildeutilde}
\be
A(\tilde{t})=i\sqrt{1+\frac{p_3^2}{p_1^2}} \;.
\ee
These error functions~\eqref{Xerfc} come from the photon propagator and are hence a new feature which we did not encounter for Schwinger or Breit-Wheeler pair production.

The probability~\eqref{Pint123} has one part
\be
|M^{12}|^2+|M^{21}|^2 
\ee
which we call the direct part, and
\be
-2\text{Re }M^{21}\bar{M}^{12}
\ee
which we call the exchange part. We treat the direct and exchange parts separately. We see from~\eqref{aveDiff12} that the exponential and error-function parts of $M^{21}$ are obtained from the corresponding parts of $M^{12}$ by changing the sign of $\delta{\bf p}\to-\delta{\bf p}$.

The exponential terms we are left with after performing the $t$, $t'$ and $X$ integrals are unchanged by $\delta{\bf p}\to-\delta{\bf p}$, so those terms are the same for both the direct and the exchange parts. By expanding these terms around the saddle point for the momenta, squaring the amplitude and summing over spins we find the momentum spectrum
\be\label{PdirSpec}
\begin{split}
P_{\rm dir}(\delta{\bf p}_1,\delta{\bf p}_2)&=\frac{1}{8}\left(\left|\text{erfc}(c)\right|^2+\left|\text{erfc}(-c)\right|^2\right)\\
&\times\frac{\alpha^2(1+3p_1^{\prime2})^2\exp\left(\mathcal{E}-\frac{Q}{\omega}\right)}{32\pi^2A^{\prime2}(\tilde{t})m_\LCperp m_\LCperp^{\prime3}p_1^{\prime2}(p_1^{\prime2}+p_3^{\prime2})} \;,
\end{split}
\ee
\be\label{PexSpec}
\begin{split}
P_{\rm ex}(\delta{\bf p}_1,\delta{\bf p}_2)&=\frac{1}{4}\text{Re}\left[\text{erfc}(c)\text{erfc}(-c^*)\right]\\
&\times(-1)\frac{81+54p_1^2+13p_1^4}{18(3+p_1^2)^2}\\
&\times\frac{\alpha^2(1+3p_1^{\prime2})^2\exp\left(\mathcal{E}-\frac{Q}{\omega}\right)}{32\pi^2A^{\prime2}(\tilde{t})m_\LCperp m_\LCperp^{\prime3}p_1^{\prime2}(p_1^{\prime2}+p_3^{\prime2})} \;,
\end{split}
\ee
where $c=b/(2\sqrt{a})$, $a$ and $b$ are given by~\eqref{ab}, $c^*$ is the complex conjugate of $c$, the exponential part is given by
\be\label{Edef}
\mathcal{E}=-\frac{2}{\omega}[3\mathcal{J}_0(p')-\mathcal{J}_0(p)] \;,
\ee
the widths of the momentum peaks are approximately Gaussian and determined by
\be\label{Qdef}
\begin{split}
&Q=\mathcal{J}_1\left(\frac{\delta p_2^2}{2}+6\delta P_2^2\right)+(\mathcal{J}_1-p_1^{\prime2}\mathcal{J}_2)\left(\frac{\delta p_1^2}{2}+6\delta P_1^2\right)\\
&+m^{\prime2}\mathcal{J}_2\left(\frac{\delta p_3^2}{2}+6\delta P_3^2\right)+2p_1'\mathcal{J}_A\left(\frac{\delta p_1\delta p_3}{2}+6\delta P_1\delta P_3\right)\\
&-\frac{\Omega}{p_1^{\prime2}}\left(\frac{1}{2}[p'_3\delta p_1-p'_1\delta p_3]^2+6[p'_3\delta P_1-p'_1\delta P_3]^2\right) \;,
\end{split}
\ee 
\be
\Omega=\left(1+\frac{m^{\prime2}p_3^{\prime2}}{p_1^{\prime4}}\right)^{-1}\frac{\omega}{p'_1 A'(\tilde{t})} \;,
\ee
the argument of $\mathcal{J}_{1,2,A}$ in $Q$ is $p'$, where $\mathcal{J}_n$ and $\mathcal{J}_A$ are a set of imaginary-time integrals defined, as in the Breit-Wheeler case~\cite{Brodin:2022dkd}, by
\be\label{JnDef}
\begin{split}
\mathcal{J}_n(p)=\frac{1}{2}\int_0^{\tilde{u}}\ud u\Big(&\left[m_\LCperp^2+(A_3-p_3)^2\right]^{\frac{1}{2}-n}\\
&+\left[m_\LCperp^2+(A_3+p_3)^2\right]^{\frac{1}{2}-n}\Big) 
\end{split}
\ee 
and
\be\label{JAdef}
\begin{split}
\mathcal{J}_A(p)=\frac{1}{2}\int_0^{\tilde{u}}\ud u\Bigg(&\frac{A_3-p_3}{\left[m_\LCperp^2+(A_3-p_3)^2\right]^{\frac{3}{2}}}\\
&-\frac{A_3+p_3}{\left[m_\LCperp^2+(A_3+p_3)^2\right]^{\frac{3}{2}}}\Bigg) \;.
\end{split}
\ee
Note that here we have functions which depend nontrivially on the longitudinal, $p_3$, and the transverse momentum component, $p_\LCperp$, separately, while in the plane-wave case we just have nontrivial dependence on the lightfront longitudinal momentum $p_0-p_3$. We will show in the next subsection that these expressions simplify considerably in the plane-wave limit.

We now turn to the integrated probability
\be\label{spec3}
P=\int\ud^3\delta{\bf p}_2\ud^3\delta{\bf p}_1 P(\delta{\bf p}_1,\delta{\bf p}_2) \;.
\ee
The momentum integrals over $\delta{\bf P}$ and $\delta p_2$ are standard Gaussians. But from~\eqref{Qdef} and~\eqref{ab} we see that the $\delta p_1$ and $\delta p_3$ integrals are more complicated. To perform these integrals we begin by changing variables from
\be\label{deltapUV}
\delta p_1=\frac{U+V}{p'_3}
\qquad
\delta p_3=\frac{U-V}{p'_1}
\ee  
to $U$ and $V$. The $U$ integral is Gaussian and we can make a shift $U\to U+\text{const.}V$ to remove the $UV$ cross term from the exponent, after which the $U$ and $V$ terms in $Q$ becomes
\be
\frac{\mathcal{T}_2}{2p_3^{\prime2}\Omega}U^2+\frac{2\Omega}{p_1^{\prime2}}\left(\frac{\mathcal{T}_1}{\mathcal{T}_2}-1\right)V^2 \;,
\ee 
where
\be\label{T1}
\mathcal{T}_1=m_\LCperp^{\prime2}(\mathcal{J}_1-p_\LCperp^{\prime2}\mathcal{J}_2)\mathcal{J}_2-p_\LCperp^{\prime2}\mathcal{J}_A^2 
\ee
and
\be\label{T2}
\mathcal{T}_2=\frac{\mathcal{J}_1-p_\LCperp^{\prime2}\mathcal{J}_2+2p'_\LCpara\mathcal{J}_A+\frac{m_\LCperp^{\prime2}p_\LCpara^{\prime2}}{p_\LCperp^{\prime2}}\mathcal{J}_2}{1+\frac{m_\LCperp^{\prime2}p_\LCpara^{\prime2}}{p_\LCperp^{\prime4}}}\frac{\omega}{p'_\LCperp A'(\tilde{t})} \;.
\ee
In contrast to the other momentum integrals, $V$ appears in the argument of the error functions, so even when expanding around $V$'s saddle point, we have something more nontrivial than a simple Gaussian integral. To perform the $V$ integral we need
\be\label{Ec1c2}
\begin{split}
E(c_1,c_2)&=\frac{1}{4}\int\ud x\,\text{erfc}(c_1x)\text{erfc}(c_2x)e^{-x^2}\\
&=\frac{1}{2\sqrt{\pi}}\text{arccos}\left(-\frac{c_1c_2}{\sqrt{(1+c_1^2)(1+c_2^2)}}\right) \;,
\end{split}
\ee
which one can prove by differentiating both sides with respect to $c_1$ or $c_2$. To calculate the first row in~\eqref{PdirSpec} we have $c_1=c_2^*\propto c$, and for~\eqref{PexSpec} we have $c_1=-c_2^*\propto c$, so for the direct part we have
\be
\frac{1}{8}\left(\left|\text{erfc}(c)\right|^2+\left|\text{erfc}(-c)\right|^2\right)
\to\frac{\text{arccos}(-Y)}{2\sqrt{\pi}}
\ee
and for the exchange term
\be
\frac{1}{4}\text{Re}\left[\text{erfc}(c)\text{erfc}(-c^*)\right]
\to\frac{\text{arccos}(Y)}{2\sqrt{\pi}} \;,
\ee
where
\be\label{YT1T2}
Y=\left(1+\frac{4\left[1+\frac{p_\LCpara^{\prime2}}{p_\LCperp^{\prime2}}\right]}{1+\frac{m_\LCperp^{\prime2}p_\LCpara^{\prime2}}{p_\LCperp^{\prime4}}}\frac{\mathcal{T}_1}{\mathcal{T}_2}\left[\frac{\mathcal{T}_1}{\mathcal{T}_2}-1\right]\right)^{-1/2} \;,
\ee
where $\mathcal{T}_1$ and $\mathcal{T}_2$ are given by~\eqref{T1} and~\eqref{T2}.
The remaining integrals over $U$, $\delta p_2$ and $\delta{\bf P}$ are Gaussian. 

Thus, for the direct part of the integrated probability we find, assuming $p_\LCperp>0$ and $p_\LCpara>0$,
\be\label{generalPdir}
\begin{split}
&P_{\rm dir}=\\
&\frac{\alpha^2\omega(1+3p_\LCperp^{\prime2})^2\text{arccos}(-Y)\exp\left\{-\frac{2}{\omega}[3\mathcal{J}_0(p')-\mathcal{J}_0(p)]\right\}}{192\sqrt{3}[A'(\tilde{t})/\omega]^2p_\LCperp^{\prime 2}(p_\LCperp^{\prime 2}+p_\LCpara^{\prime 2})m_\LCperp m_\LCperp^{\prime3}\mathcal{J}_1(\mathcal{T}_1-\mathcal{T}_2)}
\end{split}
\ee
and for the exchange term we find
\be\label{generalPex}
P_{\rm ex}=-\frac{81+54p_\LCperp^2+13p_\LCperp^4}{18(3+p_\LCperp^2)^2}\frac{\text{arccos}(Y)}{\text{arccos}(-Y)}P_{\rm dir} \;.
\ee

The general results~\eqref{generalPdir} and~\eqref{generalPex} might look quite complicated, but we can already see that in this $E\ll1$ regime the direct and exchange parts are on the same order of magnitude, unless some of the other parameters is very small or large. This is an important observation, because the exchange term has historically been neglected. It was calculated analytically for the first time in~\cite{Dinu:2017uoj,King:2018ibi} for plane waves.

\subsection{High-energy/plane-wave limit}

In the limit where either both $p_\LCperp$ and $p_\LCpara$ are large or where only $p_\LCperp$ is large, the basic time integrals~\eqref{JnDef} and~\eqref{JAdef} become
\be\label{JnPW}
\begin{split}
\mathcal{J}_n&\approx(p_1^{\prime2}+p_3^{\prime2})^{\frac{1}{2}-n}\bigg\{\tilde{u}\\
&+\frac{1-2n}{2(p_1^{\prime2}+p_3^{\prime2})}\left(\tilde{u}+\left[2n\frac{p_3^{\prime2}}{p_1^{\prime2}}-1\right]\mathcal{A}^2\mathcal{J}_{\rm PW}\right)\bigg\}
\end{split}
\ee
and
\be\label{JAPW}
\begin{split}
\mathcal{J}_A&\approx-\frac{p'_3}{(p_1^{\prime2}+p_3^{\prime2})^{3/2}}\bigg\{\tilde{u}\\
&-\frac{3}{2(p_1^{\prime2}+p_3^{\prime2})}\left(\tilde{u}+\left[2\frac{p_3^{\prime2}}{p_1^{\prime2}}-3\right]\mathcal{A}^2\mathcal{J}_{\rm PW}\right)\bigg\} \;,
\end{split}
\ee
where the first and second rows give, respectively, the leading and next-to-leading orders in $1/p$, $\mathcal{A}$ is a Lorentz invariant version of $a_0=E/\omega$,
\be\label{Adef}
\mathcal{A}=\frac{p_\LCperp}{\sqrt{p_\LCperp^2+p_\LCpara^2}}a_0 \;,
\ee
and only one nontrivial time integral is left
\be\label{JPWdef}
\mathcal{J}_{\rm PW}=\int_0^{\tilde{u}}\ud u\tilde{f}^2(u) \;.
\ee
We need the NLO terms because the LO terms cancel in several places.
For the exponential part of the spectrum we have
\be\label{EPW}
\mathcal{E}\approx-\frac{8\mathcal{A}}{\chi}(\tilde{u}-\mathcal{A}^2\mathcal{J}_{\rm PW}) \;,
\ee
where $\chi=\sqrt{-(F^{\mu\nu}p_\nu)^2}=Em_\LCperp\approx Ep_\LCperp$ is the quantum nonlinearity parameter, and
\be\label{QPW}
\begin{split}
Q\approx&\frac{\tilde{u}}{\sqrt{p_1^{\prime2}+p_3^{\prime2}}}\left(\frac{\delta p_2^2}{2}+6\delta P_2^2\right)\\
+&\frac{2(\mathcal{A}\tilde{u}\tilde{f}'-1)}{\mathcal{A}\tilde{f}'(p_1^{\prime2}+p_3^{\prime2})^{3/2}}(V^2+12\tilde{V}^2)\\
+&\frac{\tilde{u}-\mathcal{A}^2\mathcal{J}}{2p_1^{\prime2}p_3^{\prime2}\sqrt{p_1^{\prime2}+p_3^{\prime2}}}(U^2+12\tilde{U}^2) \;,
\end{split}
\ee
where $\tilde{f}'=\tilde{f}'(\tilde{u})$, $U$ and $V$ are given by~\eqref{deltapUV} and $\tilde{U}$ and $\tilde{V}$ by
\be
\delta P_1=\frac{\tilde{U}+\tilde{V}}{p'_3}
\qquad
\delta P_3=\frac{\tilde{U}-\tilde{V}}{p'_1} \;.
\ee
For the argument of the error function we find
\be\label{baPW}
c\approx\frac{1}{\sqrt{2(\mathcal{A}\tilde{u}\tilde{f}'-1)}}\left[\frac{2(\mathcal{A}\tilde{u}\tilde{f}'-1)}{\omega\mathcal{A}\tilde{f}'(p_1^{\prime2}+p_3^{\prime2})^{3/2}}V^2\right]^{1/2} \;.
\ee
Since $c$ is real we have
\be\label{erfDirPW}
|\text{erfc}(c)|^2+|\text{erfc}(-c)|^2
=2\left[1+\text{erf}^2(c)\right]
\ee
and
\be\label{erfExPW}
\text{Re}\left[\text{erfc}(c)\text{erfc}(-c^*)\right]=1-\text{erf}^2(c) \;.
\ee
The error functions in~\eqref{erfDirPW} makes the direct part of the width of $V$ wider than it would have been if it had been described by just the Gaussian factor~\eqref{QPW}, while~\eqref{erfExPW} makes the exchange term narrower. 

For the total probability we find $Y$ either by first calculating
\be\label{TPW}
\mathcal{T}_1\approx\frac{\tilde{u}(\tilde{u}-\mathcal{A}^2\mathcal{J}_{\rm PW})}{(p_1^{\prime2}+p_3^{\prime2})^2}
\qquad
\mathcal{T}_2\approx\frac{\tilde{u}-\mathcal{A}^2\mathcal{J}_{\rm PW}}{\mathcal{A}\tilde{f}'(\tilde{t})(p_1^{\prime2}+p_3^{\prime2})^2} \;.
\ee
and then plugging~\eqref{TPW} into~\eqref{YT1T2}, or by comparing~\eqref{QPW} and~\eqref{baPW} with~\eqref{Ec1c2} to see that $c_1=c_2=1/\sqrt{2(\mathcal{A}\tilde{u}\tilde{f}'-1)}$. In either case we find
\be
Y\approx\frac{1}{2\mathcal{A}\tilde{u}\tilde{f}'-1} \;.
\ee  
Performing all the integrals gives the integrated probability
\be
\begin{split}
&\lim_{p_\LCperp\sim p_\LCpara\gg1}P_{\rm dir}\\
&=\frac{\alpha^2\chi\text{arccos}(-Y)\exp\left\{-\frac{8\mathcal{A}}{\chi}(\tilde{u}-\mathcal{A}^2\mathcal{J}_{\rm PW})\right\}}{192\sqrt{3}\mathcal{A}^2\tilde{u}\tilde{f}'(\tilde{u}-\mathcal{A}^2\mathcal{J}_{\rm PW})(\mathcal{A}\tilde{u}\tilde{f}'-1)} 
\end{split}
\ee
and
\be
\lim_{p_\LCperp\sim p_\LCpara\gg1}P_{\rm ex}=-\frac{13}{18}\frac{\text{arccos}(Y)}{\text{arccos}(-Y)}P_{\rm dir} \;.
\ee
This is very similar to the high-energy limit of Breit-Wheeler. In particular, $p_\LCpara$ only enters via $\mathcal{A}$.

\subsection{Comparison with plane-wave result}

We will now show that one can obtain the results in the previous section by starting with Volkov solutions for plane-wave backgrounds. To determine what sort of plane wave we should compare with, we begin by making a Lorentz boost to the rest frame of the incoming electron,
\be
\Lambda_bp=(1,0,0,0) \;.
\ee
The argument of a field transforms as $f(\omega x^\mu)\to f(\omega[\Lambda_b^{-1}x]^\mu)$, where $\omega$ is a frequency scale in the lab frame, which we have included to make the arguments dimensionless. For a field which only depends on time in the lab frame, we have
\be
t\to p_1x+p_3z+\sqrt{p_1^2+p_3^2}t+\mathcal{O}(1/p) \;.
\ee
To simplify we rotate ($\Lambda_r$) around the $y$ axis so that the combination $(\Lambda_r\Lambda_b)^{-1}$ gives
\be
f(\omega t)\to f(\omega_{\rm PW}(t+z))
\qquad
\omega_{\rm PW}=\sqrt{p_1^2+p_3^2}\omega \;,
\ee
which shows that the frequency in the rest (or plane-wave) frame is $\mathcal{O}(p)$ times higher than in the lab frame.

The field tensor $F^{\mu\nu}$ transforms from an electric field ${\bf E}=E{\bf e}_z$ in the lab frame to ${\bf E}_{\rm PW}=-p^1E{\bf e}_x$ and ${\bf B}_{\rm PW}=p^1E{\bf e}_y$ in the rest frame. Thus, in the rest frame we have, to leading order in $\mathcal{O}(1/p)$, a plane wave which can be described by the following gauge potential
\be
a_\mu=\delta_{\mu1}a_0^{\rm PW}f(\phi)
\qquad
\phi=\omega_{\rm PW}(t+z) \;,
\ee
where 
\be
a_0^{\rm PW}=\frac{E_{\rm PW}}{\omega_{\rm PW}}=\frac{p_1 E}{\sqrt{p_1^2+p_3^2}\omega}=\mathcal{A} \;,
\ee
where $\mathcal{A}$ is the same as in the previous section~\eqref{Adef}.

To facilitate comparison of the momentum widths in~\eqref{QPW} with the plane-wave results, we rewrite~\eqref{QPW} as
\be\label{QPW2}
\begin{split}
\frac{Q}{\omega}\approx&\frac{3\mathcal{A}}{2\chi}\tilde{u}\delta p_2^2
+\frac{3}{2\chi}\left(\mathcal{A}\tilde{u}-\frac{1}{\tilde{f}'}\right)\left(\frac{p_3\delta p_1-p_1\delta p_3}{\sqrt{p_1^2+p_3^2}}\right)^2\\
+&\frac{3^3\mathcal{A}}{2\chi}(\tilde{u}-\mathcal{A}^2\mathcal{J}_{\rm PW})\frac{1}{4}\left(\frac{\delta p_1}{p_1}+\frac{\delta p_3}{p_3}\right)^2
+12(\delta p\to\delta P)\;.
\end{split}
\ee
We can now transform this expression to the rest frame using $\Lambda_b$ and $\Lambda_r$ again. Another approach, to obtain the same transformation without explicitly using $\Lambda_b$ and $\Lambda_r$, is to first rewrite~\eqref{QPW2} in a coordinate-independent way. We present this approach as it might make some features clearer. In the lab frame we have
\be
F^{\mu\nu}=\begin{pmatrix}
0&0&0&-E\\
0&0&0&0\\
0&0&0&0\\
E&0&0&0\end{pmatrix}
\qquad
G^{\mu\nu}=\begin{pmatrix}
0&0&0&0\\
0&0&E&0\\
0&-E&0&0\\
0&0&0&0\end{pmatrix} 
\ee
and the field depends on $\omega nx=\omega t$, where $n^\mu=(1,0,0,0)$. Let $q^\mu\approx(\sqrt{q_1^2+q_3^2},q^1,0,q^3)$ be the saddle-point value of the momentum of any of the three final-state particles. Then (recall the saddle-point momenta are all parallel ${\bf p}_1={\bf p}_2={\bf p}_3={\bf p}/3$)
\be
\delta q^\mu=\left(\frac{p_1\delta q_1+p_3\delta q_3}{\sqrt{p_1^2+p_3^2}},\delta q^1,\delta q^2,\delta q^3\right) \;.
\ee
The basis vectors we have in the lab frame can be expressed in a coordinate-independent way as
\be\label{coordIndepBasis}
\begin{split}
v_{(1)}^\mu&=(G^2p)^\mu=(0,-p^1E^2,0,0)\\
v_{(2)}^\mu&=G^{\mu\nu}p_\nu=(0,0,p^1E,0)\\
v_{(3)}^\mu&=(F^2p)^\mu-n^\mu\frac{v_{(1)}^2}{v_{(2)}^2}np=(0,0,0,p^3E^2) \;.
\end{split}
\ee
Instead of this $v_{(3)}$, one might be tempted to use $(Fn)^\mu=(0,0,0,E)$, as it might at first seem simpler. However, using $(Fn)^\mu$ would make it harder to take the plane-wave limit, since $(Fn)^\mu=0$ for plane waves. With $\chi=\sqrt{-(Fp)^2}\approx|p_1|E$ we have
\be
\begin{split}
\delta q_2^2=\frac{(v_{(2)}\delta q)^2}{\chi^2}&=\frac{(pG\delta q)^2}{\chi^2}\\
\left(\frac{p_3\delta p_1-p_1\delta p_3}{\sqrt{p_1^2+p_3^2}}\right)^2&=\frac{(pF\delta q)^2}{\chi^2}\\
\frac{1}{4}\left(\frac{\delta p_1}{p_1}+\frac{\delta p_3}{p_3}\right)^2&=\frac{1}{4}\left(\frac{v_{(1)}\delta q}{v_{(1)}p}+\frac{v_{(3)}\delta q}{v_{(3)}p}\right)^2 \;.
\end{split}
\ee
These coordinate-independent expressions can now be evaluated in any frame.

Applying the Lorentz transformation $\Lambda_r\Lambda_b$ gives to leading order in $1/p$
\be
F^{\mu\nu}_{\rm PW}=[\Lambda_r\Lambda_b F(\Lambda_r\Lambda_b)^T]^{\mu\nu}\approx
\begin{pmatrix}
    0&p^1E&0&0\\
    -p^1E&0&0&p^1E\\
    0&0&0&0\\
    0&-p^1E&0&0
\end{pmatrix}
\ee
and
\be
G_{\rm PW}^{\mu\nu}\approx
\begin{pmatrix}
    0&0&-p^1E&0\\
    0&0&0&0\\
    p^1E&0&0&-p^1E\\
    0&0&p^1E&0
\end{pmatrix} \;.
\ee
In the following we use $\delta q^\mu$ for the momenta as measured in the rest frame.
We still have
\be
\frac{(pG\delta q)^2}{\chi^2}\bigg|_{\rm PW}=\delta q_2^2 \;,
\ee
since the Lorentz transformation does not act on the component which is orthogonal to both ${\bf p}$ and ${\bf E}$. 
For the second term we have
\be
\frac{(pF\delta q)^2}{\chi^2}\bigg|_{\rm PW}=\delta q_1^2 \;.
\ee
$\delta q_1$ and $\delta q_2$ are the momentum components in the direction of the electric and magnetic components of the plane wave.
In the rest frame, and to leading order in $1/p$, two of the basis vectors in~\eqref{coordIndepBasis} become equal to the wave vector, $k^\mu$ such that $kx=\omega_{\rm PW}(t+z)$,
\be
[v_{(1)}^\mu]_{\rm PW}=[v_{(3)}^\mu]_{\rm PW}=\chi^2\frac{k^\mu}{\omega_{\rm PW}}=\chi^2(1,0,0,-1) \;,
\ee
which simplifies the last term
\be
\frac{1}{4}\left(\frac{v_{(1)}\delta q}{v_{(1)}p}+\frac{v_{(3)}\delta q}{v_{(3)}p}\right)^2\bigg|_{\rm PW}=\left(\frac{k\delta q}{kp}\right)^2 \;.
\ee
In the rest frame we have $k\delta q/kp=\delta q^3$, but writing it as $k\delta q/kp$ makes the results more transparent and conforms to our usual notation when dealing with plane waves. 

Thus, after the Lorentz transformation $\Lambda_r\Lambda_b$, the high-energy limit of the widths can be expressed as
\be\label{QPW3}
\begin{split}
\frac{Q}{\omega}\approx&\frac{3\mathcal{A}}{2\chi}\tilde{u}[\delta p_2^2+12\delta P_2^2]
+\frac{3}{2\chi}\left(\mathcal{A}\tilde{u}-\frac{1}{\tilde{f}'}\right)[\delta p_1^2+12\delta P_1^2]\\
+&\frac{3^3\mathcal{A}}{2\chi}(\tilde{u}-\mathcal{A}^2\mathcal{J}_{\rm PW})[\delta s^2+12\delta S^2]\;,
\end{split}
\ee
where we have introduced $\delta s=k\delta p/kp$ and $\delta S=k\delta P/kp$ to further facilitate comparison with the plane-wave results. In general, for an intermediate particle (e.g. the photon in trident) or a final state particle with momentum $q^\mu$, we refer to $kq=\omega(q_0-q_3)$ as a lightfront longitudinal momentum, and we usually work with the ratio $s=kq/kp$ (or $q=kl/kp$ for photons).

The argument of the error function becomes
\be\label{cPW}
c=\frac{\sqrt{3}\delta p_1}{2\sqrt{\chi\tilde{f}'}} \;.
\ee

At this point we have taken the high-energy limit of our results for $E(t)$ and expressed it in a way that looks like what one might expect if one started with a plane wave instead of $E(t)$. To show that this high-energy limit is in fact exactly equal to the corresponding plane-wave result, we start with Eq.~(11) in~\cite{Dinu:2017uoj}, which gives the amplitude for trident in a plane wave. That equation has two lightfront-time ($\phi=kx$) integrals: $\phi_1$ for the photon-emission vertex, and $\phi_2$ for the pair-production vertex. (There is also a lightfront-instantaneous term, with only one $\phi$ integral, but it does not contribute to leading order in $\chi\ll1$.) In~\cite{Dinu:2017uoj} we squared the amplitude and then performed the transverse momentum integrals exactly/analytically, which is possible since they are Gaussian. Then on the probability level there are up to four $\phi$ integrals. We showed how to perform those integrals with the saddle-point method. However, here we would like to study not just the integrated probability or the longitudinal momentum dependence, but also the dependence on the transverse momenta. Since we will not perform the integrals over the transverse momenta, there is no reason to first square the amplitude before performing the $\phi$ integrals. Thus, we will perform the $\phi$ integrals directly on the amplitude level starting with Eq.~(11) in~\cite{Dinu:2017uoj}.  

For a symmetric, linearly polarized plane wave with a single peak, given by a potential $a_\mu=\delta_{\mu,1}\mathcal{A}f(kx)$, where $f(-u)=f(u)$, we have a saddle-point at
\be\label{phi1phi2tildeu}
\phi_1=\phi_2=i\tilde{u}
\qquad
\tilde{f}(\tilde{u})=\frac{1}{\mathcal{A}} \;,
\ee
where $f(iy)=:i\tilde{f}(y)$, and for the momenta ${\bf p}_1={\bf p}_2={\bf p}_3={\bf p}/3$. Although we will not perform the momentum integrals, we consider only the region close to the saddle-point values of the momenta. This is not really a severe restriction, since the spectrum is anyway exponentially suppressed further away from the momentum saddle point. If we did not make this restriction, then the saddle point for the $\phi$ integrals would have been more complicated. One hint that this would be the case can be seen in the following.

As before, we change variable from ${\bf p}_1$ to $\delta{\bf p}_1={\bf p}_1-{\bf p}_{1,\text{saddle}}$ etc., afterwards we use ${\bf p}_1$ to denote ${\bf p}_{1,\text{saddle}}={\bf p}/3$, and similarly for the other variables. Because of the lightfront-time ordering step function, $\theta(\phi_2-\phi_1)$, we change from $\phi_1=\sigma-\theta/2$ and $\phi_2=\sigma+\theta/2$ to $\sigma$ and $\theta$. Expanding the exponential part of the integrand to quadratic order around the saddle points gives $e^{\varphi_0+\varphi_1+\varphi_2}$
\be
\varphi_0=-\frac{4\mathcal{A}}{\chi}(\tilde{u}-\mathcal{A}^2\mathcal{J}_{\rm PW}) \;,
\ee
\be
\varphi_1=\frac{12i\mathcal{A}^2}{\chi}\mathcal{J}_{\rm lin}\delta P_1
\ee
and
\be\label{varphi2PW}
\begin{split}
\varphi_2&=-\frac{\mathcal{A}^2\tilde{f}'}{\chi}\left[4\left(\delta\sigma-\frac{3\delta P_1}{2\mathcal{A}\tilde{f}'}+\frac{\delta\theta}{4}\right)^2
+\frac{3}{4}\left(\delta\theta-\frac{\delta p_1}{\mathcal{A}\tilde{f}'}\right)^2\right]\\
&-\frac{3\mathcal{A}\tilde{u}}{4\chi}\bigg[\delta p_2^2+12\delta P_2^2+\left(1-\frac{1}{\mathcal{A}\tilde{u}\tilde{f}'}\right)(\delta p_1^2+12\delta P_1^2)\\
&\hspace{1cm}+9\left(1-\frac{\mathcal{A}^2\mathcal{J}_{\rm PW}}{\tilde{u}}\right)(\delta s^2+12\delta S^2)\bigg]\\
&+\frac{9i\mathcal{A}^2\mathcal{J}_{\rm lin}}{\chi}\left(-\frac{1}{2}\delta s\delta p_1+2\delta S\delta P_1\right) \;,
\end{split}
\ee
where $\delta p$ and $\delta P$ are defined as in~\eqref{aveDiff12}, $\delta\sigma=\sigma-\sigma_{\rm saddle}=\sigma-i\tilde{u}$, $\delta\theta=\theta-\theta_{\rm saddle}=\theta$, and
\be
\mathcal{J}_{\rm lin}=\int_0^{\tilde{u}}\ud u\,\tilde{f}(u) \;.
\ee

We can already see that squaring the amplitude gives $2\varphi_0$ which agrees with~\eqref{EPW}, and from the second square brackets in~\eqref{varphi2PW} we recover~\eqref{QPW3}. Since $\varphi_1$ is purely imaginary, does not depend on the amplitude-level integration variables, and is invariant under swapping the two final-state electrons (which changes the sign of $\delta p$ and $\delta s$), it is just an overall phase for both amplitude terms and therefore drops out upon squaring the amplitude. The same holds for the last term in~\eqref{varphi2PW}.

The first row in~\eqref{varphi2PW} gives a hint that things would be more complicated if we did not assume that $\delta p$ and $\delta P$ are small. We would still be able to perform the $\phi$ integrals with the saddle-point method, but the saddle points would be nontrivial functions of the momenta, $\phi_s(\delta p,\delta P)$. The saddle point in~\eqref{phi1phi2tildeu} is actually for $\phi_s(0,0)$. By diagonalizing the first row in~\eqref{varphi2PW}, we see that the corrections to $\phi_s(0,0)$ are $\mathcal{O}(\delta p,\delta P)$. We have included this first-order correction not in $\sigma_{\rm saddle}$ and $\theta_{\rm saddle}$ but instead by allowing cross terms between $\delta\sigma$ and $\delta\theta$ with $\delta p$ and $\delta P$.  
To obtain~\eqref{varphi2PW} we have assumed that all the $\delta...$ variables are $\mathcal{O}(\sqrt{\chi})$, which includes both the ones we always integrate over, i.e. $\delta\sigma$ and $\delta\theta$, and the free variables for the spectrum, $\delta p$ and $\delta P$.   

To leading order in the saddle-point expansion we can just replace all the variables in the pre-exponential factors with their saddle-point values. This gives a simple integral for the average lightfront time,
\be
\int_{-\infty}^\infty\ud\sigma\exp\left\{-\frac{\mathcal{A}^2\tilde{f}'}{\chi}4(\delta\sigma-\dots)^2\right\}=\frac{1}{2\mathcal{A}}\sqrt{\frac{\pi\chi}{\tilde{f}'}} \;.
\ee
For the lightfront-time difference $\theta$ we have, because of lightfront-time ordering, a more nontrivial integral,
\be
\begin{split}
&\int_0^\infty\ud\theta\exp\left\{-\frac{\mathcal{A}^2\tilde{f}'}{\chi}\frac{3}{4}\left(\delta\theta-\frac{\delta p_1}{\mathcal{A}\tilde{f}'}\right)^2\right\}\\
&=\frac{2}{a_0}\sqrt{\frac{\pi\chi}{3\tilde{f}'}}\frac{1}{2}\text{erfc}(-c) \;,
\end{split}
\ee
where $c$ is the same as in~\eqref{cPW}.

One could of course calculate the spin-part of the prefactor without choosing any particular representation for the spinors and Dirac matrices. However, it is convenient to use the basis described in~\cite{Dinu:2019pau}. We average over the initial spin and sum over all the final spins. 

We finally obtain the spectrum (note that the normalization here is different from~\eqref{spec3}, since we prefer to use the ratio $s=kq/kp$ when dealing directly with plane waves)
\be\label{PspecPW}
P=\int\ud^2\delta p_\LCperp\ud\delta s\,\ud^2\delta P_\LCperp\ud S\, P(\delta p,\Delta P,\delta s,\delta S) \;,
\ee
where the direct and exchange parts, $P=P_{\rm dir}+P_{\rm ex}$, are given by
\be\label{specDirPW}
P_{\rm dir}=\frac{3^5}{2^6}\left(\frac{\alpha}{\pi\chi\tilde{f}'}\right)^2\frac{1}{4}[\text{erfc}^2(c)+\text{erfc}^2(-c)]e^\varphi
\ee
and
\be
P_{\rm ex}=-\frac{13}{18}\frac{3^5}{2^6}\left(\frac{\alpha}{\pi\chi\tilde{f}'}\right)^2\frac{1}{2}\text{erfc}(c)\text{erfc}(-c)e^\varphi \;,
\ee
where $c$ is given by~\eqref{cPW} and the exponential part of both contributions is given by
\be
\varphi=\eqref{EPW}-\eqref{QPW3} \;.
\ee
The results agree perfectly with the high-energy limit of the $E(t)$ results.

\subsection{Locally-constant-field limit}

Another important limit is the locally-constant-field (LCF) limit, where the field varies slowly compared to the field strength, in the sense $\gamma\ll1$. Here we begin by solving~\eqref{ftildeutilde} perturbatively in $\gamma\ll1$,
\be\label{tildeuLCF}
\tilde{u}\approx\frac{1}{f'}\left(1+\frac{p_3^2}{p_1^2}\right)^{1/2}\gamma+\frac{f'''}{6f^{\prime4}}\left(1+\frac{p_3^2}{p_1^2}\right)^{3/2}\gamma^3 \;,
\ee 
where $f'=f'(0)>0$ and $f'''=f'''(0)<0$. We need the NLO because the LO terms cancel in some places. To approximate the basic time integrals~\eqref{JnDef} and~\eqref{JAdef}, we first change integration variable from $u=\tilde{u}v$ to $v$ and then expand the integrand in $\gamma\ll1$ with $v=\mathcal{O}(1)$. We need to perform one nontrivial time integral,
\be
I(a,b,c)=\int_0^1\frac{\ud v}{2}\left(\sqrt{a+ibv-cv^2}+\sqrt{a-ibv-cv^2}\right) \;.
\ee
The (analytical) result, which involves $\text{arccot}$ and square roots, is somewhat complicated when expressed in terms of general $a$, $b$ and $c$. However, all the other $v$ integrals can be obtained by taking derivatives of $I$ with respect to $a$, $b$ or $c$. After taking the derivatives, the results simplify greatly for the actual values of $a$, $b$ or $c$.
We find for the basic time integrals (with momentum argument $p'$)
\be
\mathcal{J}_0\approx\frac{\gamma}{2f'}\left(\frac{2p_3^{\prime2}}{p_1'}+p_1'+m_\LCperp^{\prime2}\text{arccot }p_1'\right) \;,
\ee
\be
\begin{split}
\mathcal{J}_1\approx&\frac{\gamma}{f'}\text{arccot }p_1'\\
+&\frac{f'''\gamma^3}{4f^{\prime4}}\left[\frac{2p_3^{\prime2}}{p_1'}-p_1'+(m_\LCperp^{\prime2}-2p_3^{\prime2})\text{arccot }p_1'\right] \;,
\end{split}
\ee
\be
\begin{split}
\mathcal{J}_2&\approx\frac{\gamma}{f'}\frac{1}{p_1' m_\LCperp^{\prime2}}\left(1+\frac{m_\LCperp^{\prime2}p_3^{\prime2}}{p_1^{\prime4}}\right)^{-1}\\
+&\frac{f'''\gamma^3}{2f^{\prime4}}\left[\frac{p_1'}{m_\LCperp^{\prime2}}\left(1+\frac{p_1^{\prime2}+p_3^{\prime2}}{p_1^{\prime4}+m_\LCperp^{\prime2}p_3^{\prime2}}\right)-\text{arccot }p_1'\right] 
\end{split}
\ee
and
\be
\begin{split}
\mathcal{J}_A&\approx-\frac{\gamma}{f'}\frac{p_3'}{p_1^{\prime3}}\left(1+\frac{m_\LCperp^{\prime2}p_3^{\prime2}}{p_1^{\prime4}}\right)^{-1}\\
+&\frac{f'''\gamma^3p_3'}{f^{\prime4}}\left[\frac{1}{p_1'}\left(1-\frac{1}{2}\frac{p_1^{\prime2}+p_3^{\prime2}}{p_1^{\prime4}+m_\LCperp^{\prime2}p_3^{\prime2}}\right)-\text{arccot }p_1'\right] \;.
\end{split}
\ee

For the exponential part of the spectrum we thus find
\be
\mathcal{E}\approx-\frac{1}{Ef'}[3m_\LCperp^{\prime2}\text{arccot}(p'_\LCperp)-m_\LCperp^2\text{arccot}(p_\LCperp)] \;.
\ee
To approximate the momentum widths $Q$ we expand in $\gamma$ with $\delta p_1$, $\delta P_1$, $\delta p_2$ and $\delta P_2$ as $\mathcal{O}(\gamma^0)$, while $\delta p_3$ and $\delta P_3$ are $\mathcal{O}(\gamma^{-1})$. The inverse scaling means the spectrum becomes wider as $\gamma\to0$. For the coefficient in front of $\delta p_3^2+12\delta P_3^2$, we need 
\be\label{NLOcomb}
m_\LCperp^{\prime2}\mathcal{J}_2^{\rm NLO}-\Omega^{\rm NLO} \;,
\ee
where $\mathcal{J}_2\approx\gamma\mathcal{J}_2^{\rm LO}+\gamma^3\mathcal{J}_2^{\rm NLO}$ and $\Omega\approx\gamma\Omega^{\rm LO}+\gamma^3\Omega^{\rm NLO}$. $\Omega^{\rm NLO}$ cancels the part of $\mathcal{J}_2^{\rm NLO}$ which involves $p_3'$. We find the LCF limit of the widths
\be\label{LCFQ}
\begin{split}
\frac{Q}{\omega}&\approx\frac{1}{Ef'}\text{arccot}p_1'\left(\frac{\delta p_2^2}{2}+6\delta P_2^2\right)\\
&+\frac{1}{Ef'm_\LCperp^{\prime2}}(m_\LCperp^{\prime2}\text{arccot}p_1'-p_1')\left(\frac{\delta p_1^2}{2}+6\delta P_1^2\right)\\
&+\frac{-f'''\gamma^2}{2Ef^{\prime4}}(m_\LCperp^{\prime2}\text{arccot}p_1'-p_1')\left(\frac{\delta p_3^2}{2}+6\delta P_3^2\right) \;.
\end{split}
\ee
From the first two rows we can see that the widths in the plane transverse to the field indeed scales as $\mathcal{O}(\gamma^0)$, while the third row shows that the longitudinal widths scale as $\mathcal{O}(\gamma^{-1})$. 

The argument of the error function is $c=\mathcal{O}(\gamma^{-1})$, so
\be\label{erfcLCF12}
\frac{1}{8}\left(\left|\text{erfc}(c)\right|^2+\left|\text{erfc}(-c)\right|^2\right)\approx\frac{1}{2} \;,
\ee
while $\text{Re}\left[\text{erfc}(c)\text{erfc}(-c^*)\right]\approx0$. Thus, to leading order we can neglect the exchange term. This is not surprising, because in the plane-wave case we know that the incoherent product gives the leading-order in the LCF limit, but the exchange term does not contribution to the incoherent product and therefore not to the leading order. 

For the integrated probability we calculate both the leading ($\mathcal{O}(\gamma^{-2})$) and the NLO ($\mathcal{O}(\gamma^{-1})$) terms. In the literature on trident in plane waves, the LO has been referred to as a two-step process, because it is equal to the incoherent product of the probabilities of nonlinear Compton scattering followed by Breit-Wheeler pair production. The NLO term is then referred to as a one-step process. The reason that NLO is only $\gamma$ rather than $\gamma^2$ times smaller than LO is due to $\text{arccos}(\pm Y)$, which came from the error function in~\eqref{Ec1c2}, which in turn came from the photon propagator. We first find
\be
\frac{\mathcal{T}_1}{\mathcal{T}_2}\approx1+\gamma^2f' p_1'\left(1+\frac{m_\LCperp^{\prime2}}{p_1^{\prime4}}\right)(m_\LCperp^{\prime2}\mathcal{J}_2^{\rm NLO}-\Omega^{\rm NLO}) \;,
\ee  
where we see the same combination of NLO terms as in~\eqref{NLOcomb}. From this we obtain
\be
Y\approx1-\frac{-f''' p_1'}{f^{\prime3}\mathcal{A}^2}(m_\LCperp^{\prime2}\text{arccot}p_1'-p_1') \;.
\ee
Then from
\be\label{arccosLCF}
\text{arccos}[-(1-\delta^2)]\approx\pi-\sqrt{2}\delta
\qquad
\text{arccos}[1-\delta^2]\approx\sqrt{2}\delta
\ee
we obtain terms which scale as $\gamma$ rather than $\gamma^2$ with respect to LO.
We thus find 
\be\label{PdirTwoOne}
P_{\rm dir}=P_{\rm two}+P_{\rm one}^{\rm dir}
\qquad
P_{\rm ex}=P_{\rm one}^{\rm ex} \;,
\ee
where the two-step term (which only has a direct part) is given by
\be\label{PtwoGeneral}
\begin{split}
P_{\rm two}=&\frac{\pi\alpha^2\mathcal{A}^2Ef^{\prime4}(1+3p_\LCperp^{\prime2})^2}{96\sqrt{3}(-f''')m_\LCperp m'_\LCperp p_\LCperp^{\prime4}\text{arccot}(p'_\LCperp)}\\
\times&\frac{e^{\mathcal{E}_{\rm LCF}}}{(m_\LCperp^{\prime2}\text{arccot}(p'_\LCperp)-p'_\LCperp)^2} \;,
\end{split}
\ee
the direct part of the one-step term by
\be\label{PoneDirGeneral}
\begin{split}
P_{\rm one}^{\rm dir}=&-\frac{\alpha^2\mathcal{A}E f^{\prime5/2}(1+3p_\LCperp^{\prime2})^2}{48\sqrt{-6f'''}m_\LCperp m'_\LCperp p_\LCperp^{\prime7/2}\text{arccot}(p'_\LCperp)}\\
\times&\frac{e^{\mathcal{E}_{\rm LCF}}}{(m_\LCperp^{\prime2}\text{arccot}(p'_\LCperp)-p'_\LCperp)^{3/2}}	
\end{split}	
\ee
and the exchange part of the one-step by
\be
P_{\rm one}^{\rm ex}=R(p_\LCperp)P_{\rm one}^{\rm dir} \;,
\ee
where
\be
R(p_\LCperp)=\frac{81+54p_\LCperp^2+13p_\LCperp^4}{18(3+p_\LCperp^2)^2} \;.
\ee
We note again that $p_\LCpara$ only appears via $\mathcal{A}$~\eqref{Adef}. Thus, in order to determine how important the correction to the leading order (two-step) is, we need to consider both the direct and the exchange parts of the one-step, because they are both on the same order of magnitude, which we see from the fact that the ratio $R$ increases monotonically from $R(0)=1/2$ to $R(\infty)=13/18$. 

\subsection{High energy and LCF}

In the previous subsections we considered first the high-energy (plane-wave) limit, and then the LCF ($\gamma\ll1$) limit. Is the $\gamma\ll1$ limit of the $p_0\gg1$ result equal to the $p_0\gg1$ limit of the $\gamma\ll1$ limit? Consider the exact result for a Sauter pulse, $f(u)=\tanh(u)$,
\be\label{J0sauter}
\begin{split}
\mathcal{J}_0&=\frac{1}{2\gamma}\sqrt{(1+\gamma p_3')^2+\gamma^2(1+p_1^{\prime2})}\\
&\times\text{arccos}\left[\frac{p_1^{\prime2}-\gamma p_3'}{\sqrt{(1+p_1^{\prime2})([1+\gamma^2]p_1^{\prime2}+[\gamma p_3']^2)}}\right]\\
&+\frac{1}{2\gamma}\sqrt{(1-\gamma p_3')^2+\gamma^2(1+p_1^{\prime2})}\\
&\times\text{arccos}\left[\frac{p_1^{\prime2}+\gamma p_3'}{\sqrt{(1+p_1^{\prime2})([1+\gamma^2]p_1^{\prime2}+[\gamma p_3']^2)}}\right]\\
&-\frac{1}{\gamma}\text{arctan}\frac{1}{p_1'} \;.
\end{split}
\ee
The other integrals can be obtained by differentiating $\mathcal{J}_0$ as in Eqs. (48)-(50) in~\cite{Brodin:2022dkd}. If we first make a $\gamma\ll1$ expansion then we would treat $\gamma p_3'$ and $\gamma p_1'$ as small, but if we first make a $p_0\gg1$ expansion we would instead treat them as large. This suggests that the $p_0\gg1$ limit might not commute with the $\gamma\ll1$ limit.

To investigate this, we consider the regime where $\mathcal{O}(L):=\mathcal{O}(p_1)=\mathcal{O}(p_3)=\mathcal{O}(1/\gamma)$ and $L\gg1$ is a bookkeeping parameter. 
Apart from the explicit appearance of $\gamma p_3'$ and $\gamma p_1'$ in~\eqref{J0sauter}, studying $p\sim1/\gamma$ is also motivated by cases where the acceleration of the fermions by the background field leads to changes in the momenta which are on the same order of magnitude as, or larger than, the incoming momentum.

We consider arbitrary field shapes, $f(u)$. We can still expand the turning point as in~\eqref{tildeuLCF}, and we again change integration variable from $u=\tilde{u}v$ to $v$. In contrast to the previous section, though, now we can further expand the integrands since $p$ is large. We need at least the next-to-leading order in $1/L$, since the leading order cancels in various places. We normalize the field strength $E$ such that $f'(0)=1$, and the frequency scale $\omega$ such that $f'''(0)=-2$. The reason for choosing $-2$ is because we want the Sauter pulse to be described by $f(u)=\tanh(u)$ rather than, say, $\tanh(u/2^{1/3})$. For the leading and next-to-leading orders ($\mathcal{O}(L^0)$ and $\mathcal{O}(L^{-2})$) we find
\be
\mathcal{J}_0=\frac{\sqrt{p_1^{\prime2}+p_3^{\prime2}}}{\mathcal{A}}\left[1+\frac{1}{3}\left(\frac{1}{p_1^{\prime2}+p_3^{\prime2}}-\frac{1}{\mathcal{A}^2}\right)\right]+\mathcal{O}(L^{-4}) \;.
\ee
Using this and similar results for the other integrals, we find, with the spectrum normalized as in~\eqref{spec3},
\be\label{specLarge}
P\approx P_{\rm dir}\approx\frac{3}{p_0^{\prime2}}\left(\frac{3\alpha}{8\pi\chi}\right)^2 e^{\varphi_0+\varphi_2} \;,
\ee
where $p_0'\approx\sqrt{p_1^{\prime2}+p_3^{\prime2}}$,
\be
\varphi_0=-\frac{16}{3\chi}+\frac{1}{\chi}\left(\frac{16}{15\mathcal{A}^2}+\frac{8}{9p_1^{\prime2}}\right)+\mathcal{O}\left(\frac{1}{\chi L^4}\right)
\ee
and
\be\label{varphi2large}
\begin{split}
\varphi_2=&-\frac{1}{\chi}\left[\frac{3\delta p_2^2}{2}+\frac{1}{\mathcal{A}^2}\left(\frac{p_3\delta p_1-p_1\delta p_3}{\sqrt{p_1^2+p_3^2}}\right)^2+\frac{\delta p_1^2}{p_1^{\prime2}}\right]\\
&+12(\delta p_j\to\delta P_j) \;.
\end{split}
\ee

The factor of $1/p_0^{\prime2}$ makes the result Lorentz invariant, given the normalization in~\eqref{spec3}. Using~\eqref{erfcLCF12} to simplify the prefactor in the plane-wave result~\eqref{specDirPW}, and recalling that we used a different normalization in the plane-wave case~\eqref{PspecPW}, we immediately see that the prefactor in~\eqref{specLarge} agrees with the LCF ($\gamma\ll1$) limit of the plane-wave result.  

The leading order in $\varphi_0$ agrees with the result for trident in a constant-crossed field~\cite{Ritus:1972nf}.
We have included the NLO $\mathcal{O}(1/[\chi L^2])$ in $\varphi_0$ since $\exp(-\text{``something small''}/\chi)$ is not necessarily $\approx1$. But actually, if one considers a regime where the $1/[\chi L^2]$ term is necessary for good precision, then $\chi\sim1/L^2$ would likely be so small that the overall exponential suppression, $\exp(-16/[3\chi])$, would make the process irrelevant for experiments. 

From $\varphi_2$ we see that the width of the spectrum in the plane spanned by the electric field and the fermion momenta scales as $\mathcal{O}(\delta p_1)=\mathcal{O}(\delta p_3)=\mathcal{O}(\sqrt{\chi}L)$, while in the perpendicular direction we have $\mathcal{O}(\delta p_2)=\mathcal{O}(\sqrt{\chi})$. Thus, the spectrum is relatively narrow in the perpendicular direction.

Next we compare~\eqref{varphi2large} with the $\gamma\ll1$ (or $\mathcal{A}\gg1$) limit of the $p_0\gg1$ (plane-wave) result~\eqref{QPW2},
\be\label{LCFofHighQ}
\begin{split}
&\lim_{\gamma\ll1}\lim_{p_0\gg1}\frac{Q}{\omega}=\lim_{\gamma\ll1}\eqref{QPW2}=\\
&\frac{1}{\chi}\left[\frac{3\delta p_2^2}{2}+\frac{1}{\mathcal{A}^2}\left(\frac{p_3\delta p_1-p_1\delta p_3}{\sqrt{p_1^2+p_3^2}}\right)^2+\frac{1}{4}\left(\frac{\delta p_1'}{p_1}+\frac{\delta p_3}{p_3'}\right)^2\right] \\
&+12(\delta p\to\delta P) \;.
\end{split}
\ee
The last term in the square brackets does not look like the corresponding term in~\eqref{varphi2large}. This is related to the question of whether the $\gamma\ll1$ and $p_0\gg1$ limits commute. But in this case we can actually get agreement, although it is somewhat subtle. If the dependence of the spectrum on a single momentum variable $\delta q$ scales as $e^{-\delta q^2/\epsilon}$, then a natural interval, e.g. for plotting the spectrum, would be something like $-\sqrt{5\epsilon}\lesssim\delta q\lesssim\sqrt{5\epsilon}$. In the present case we have $\epsilon\propto\chi$ for all momentum variables. However, the scaling of $\epsilon$ with respect to $p$ is not the same for the two round brackets in~\eqref{QPW2} for $\delta_1$ and $\delta p_3$. To see how this can be used, we first define 
\be
x:=\frac{p_3\delta p_1-p_1\delta p_3}{\sqrt{p_1^2+p_3^2}}
\qquad
y:=\frac{1}{2}\left(\frac{\delta p_1}{p_1}+\frac{\delta p_3}{p_3}\right) \;,
\ee
so that the natural range is independent of $p$ for both $x$ and $y$. One of the two $\delta p_{1,3}$ terms in~\eqref{varphi2large} is already $x^2$. For the other we have
\be
\left(\frac{\delta p_1}{p_1'}\right)^2=\left(y+\frac{\sqrt{p_1^{\prime2}+p_3^{\prime2}}}{p_1'p_3'}\frac{x}{2}\right)^2=y^2+\mathcal{O}(p^{-1}) \;.
\ee
This result can also be seen immediately by noticing that $x$ scales as $p^0$, while $y$ scales as $1/p$, so the $x$ term is more exponentially suppressed for generic values of $\delta p_1$ and $\delta p_3$, which means that the $x$ term basically serves as a delta function for the $y$ term, i.e. one can set $p_3\delta p_1-p_1\delta p_3=0$ in $y$. 
Since~\eqref{QPW2} has been obtained by expanding to leading order in $1/p$ only, one can replace $y^2$ in~\eqref{QPW2} with $(\delta p_1/p_1')^2$, since the difference is higher order in $1/p$ in the natural range of $\delta p_1$ and $\delta p_3$. But if one only looks at~\eqref{QPW2} then one might not see any obvious reason for making that replacement. And in~\eqref{LCFofHighQ} one definitely does not see any reason for the replacement, because both $x^2/\mathcal{A}^2$ and $y^2$ scale as $1/L^2$ for $\mathcal{A}\sim p\sim L$, so the $x$ term does not serve as a delta function for the $y$ term.
Thus, it is possible to make a replacement in the $p_0\gg1$ limit~\eqref{QPW2} such that the $\gamma\ll1$ limit of that ``improved'' $p_0\gg1$ limit agrees with the correct $p_0\sim1/\gamma\gg1$ limit~\eqref{varphi2large}. However, if we did not already know~\eqref{varphi2large}, then there would not have been any obvious reason for making that replacement. And if one does not make that replacement, then one would be led to~\eqref{LCFofHighQ} instead of~\eqref{varphi2large}. 
This raises some questions about how much one can rely on plane-wave results if one is interested in the $p_0\sim1/\gamma\gg1$ regime.   

If we take the limits in the opposite order we find
\be
\begin{split}
&\lim_{p_0\gg1}\lim_{\gamma\ll1}\frac{Q}{\omega}=\lim_{p_0\gg1}\eqref{LCFQ}=\\
&\frac{1}{\chi}\left[\frac{3}{2}\delta p_2^2+\gamma^2\delta p_3^2+\frac{\delta p_1^2}{p_1^{\prime2}}\right]+12(\delta p\to\delta P) \;,
\end{split}
\ee
which is also different from $p_0\sim1/\gamma\gg1$ limit~\eqref{varphi2large}. It is again possible to ``fix''~\eqref{LCFQ} so that the $p_0\gg1$ limit agrees with~\eqref{varphi2large}. Since~\eqref{LCFQ} is only the leading order in $\gamma\ll1$ we can replace
\be
\gamma^2\delta p_3^2=\frac{1}{\mathcal{A}^2}\left(\frac{p_1\delta p_3}{\sqrt{p_1^2+p_3^2}}\right)^2\approx\frac{1}{\mathcal{A}^2}\left(\frac{p_3\delta p_1-p_1\delta p_3}{\sqrt{p_1^2+p_3^2}}\right)^2 \;,
\ee
where the approximate sign is justified because the natural range for $\delta p_1$ scales as $\gamma^0$, which is much shorter than the natural range for $\delta p_3$, which scales as $\gamma^{-1}$. Making this replacement in~\eqref{LCFQ} and then taking the $p_0\gg1$ limit gives agreement with the $p_0\sim1/\gamma\gg1$ limit~\eqref{varphi2large}. But, again, if we did not already have~\eqref{varphi2large}, it would not have been obvious that~\eqref{LCFQ} ``should'' be improved like that.  

Thus, naively taking the high-energy $p_0\gg1$ limit of the LCF $\gamma\ll1$ limit, or the LCF limit of the high-energy limit, does not agree with the $p_0\sim1/\gamma\gg1$ limit.

\subsection{Two-step term}

We can confirm our claim that $P_{\rm two}$ in~\eqref{PdirTwoOne} comes from a two-step process as follows. The Feynman propagator~\eqref{photonPropagator} for the intermediate photon can be expressed as
\be\label{photonPropagatorTwo}
\begin{split}
&D_{\mu\nu}(y-x)=-g_{\mu\nu}\int\frac{\ud^3{\bf l}}{(2\pi)^32l_0}e^{-il_j(y-x)^j}\\
&\times\left[\theta(y^0-x^0)e^{-il_0(y-x)^0}+\theta(x^0-y^0)e^{il_0(y-x)^0}\right] \;,
\end{split}
\ee
where $l_0=+|{\bf l}|$ is the on-shell momentum and $\theta(.)$ is the step function. The term with $\theta(x^0-y^0)$ corresponds to a process where the pair-production vertex happens earlier than the Compton vertex, i.e. an electron, a positron and a photon are first produced and then the intermediate photon is later absorbed by the incoming electron. This gives a nonzero contribution, but from semiclassical reasoning we expect it to be subdominant compared to the contribution coming from the $\theta(y^0-x^0)$ term, where the intermediate photon is emitted by the initial electron and later decays into a pair. In fact, by redoing the calculation with
\be\label{photonPropagatorTwoSimp}
\begin{split}
&D_{\mu\nu}(y-x)\approx-g_{\mu\nu}\theta(y^0-x^0)\int\frac{\ud^3{\bf l}}{(2\pi)^32l_0}e^{-il(y-x)} 
\end{split}
\ee
and using 
\be
\left(\sqrt{\frac{\pi}{a}}\right)^{-1}\int_0^\infty\ud\tau e^{-a(\tau-\tau_s)^2}=\frac{1}{2}\text{erfc}(-\sqrt{a}\tau_s)
\ee
to perform the integral over $\tau=t'-t$, we recover exactly the full results in~\eqref{PdirSpec} and~\eqref{PexSpec}. This shows clearly that the dominant contribution in this semi-classical regime comes from an on-shell intermediate photon which is emitted by the incoming electron and later decays into an electron positron pair. However, the full results still contain coherent contributions, so we want to know if or how these contributions can be neglected, which should be useful for finding more general incoherent-product approximations for more complicated higher-order processes. We will now show that the dominant contribution in the LCF regime can be written as an incoherent product of the probabilities of Compton scattering and Breit-Wheeler pair production.
We will follow the ideas in~\cite{Dinu:2017uoj} for the construction of a two-step in plane-wave backgrounds. So, we square the amplitude before we perform the $t=x^0$ and $t'=y^0$ integrals. We use $t$ and $t'$ for the time integrals coming from the amplitude $M$, and $\bar{t}$ and $\bar{t}'$ for the time integrals coming from the complex conjugate $M^*$. Our first step in constructing a two-step term is to drop the $\theta(t-t')$ term from~\eqref{photonPropagatorTwo}. The other step function gives on the probability level
\be\label{twoStepFun}
\theta(t'-t)\theta(\bar{t}'-\bar{t}) \;.
\ee
Following~\cite{Dinu:2017uoj} we expect that we can still obtain the leading order even if we replace~\eqref{twoStepFun} with a single step function
\be\label{oneStepFun}
\eqref{twoStepFun}\to\theta(\sigma'-\sigma) \;,
\ee  
where $\sigma=(t+\bar{t})/2$ and $\sigma'=(t'+\bar{t}')/2$. This simplifies the calculations because now there is no step function for the integrals over $\theta=t-\bar{t}$ and $\theta'=t'-\bar{t}'$.
We define $\tilde{P}_{\rm two}$ as what one gets by first dropping the $\theta(t-t')$ term on the amplitude level, and then replacing the step-function combination in~\eqref{twoStepFun} with~\eqref{oneStepFun}. 

We can still perform the time integrals by expanding the exponent around the same saddle point as before. Because of the step function~\eqref{oneStepFun} we find that, relative to the result without any step functions, we have
\be\label{twoErf}
\begin{split}
&\left[\frac{\pi}{\sqrt{a' a}}\right]^{-1}\int\ud\sigma'\ud\sigma\theta(\sigma'-\sigma)e^{-a(\sigma-\sigma_s)^2-a'(\sigma'-\sigma_s')^2}\\
&=\frac{1}{2}\left(1+\text{erf}\left[\sqrt{\frac{a'a}{a'+a}}(\sigma'_s-\sigma_s)\right]\right) \;,
\end{split}
\ee
where $a$, $a'$, $\sigma'_s$ and $\sigma_s$ are obtained by expanding around the same saddle point as before. What is important here is that $\sigma'_s-\sigma_s\propto p'_1\delta p_3-p'_3\delta p_1$ is linear in $\delta{\bf p}$. The two-step term should be constructed so that it is an approximation of $|M_{12}|^2+|M_{21}|^2$, rather than e.g. just $|M_{12}|^2$. If we denote $\tilde{P}_{\rm two}^{12/21}$ as a two-step of only $|M_{12/21}|^2$, then the true/full result is obtained from $\tilde{P}_{\rm two}=(\tilde{P}_{\rm two}^{12}+\tilde{P}_{\rm two}^{21})/2$. We sum over spins, and then the only part of $\tilde{P}_{\rm two}^{12}$ which is not already symmetric comes from~\eqref{twoErf}. Since $\delta{\bf p}\to-\delta{\bf p}$ when swapping the two final-state electrons, we have
\be\label{step12}
\frac{1}{2}\left\{\frac{1}{2}[1+\text{erf}(x)]+\frac{1}{2}[1+\text{erf}(-x)]\right\}=\frac{1}{2} \;.
\ee  
Thus, we find the same result as if we had replaced from the start the step-function in~\eqref{oneStepFun} as $\theta(\sigma'-\sigma)\to1/2$. Thus, the time integrals factorizes so that we can perform the time integrals for the Compton step independently of the time integrals for the Breit-Wheeler step, except there is a factor of $1/2$ because the pair is produced after and not before the photon emission. We will next look for a way to write $\tilde{P}_{\rm two}({\bf p};{\bf p}_1,{\bf p}_2)$ as a product of the spectrum for Compton scattering $P_{\rm C}({\bf p};{\bf l})$ and the spectrum for Breit-Wheeler $P_{\rm BW}({\bf l}; {\bf p}_3)$.

$P_{\rm C}({\bf p};{\bf l})$ can be large (or diverge) for soft photons, but the probability that a soft photon produces a pair is exponentially suppressed. In fact, we already know from the calculation above that the dominant contribution to trident comes from a region around ${\bf l}\sim(2/3){\bf p}$, i.e. the photon takes away two thirds of the initial electron's momentum, so it is definitely a hard photon. We will therefore calculate an approximation of $P_{\rm C}({\bf p};{\bf l})$ for ${\bf l}=(2/3){\bf p}+\delta{\bf l}$ with small $\delta{\bf l}$. In the pre-exponential factors we can set $\delta{\bf l}\to0$.

Another lesson from the plane-wave case is that the incoherent-product can in general be obtained by treating spin and polarization of intermediate particles using Stokes vectors and Mueller matrices. Here we only have a photon as intermediate particle, so we only need to consider photon polarization. A general polarization state can be written as
\be
\epsilon_\mu=\cos\left(\frac{\rho}{2}\right)\epsilon_\mu^{(\LCpara)}
+\sin\left(\frac{\rho}{2}\right)e^{i\lambda}\epsilon_\mu^{(\LCpara)} \;,
\ee
where the two basis vectors are given by
\be\label{paraPerpPol}
\epsilon_\mu^{(\LCpara)}=\frac{1}{k_0}\{0,-k_3,0,k_1\}
\qquad
\epsilon_\mu^{(\LCperp)}=\{0,0,1,0\} \;,
\ee
which corresponds to parallel and perpendicular polarization for a photon with $k_\mu=\{k_0,k_1,0,k_3\}$. For arbitrary constants $\rho$ and $\lambda$, we can write the spectrum as (cf.~\cite{Dinu:2019pau})
\be
P_{\rm C}({\bf p};{\bf l})={\bf N}\cdot{\bf M}_{\rm C}({\bf p};{\bf l}) \;,
\ee
where ${\bf N}$ is a Stokes vector,
\be
{\bf N}=\{1,\cos\lambda\sin\rho,\sin\lambda\sin\rho,\cos\rho\} \;,
\ee
and we refer to ${\bf M}_{\rm C}$ as a Mueller vector. (It would be a $4\times4\times4$ Mueller matrix if we also considered the spin of the fermions.)
Note that ${\bf M}_{\rm C}$ does not depend on the photon polarization. The calculation is very similar to the trident case. We find, for small $\delta{\bf l}$,
\be\label{MuellerC}
{\bf M}_{\rm C}=\frac{\alpha\exp\left[\mathcal{E}_{\rm C}-\frac{Q_{\rm C}}{\omega}+\frac{2L_{\rm C}}{\omega}\right]}{8\pi A'(\tilde{t})l_0 p'_1 m_\LCperp m'_\LCperp}\{1+7p_1^{\prime2},0,0,1+3p_1^{\prime2}\} \;,
\ee
where the exponential scaling of the probability is given by
\be
\mathcal{E}_{\rm C}=-\frac{2}{\omega}[l_0+\mathcal{J}(p')-\mathcal{J}(p)] \;,
\ee
the quadratic terms by
\be
\begin{split}
Q_{\rm C}=&\mathcal{J}_1\delta l_2^2+(\mathcal{J}_1-p_1^{\prime2}\mathcal{J}_2)\delta l_1^2+m_\LCperp^{\prime2}\mathcal{J}_2\delta l_3^2\\
&+2p_1'\mathcal{J}_A\delta l_1\delta l_3+\frac{\delta l_2^2}{l_0}\\
&+\left[\frac{4}{l_0^3}-\left(\frac{3}{2}+\frac{2p_3^{\prime2}}{l_0^2p_1^{\prime2}}\right)\frac{\Omega}{p_1^{\prime2}}\right](p_3'\delta l_1-p_1'\delta l_3)^2
\end{split}
\ee
and linear terms (which we did not have before)
\be
L_{\rm C}=\mathcal{J}_1p_1'\delta l_1-\mathcal{J}_A^1\delta l_3-\frac{p_1'\delta l_1+p_3'\delta l_3}{\sqrt{p_1^{\prime2}+p_3^{\prime2}}} \;,
\ee
where, in addition to the previous time integrals in~\eqref{JnDef} and~\eqref{JAdef}, we now also have
\be
\begin{split}
\mathcal{J}_A^1(p')=\frac{1}{2}\int_0^{\tilde{u}}\ud u\Bigg(&\frac{A-p'_3}{\left[m_\LCperp^{\prime2}+(A-p'_3)^2\right]^{\frac{1}{2}}}\\
&-\frac{A+p'_3}{\left[m_\LCperp^{\prime2}+(A+p'_3)^2\right]^{\frac{1}{2}}}\Bigg) \;,
\end{split}
\ee
$l_0=(2/3)\sqrt{p_1^2+p_3^2}$ and ${\bf p}'={\bf p}/3$ as before. The linear terms, $L_{\rm C}$, are nonzero because ${\bf l}=(2/3){\bf p}$ is not a saddle point for $P_{\rm C}$ alone. It is only a saddle point when $P_{\rm C}$ is combined with $P_{\rm BW}$. We will now show that there is a corresponding linear term in $P_{\rm BW}$ which cancels the linear term in $P_{\rm C}$. 

For the Breit-Wheeler step we find
\be\label{MuellerBW}
\begin{split}
{\bf M}_{\rm BW}=&
\frac{\alpha\exp\left(\mathcal{E}_{\rm BW}-\frac{Q_{\rm BW}}{\omega}+\frac{2L_{\rm BW}}{\omega}\right)}{4\pi A' l_0 p_1' m_\LCperp^{\prime2}}\\
&\times\{1+3p_1^{\prime2},0,0,1-p_1^{\prime2}\} \;,
\end{split}
\ee
where $L_{\rm BW}=-L_{\rm C}$,
\be
\mathcal{E}_{\rm BW}=-\frac{2}{\omega}[-l_0+2\mathcal{J}_0(p')] \;,
\ee
and
\be
\begin{split}
&Q_{\rm BW}=\mathcal{J}_1(\delta p_{22}^2+\delta p_{32}^2)+(\mathcal{J}_1-p_1^{\prime2}\mathcal{J}_2)(\delta p_{21}^2+\delta p_{31}^2) \\
&+m_\LCperp^{\prime2}\mathcal{J}_2(\delta p_{23}^2+\delta p_{33}^2)
+2p_1'\mathcal{J}_A(\delta p_{21}\delta p_{23}+\delta p_{31}\delta p_{33})\\
&-\frac{\delta l_2^2}{l_0}-\left[\frac{4}{l_0^3}-\left(\frac{3}{2}+\frac{2p_3^{\prime2}}{l_0^2 p_1^{\prime2}}\right)\frac{\Omega}{p_1^{\prime2}}\right](p_3'\delta l_1-p_1'\delta l_3)^2\\
&-\frac{\Omega}{p_1^{\prime2}}\left[\frac{1}{2}(p_3'\delta p_1-p_1'\delta p_3)^2+6(p_3'\delta P_1-p_1'\delta P_3)^2\right] \;,
\end{split}
\ee
where $\delta p_{ij}={\bf e}_j\cdot\delta{\bf p}_i$, $\delta{\bf l}=-\delta{\bf p}=\delta{\bf p}_2+\delta{\bf p}_3$.

From the two Mueller vectors, ${\bf M}_{\rm C}$ and ${\bf M}_{\rm BW}$, we now construct a two-step approximation as (cf.~\cite{Dinu:2019pau})
\be
\tilde{P}_{\rm two}({\bf p}_1,{\bf p}_2)=\frac{2}{2}{\bf M}_{\rm C}\cdot{\bf M}_{\rm BW} \;,
\ee 
where the factor of $1/2$ comes from~\eqref{step12}, and the factor of $2$ is part of how in two Mueller vectors (or matrices in general) are glued together. For the particular form of~\eqref{MuellerC} and~\eqref{MuellerBW}, it is possible to obtain the two-step as 
\be\label{twoSum}
\tilde{P}_{\rm two}=\tilde{P}_{\rm two}(\epsilon_\mu^{(\LCpara)})+\tilde{P}_{\rm two}(\epsilon_\mu^{(\LCperp)}) \;,
\ee
where $\tilde{P}_{\rm two}(\epsilon_\mu^{(\LCpara)})$ is obtained by only considering an intermediate photon with parallel polarization, and similarly for $\tilde{P}_{\rm two}(\epsilon_\mu^{(\LCperp)})$. Parallel and perpendicular polarization corresponds to ${\bf N}=\{1,0,0,\pm1\}$. We have
\be
\begin{split}
&({\bf M}_{\rm C}\cdot\{1,0,0,-1\})({\bf M}_{\rm BW}\cdot\{1,0,0,-1\})+\\
&({\bf M}_{\rm C}\cdot\{1,0,0,1\})({\bf M}_{\rm BW}\cdot\{1,0,0,1\})=2{\bf M}_{\rm C}\cdot{\bf M}_{\rm BW} \;.
\end{split}
\ee 
This helps to explain why there is a factor of $2$. Note that, while on the amplitude level one can use any basis, it is not possible to obtain the two step if one sums the incoherent product of probabilities if one chooses some other basis. For example, if one tried to sum the incoherent product over the two circular polarization states, i.e. ${\bf N}=\{1,0,\pm1,0\}$, one would find
\be
\begin{split}
&({\bf M}_{\rm C}\cdot\{1,0,-1,0\})({\bf M}_{\rm BW}\cdot\{1,0,-1,0\})+\\
&({\bf M}_{\rm C}\cdot\{1,0,1,0\})({\bf M}_{\rm BW}\cdot\{1,0,1,0\})\ne2{\bf M}_{\rm C}\cdot{\bf M}_{\rm BW} \;.
\end{split}
\ee   
Using $2{\bf M}_{\rm C}\cdot{\bf M}_{\rm BW}$ directly is more convenient and means we do not have to figure out a special polarization basis. Moreover, $2{\bf M}_{\rm C}\cdot{\bf M}_{\rm BW}$ works in general, while it is not always possible to find two spin/polarization states such that one can obtain the entire two-step by summing as in~\eqref{twoSum}.

In any case, we find
\be
\tilde{P}_{\rm two}=\left[\frac{1}{4}\left(\left|\text{erfc}(c)\right|^2+\left|\text{erfc}(-c)\right|^2\right)\right]^{-1}P_{\rm dir} \;,
\ee 
where $P_{\rm dir}$ is given by~\eqref{PdirSpec}. Thanks to~\eqref{erfcLCF12} we have $\tilde{P}_{\rm two}\approx P_{\rm dir}$ in the LCF limit. This confirms that we have correctly constructed the two-step.

\subsection{Longitudinal limit}

In this section we will consider a limit which is not usually considered. It was considered in~\cite{Brodin:2022dkd} when considering backreaction on the field in Schwinger pair production, and in~\cite{DegliEsposti:2023qqu} when considering the trajectories of particles produced by Schwinger pair production in 4D e-dipole fields. In such cases, the electron can have a large momentum because it has been accelerated by an electric field. But, if the electron trajectory is parallel to the electric field and if the magnetic field vanishes along this trajectory, then regardless of how high the energy is, the electron will not see the field as a plane wave. Instead it will see a field of the form $E_3(t+z)$, which is very different from a plane wave even though they both depend on $t+z$. See~\cite{Tomaras:2001vs,Ilderton:2014mla} for Schwinger pair production in such fields.

If the energy of the electron mainly comes from the acceleration by the field, then we expect the longitudinal momentum to be on the order of $p_3\sim(E/\omega)=a_0$. We will refer to $p_3\sim a_0\gg1$ as the longitudinal limit. For the integrated probability we find
\be
\lim_{p_\LCpara\sim a_0\gg1}P_{\rm dir}=\frac{\alpha^2E(1+3p_\LCperp^{\prime2})^2J_{\rm dir}e^{\mathcal{E}_{\rm long}}}{96\sqrt{3}\tilde{f}^{\prime2}m_\LCperp m_\LCperp^{\prime3}p_\LCperp^{\prime2}\hat{p}_\LCpara^{\prime2}\mathcal{S}_1(\hat{\mathcal{T}}_1-\hat{\mathcal{T}}_2)} \;,
\ee
where $\hat{p}'_\LCpara=p'_\LCpara/a_0$ is considered as $\mathcal{O}(1)$,
\be
\mathcal{E}_{\rm long}=-\frac{1}{E}[3m_\LCperp^{\prime2}\mathcal{S}_1(p')-m_\LCperp^2\mathcal{S}_1(p)] \;,
\ee
\be
\hat{\mathcal{T}}_1=m_\LCperp^{\prime2}\mathcal{S}_1\mathcal{S}_2-p_\LCperp^{\prime2}\mathcal{S}_A^2
\ee
and
\be
\hat{\mathcal{T}}_2=\left(\mathcal{S}_2+\frac{p_\LCperp^{\prime2}}{m_\LCperp^{\prime2}\hat{p}_\LCpara^{\prime2}}[\mathcal{S}_1+2\hat{p}'_\LCpara\mathcal{S}_A]\right)\frac{p'_\LCperp}{\tilde{f}'} \;,
\ee
where the time integrals $\mathcal{S}_{1,2,A}=\mathcal{S}_{1,2,A}(p')$ are given by (\eqref{S1long}, \eqref{S2long} and~\eqref{SAlong} are the same function as in the Breit-Wheeler case~\cite{Brodin:2022dkd})
\be\label{S1long}
\mathcal{S}_1=\int_0^{\tilde{u}}\ud u\frac{\hat{p}_\LCpara}{\hat{p}_\LCpara^2+\tilde{f}^2(u)} \;,
\ee
\be\label{S2long}
\mathcal{S}_2=\int_0^{\tilde{u}}\ud u\frac{\hat{p}_\LCpara(\hat{p}_\LCpara^2-3\tilde{f}^2(u))}{(\hat{p}_\LCpara^2+\tilde{f}^2(u))^3} \;,
\ee
and
\be\label{SAlong}
\mathcal{S}_A=\int_0^{\tilde{u}}\ud u\frac{\tilde{f}^2(u)-\hat{p}_\LCpara^2}{(\hat{p}_\LCpara^2+\tilde{f}^2(u))^2} \;,
\ee
and
\be
Y\approx\left(1+\frac{4p_\LCperp^{\prime2}}{m_\LCperp^{\prime2}}\frac{\hat{\mathcal{T}}_1}{\hat{\mathcal{T}}_2}\left[\frac{\hat{\mathcal{T}}_1}{\hat{\mathcal{T}}_2}-1\right]\right)^{-1/2} \;.
\ee

We see that, just as in the Breit-Wheeler case, the probability approaches a constant in this limit, i.e. it is independent of the large expansion parameter $a_0$ (with $\hat{p}'_\LCpara\sim1$ as independent of $a_0$). Thus, we have the same conclusion: in this limit there is no large parameter to compensate for $\alpha^2\ll1$.

The probability would still be enhanced by a volume/pulse length factor due to the number of field maxima (here we have only considered the contribution from a single field maximum), but such a factor also contributes to the Schwinger pair production probability, so the trident process should still be suppressed relative to Schwinger pair production. However, this argument assumes that the field is not very weak and that we actually do not have much exponential suppression. Strictly, the saddle-point approximations are only precise for sufficiently low field strengths. If we actually consider a regime where the field is sufficiently weak then the trident probability can be much larger than the Schwinger probability, because the exponential part of $P_{\rm trident}$ is much larger and the exponential enhancement can be more than enough to compensate for $\alpha^2\ll1$, i.e. the initial electron assists the pair-production process. However, the structure of these saddle-point approximations also gives us some insights into what parameters are important even beyond the saddle-point regime. They confirm the conclusions based on the Lorentz boost arguments presented in~\cite{Brodin:2022dkd}. So, if we are in a regime where the probability for Schwinger pair production is not exponentially suppressed, then it does seem like the trident probability can to a first approximation be neglected because of the suppression with respect to $\alpha$.

\section{Space-time dependent fields with WKB}\label{space time WKB sec}

So far we have considered fields which only depend on time. Now we will consider fields which also depend on $z$, but as a perturbation
\be
A_3(t,z)=A_3(t)+\delta A_3(t,z) \;.
\ee
To clearly see the justification for the semi-classical/WKB approximation, which is valid for weak fields, we rescale $x^\mu\to x^\mu/E$. 
We will calculate the pre-exponential factors to LO, so most of them will be exactly the same as what we have already calculated. In particular, the correction we are interested in is the same for both spinor and scalar particles. It therefore suffices to consider the Klein-Gordon equation,
\be
(D^2+m^2)\phi=0
\qquad D_\mu=\partial_\mu+iA_\mu(t,z) \;.
\ee
For a double Sauter pulse we would have 
\be\label{doubleSauter}
A_3(t,z)=\frac{E}{\gamma_t}\tanh(\gamma_t t)\text{sech}^2(\gamma_z z) \;.
\ee
We can solve this approximately with a WKB ansatz $\phi=\rho e^{-iS}$, where (cf.~\cite{Kohlfurst:2021skr})
\be
(\partial_\mu S-A_\mu(t,z))^2=m^2 \;.
\ee
To LO we have 
\be
S_{\rm LO}=p_jx^j+\int_0^t\pi_0 \;,
\ee
so, writing $S=S_{\rm LO}+\delta S$ with $\delta S$ on the same order of magnitude as $\delta A$, we have to first order
\be\label{deltaSeq}
\pi^\mu(t)\partial_\mu\delta S(t,z)=\pi^\mu\delta A_\mu \;.
\ee
To solve~\eqref{deltaSeq} we change variables from $t$ and $z$ to $t$ and
\be\label{zetaDef}
\zeta=z+\int_0^t\ud t'\frac{\pi_3}{\pi_0} \;.
\ee
If the LO field is constant, $A_3(t)=t$, then the integral is a total derivative, $\pi_3/\pi_0=-\partial_t\pi_0$, so
\be
\zeta=z-\pi_0(t) \;.
\ee
In terms of the new variables~\eqref{deltaSeq} becomes
\be
\partial_t\delta S(t,\zeta)=\frac{1}{\pi_0}\pi\delta A\left(t,z=\zeta-\int_0^t\frac{\pi_3}{\pi_0}\right) \;.
\ee
We choose a gauge where $\delta A_\mu=\delta_\mu^3\delta A_3$.
A general solution is given by the sum of a homogeneous and a particular solution,
\be\label{Stzh}
\delta S(t,z)=\delta S_h(\zeta)-\int_0^t\ud t'\frac{\pi_3(t')}{\pi_0(t')}\delta A_3\left(t',\zeta-\int_0^{t'}\frac{\pi_3}{\pi_0}\right) \;.
\ee

To interpret the homogeneous solution $\delta S_h(\zeta[t,z])$, we consider the momentum
\be
\Pi_\mu:=\partial_\mu S-A_\mu=\pi_\mu+\delta\Pi_\mu
\qquad
\delta\Pi_\mu=\partial_\mu\delta S-\delta A_\mu \;.
\ee  
We have
\be\label{ddeltaS}
\begin{split}
&\frac{\pi_0}{\pi_3}\delta\Pi_0=\delta\Pi_z=\\
&\left[\delta S_h'-\delta A_3(0,\zeta)-\int_0^t\ud t'\delta E\left(t',\zeta-\int_0^{t'}\frac{\pi_3}{\pi_0}\right)\right] \;,
\end{split}
\ee
where $\delta E(t,z)=\partial_0\delta A_3(t,z)$. We choose a gauge where $\delta A_3(0,z)=0$, so the second term in~\eqref{ddeltaS} vanishes. 

We can compare~\eqref{ddeltaS} with the solution to the Lorentz-force equation,
\be
\frac{\ud\Pi^z}{\ud t}=E(t,z)\approx E(t)+\delta E(t,z) \;,
\ee
where $\Pi^z=\ud z/\ud u$ and $\ud t/\ud u=\sqrt{m_\LCperp^2+\Pi_z^2}$. With $z\approx z_0+\delta z$, we have to leading order $\Pi_0^z=-\pi_3=-[p_3-A(t)]$ and to NLO
\be\label{deltaPiLorentz}
\delta\Pi^z=c+\int_0^t\ud t'\delta E(t,z_0) \;,
\ee 
where $c$ is a constant and (using $\ud t/\ud u\approx\pi_0$)
\be
z_0\approx b-\int_0^t\frac{\pi_3}{\pi_0} \;,
\ee
where $b$ is another constant, which we can think of as an impact parameter. Comparing this with~\eqref{zetaDef} shows that, if we evaluate the wave function $\phi(x^\mu[u])$ on the solution to the Lorentz-force equation, then $\zeta$ corresponds to $b$, so $\zeta$ would be approximately constant on a classical trajectory, and different values of $\zeta$ correspond to different impact parameters. 
With $\zeta$ as a constant of motion, we find agreement between~\eqref{ddeltaS} and~\eqref{deltaPiLorentz} if we identify the derivative of the homogeneous solution $\delta S_h'(\zeta)$ with the constant $-c$ which appears in the momentum. 

We determine $\delta S_h'(\zeta)$ by demanding that the zeroth order already gives the correct asymptotic momentum, which gives
\be\label{deltaSfin}
\delta S(t,z)=-\int_{t_a}^t\ud t'\frac{\pi_3(t')}{\pi_0(t')}\delta A_3\left(t',z+\int_{t'}^t\frac{\pi_3}{\pi_0}\right) \;,
\ee
where $t_a\to\infty$ ($t_a\to-\infty$) for an outgoing (incoming) electron.
By comparing this with the zeroth order~\eqref{UandV}, we see that the positron state is obtained by replacing $\pi_3=p_3-A\to-p_3-A$ and $\pi_0=\sqrt{m_\LCperp^2+(p_3-A)^2}\to-\sqrt{m_\LCperp^2+(p_3+A)^2}$.

If we take $t\to-t_a$ and $z$ to infinity with $\zeta$ kept finite, then
\be\label{deltaSinA}
\delta S(t,z)\to-\int_{t_a}^{-t_a}\ud t'\frac{\pi_3(t')}{\pi_0(t')}\delta A_3\left(t',\zeta-\int_0^{t'}\frac{\pi_3}{\pi_0}\right) \;,
\ee
which is still a nontrivial function of $\zeta$. For a gauge field such as~\eqref{doubleSauter}, which always vanishes if $|z|\to\infty$ regardless of whether $\zeta$ is kept fixed, one might naively have expected that $\delta S\to0$ in this limit. However, $\delta S$ is nonzero because of a memory effect, which we can see from the classical momentum~\eqref{deltaPiLorentz}. We already have a memory effect at zeroth order, since $\pi_3(-\infty)\ne\pi_3(+\infty)$ for $A_3(+\infty)\ne A_3(-\infty)$, so it is not strange that $\delta A_3(t,z)$ gives an additional change to the asymptotic momentum. But in order to find agreement between this classical prediction and $\partial_\mu\delta S$, the asymptotic $\delta S$ cannot vanish or be constant. While the asymptotic $\delta S$ has to solve~\eqref{deltaSeq} with $\delta A\to0$, as we showed in~\eqref{Stzh}, this still allows $\delta S$ to be a nontrivial function of $\zeta$. Moreover, it is not surprising that the classical asymptotic momentum depends nontrivially on the impact parameter, but, for such a nontrivial dependence to emerge from the wave function, and since $\zeta$ corresponds to the impact parameter, $\delta S$ must have a nontrivial dependence on $\zeta$.   

A simple check of the above results is to consider the trivial case where $\delta A_3(t)$ does not depend on $z$, for which we can just use~\eqref{UandV} directly. Expanding the integrand in the exponent in~\eqref{UandV} gives
\be
\pi_0\to\sqrt{m_\LCperp^2+(p_3-A_3-\delta A_3)^2}\approx\pi_0-\frac{\pi_3}{\pi_3}\delta A_3 \;,
\ee
which agrees with~\eqref{deltaSfin}. We turn now to a nontrivial application.

\subsection{$\gamma_z\ll1$ correction}

For a purely time dependent field, $A_3(t)$, the exponential part of the trident probability is given by~\eqref{Edef}. We will now calculate the leading order correction due to a spatial inhomogeneity, $A_3(t)\to A_3(t)+\delta A_3(t,z)$. We consider a field which is slowly varying in $z$. Then it makes sense to expand around the maximum, which we choose to be at $z=0$. We further assume that the field is a product of a temporal and a spatial function, $A_3(t,z)=A(t)F(z)$, where $F(0)=1$, $F'(0)=0$ and we normalize $\gamma_z$ such that $F''(0)=-2\gamma_z^2$. We then have
\be
\delta A_3(t,z)=-(\gamma_z z)^2A_3(t) \;.
\ee
However, after making this expansion, we no longer have $\delta A_3(t,z)\to0$ as $|z|\to\infty$. So, before we expand, we perform a partial integration in~\eqref{deltaSfin},
\be\label{deltaSfin2}
\delta S(t,z)=-\int_{t_a}^t\ud t'E(t')f\left(z+\int_{t'}^t\frac{\pi_3}{\pi_0}\right) \;,
\ee
where $E(t)=A_3'(t)$ and $f'(z)=F(z)-1$. This is also what we would have obtained if we had started in a gauge with $\delta A_0\ne0$ rather than $\delta A_3\ne0$. Now we can expand,
\be\label{deltaSfinz}
\delta S=\frac{\gamma_z^2}{3}\int_{t_a}^t\ud t'E(t')\left(z+\int_{t'}^t\frac{\pi_3}{\pi_0}\right)^3 \;.
\ee

The exponential part on the amplitude level can therefore be written as 
\be
\exp\left\{\frac{i}{E}[\varphi_0+\gamma_z^2\varphi_1]\right\} \;,
\ee
where $\varphi_0$ is the part we considered previously
\be\label{varphi01}
\begin{split}
&\varphi_0=z(l_3+p_{13}-p_3)+z'(p_{23}+p_{33}-l_3)\\
&+l_0(t-t')-I({\bf p},t)+I({\bf p}_1,t)+I({\bf p}_2,t')+I(-{\bf p}_3,t') \;,
\end{split}
\ee
$I$ is given by~\eqref{Iint} except that a factor of $1/E$ has been factored out, and $\varphi_1$ comes from $\delta S$. When we perform the time integrals with the saddle-point method, the saddle points will change due to $\varphi_1$, $t_s\approx t_0+\gamma_z^2 t_1$. However, since $\varphi_0'(t_0)=0$, the correction $t_1$ drops out,
\be
(\varphi_0+\gamma_z^2\varphi_1)(t_0+\gamma_z^2 t_1)=\varphi_0(t_0)+\gamma_z^2\varphi_1(t_0)+\mathcal{O}(\gamma_z^4) \;.
\ee
Thus, the time integrals are the same as before. The $z$ integrals, on the other hand, change significantly, because the $\varphi_1$ term means that we no longer obtain delta functions for the $z$ component of the momenta. Instead, we will perform these integrals with the saddle-point method. We begin by changing variables as
\be
z\to z-\frac{r}{2}
\qquad
z'\to z+\frac{r}{2}
\ee
and then from
\be
r=u-\frac{v}{2}
\qquad
l_3=\frac{1}{2}(p_3-p_{13}+p_{23}+p_{33})+u+\frac{v}{2}
\ee
to $u$ and $v$. The first line in~\eqref{varphi01} then becomes
\be\label{firstLine}
-u^2+\frac{v^2}{4}+\Sigma z \;,
\ee
where
\be
\Sigma=p_{13}+p_{23}+p_{33}-p_3 \;.
\ee
Although $u$ and $v$ also appear in $\varphi_1$, we can perform these integrals in the same way as the time integrals, so the saddle point is simply $u=v=0$. 

If we had started with $\gamma_z=0$, then we would have found $\delta(\Sigma)$, so we expect that performing the $z$ integral for $\gamma_z>0$ will force $\Sigma$ to be small for $\gamma_z\ll1$. Now the dependence on $z$ is still simple,
\be
\exp\left\{\frac{1}{E}[i\Sigma z-\gamma_z^2(az^2+bz)]\right\} \;,
\ee  
where $a$ and $b$ are obtained from the coefficients in
\be\label{deltaSabc}
\begin{split}
&-i\delta S({\bf p})+i\delta S({\bf p}_1)+i\delta S({\bf p}_2)-i\delta S(-{\bf p}_3,\pi_0\to-\pi_0)\\
&=-\frac{\gamma_z^2}{E}(az^2+bz+c) \;,
\end{split}
\ee
and therefore depend on $\Sigma$ but not on $z$. We have a saddle point at
\be\label{zsaddle}
z_s=-\frac{b}{2a}+\frac{i\Sigma}{2a\gamma_z^2} \;,
\ee
and performing the $z$ integral and squaring the amplitude gives
\be\label{expgammaz2}
\exp\left\{-\frac{\gamma_z^2}{E}\left(2c_r-\frac{b_r^2}{2a_r}+\frac{a_r}{2|a|^2}\left[\frac{\Sigma}{\gamma_z^2}-\frac{a_rb_i-a_ib_r}{a_r}\right]^2\right)\right\} \;,
\ee
where $a_r=\Re a$ and $a_i=\Im a$ etc. 

As $\gamma_z\to0$, \eqref{expgammaz2} gives a regularized delta function which implies an approximate conservation of the longitudinal momentum, $\delta_{\rm reg}(p_{13}+p_{23}+p_{33}-p_3)$. We also see that for $\gamma_z\ne0$, the saddle point is nonzero, 
\be\label{SigmaSaddle}
\Sigma_s=\gamma_z^2\frac{a_rb_i-a_ib_r}{a_r} \;.
\ee 
Plugging~\eqref{SigmaSaddle} into~\eqref{zsaddle} gives a real saddle point for $z_s$, 
\be
z_s(\Sigma_s)=-\frac{b_r}{2a_r} \;.
\ee
Note that $z_s$ is nonzero and does not depend on $\gamma_z$. Since we consider symmetric fields, one might naively have expected the process to be centered in the middle of the field, i.e. with a saddle point at $z_s=0$. However, the initial state contains a single electron which is accelerated by the field in the $z$ direction, so the process is inherently non-symmetric. But it is only now when we consider $z$ dependent fields that we can fully see this non-symmetry.  

For $\gamma_z=0$ we have a saddle point at ${\bf p}_1={\bf p}_2={\bf p}_3={\bf p}/3$. But for $\gamma_z>0$, this symmetry is broken. For the longitudinal component, we see this from~\eqref{SigmaSaddle}. For the $x$ component, it follows because if we expand~\eqref{expgammaz2} around what would have been a saddle point for $\gamma_z=0$, then we find a linear term,
\be\label{deltaP1shift}
\exp\left\{-\frac{Y}{\omega}\delta P_1^2+\frac{2\gamma_z^2}{E}\frac{p_{11}}{m_1}\frac{b_r}{a_r}\frac{\partial b_r}{\partial m_1}\delta P_1\right\} \;,
\ee
where $Y$ is the coefficient of $\delta P_1^2$ in the first row of~\eqref{Qdef}. This linear term implies that the saddle point is shifted from $\delta P_1^s=0$ to $\delta P_1^s\ne0$. Since there is no linear term in $\delta p_1$, we still have $\delta p_1^s=0$, so the saddle points for the two final-state electrons are still equal, $p_{11}^s=p_{21}^s$, but they are no longer equal to the positron momentum, $p_{11}^s=p_{21}^s\ne p_{31}^s$. We find
\be
p_{11}^s=p_{21}^s=\frac{p_1}{3}+\delta P_1^s
\qquad
p_{31}^s=\frac{p_1}{3}-\delta P_1^s\;,
\ee
where $p'_1=p_1/3$ is the saddle point to zeroth order, and
\be\label{deltaP1saddle}
\delta P_1^s=\frac{p'_1}{m'}\frac{b_r}{a_r}\frac{\partial b_r}{\partial m'}\frac{\gamma_t\gamma_z^2}{Y} \;.
\ee

We will compare the $\mathcal{O}(\gamma_z^2)$ results in this section with the results obtained in the next section with the worldline instanton formalism, in which we do not have to expand in $\gamma_z$.

\section{Worldline instanton approach}

We have developed an instanton approach based on open worldlines, which has allowed us to calculate the momentum spectrum of Schwinger pair production in space-time dependent fields~\cite{DegliEsposti:2022yqw,DegliEsposti:2023qqu}, nonlinear Compton scattering and Breit-Wheeler pair production~\cite{DegliEsposti:2021its,DegliEsposti:2023fbv}. In all these cases there has been a single worldline. In the pair-production case we have a worldline that starts in the future as a positron, goes backwards in time, turns and goes to the future again as an electron. In the Compton case we have a worldline that starts in the past, as the incoming electron, and ends in the future as the outgoing electron. For trident we will have one worldline for the photon-emission step and another worldline for the pair-production step. Several steps are similar to the Breit-Wheeler case, so more details about those steps can be found in~\cite{DegliEsposti:2023fbv}. 

Both instantons solve the Lorentz-force equation,
\be\label{LFeq}
t''=E(t,z)z' \qquad z''=E(t,z)t' \;,
\ee 
where $t'=\ud t/\ud u$ and $u$ is proper time, except at $u=0$ where the photon is emitted or absorbed. When necessary in order to distinguish the two solutions, we use $q^\mu_C$ for the photon-emission worldline, and $q^\mu_B$ for the pair-production worldline. The trajectories are continuous at $u=0$, but the velocities are not. The discontinuities are given by
\be\label{dqDiff}
\begin{split}
t_C'(0+)-t_C'(0-)&=-l_0 \qquad z_C'(0+)-z_C'(0-)=l_3\\
t_B'(0+)-t_B'(0-)&=l_0 \qquad z_B'(0+)-z_B'(0-)=-l_3 \;.
\end{split}
\ee 
These conditions tell us that the energy of the initial electron drops discontinuously when emitting a photon with energy $l_0$, which is then given to the electron-positron pair. 

As explained in~\cite{DegliEsposti:2021its}, it is possible to choose different complex einbeins, i.e. different contours in the complex $u$ plane. One einbein that we found to be especially useful~\cite{DegliEsposti:2023qqu} starts (for the electron half of the B worldline) at $u=0$ and initially follows the imaginary axis until $u_c=-i|u_c|$, and afterwards goes parallel to the real axis towards $-i|u_c|+\infty$. To make the corner at $u_c$ smooth, we use the following einbein~\cite{DegliEsposti:2022yqw},
\be\label{dudr}
\frac{\ud u}{\ud r}=1-(\epsilon i+1)\psi(r) \;,
\ee
where $\psi(r)$ is a bump function,
\be
\psi(r)=\frac{1}{2\tanh(L/W)}\left(\tanh\left[\frac{r+L}{W}\right]-\tanh\left[\frac{r-L}{W}\right]\right) \;,
\ee
where $L$ is approximately the length of the interval in which $\psi\approx1$, and $W$ is the width of the regions where $\psi$ goes from $0$ to $1$ or from $1$ to $0$. $\epsilon=1$ for the final state particles, and $\epsilon=-1$ for the initial electron. We choose $W$ to be somewhat small, e.g. $W=0.1$. This is essentially just a choice of the shape of the bump function. $L$, on the other hand, is chosen such that the instanton becomes real asymptotically, so determining $L$ then requires a nontrivial computation. This choice of $L$ is not necessary. We would obtain the same probability and the same, real asymptotic momenta for other values of $L$, in particular for values which are held fixed even when we vary $\gamma$ or some other parameter, or even for different einbeins all together. But the asymptotic $q^\mu$ would not be real for general einbeins. This is not actually a problem, because we are only interested in the probability that the particles are produced with a certain (real) momenta, not that the particles appear at some position at some asymptotic time (though that could be an interesting problem for future studies). It can in fact be numerically convenient to choose such ``non-physical'' einbeins~\cite{DegliEsposti:2022yqw}. However, choosing $L$ such that $q^\mu$ is real asymptotically is not just conceptually appealing, it can also be useful for the calculations, e.g. to find approximations~\cite{DegliEsposti:2023qqu}. We will therefore choose such a ``physical'' einbein here, even though there is some extra numerical work to determine the values of $L$, which change when we change various parameters, such as $\gamma$. 

$u=0$ corresponds to the middle of the instanton, where the emission or absorption happens. $u_0$ has a large and negative real part and corresponds to the incoming electron in the asymptotic past, or the outgoing positron in the asymptotic future. $u_1$ has a large and positive real part and corresponds to the outgoing electrons. The instantons are specified by asymptotic conditions at $u_0$ and $u_1$, which say that $z'(\pm\infty)$ should be the momenta we measure and (for the physical einbein) $q^\mu$ should be real. We can find instantons for some arbitrary values of the momenta by imposing the asymptotic conditions,
\be\label{asymptoticConditions}
z_C'(\infty)=-p_{13}
\qquad
z_B'(\infty)=-p_{23}
\qquad
z_B'(-\infty)=p_{33} \;.
\ee
However, since we do not have a good reason to choose otherwise, we will consider the saddle-point values of the final-state momenta, and calculate the Hessian matrix to determine the widths of the spectrum around the saddle point. As shown in~\cite{DegliEsposti:2022yqw}, the corresponding instantons have the following four conditions on the asymptotic velocities,
\be\label{Imdz}
\text{Im }z_{C,B}'(u_{0,1})=0 \;.
\ee
To obtain the dependence on the initial momentum, we would replace $\text{Im }z_C'(u_0)=0$ with $z_C'(u_0)=-p_3$. The asymptotic energy, $t'(u_{0,1})$, is automatically real since the instantons are on shell. 

To numerically integrate~\eqref{LFeq} it is easier to impose initial conditions at $u=0$. As we showed in~\cite{DegliEsposti:2023fbv}, it is possible to find analytical conditions for $q'_\mu(0\pm)$. Here we find
\be\label{dz0}
\begin{split}
z_C'(0-)&=(i,p_1)\cdot\frac{(l_0,-l_3)}{l_1}
\quad
z_C'(0+)=(i,p_{11})\cdot\frac{(l_0,-l_3)}{l_1} \\
z_B'(0-)&=(i,-p_{31})\cdot\frac{(l_0,-l_3)}{l_1}
\quad
z_B'(0+)=(i,p_{21})\cdot\frac{(l_0,-l_3)}{l_1}
\end{split}
\ee
and
\be\label{dt0}
\begin{split}
t_C'(0-)&=(p_1,i)\cdot\frac{(l_0,-l_3)}{l_1}
\quad
t_C'(0+)=(p_{11},i)\cdot\frac{(l_0,-l_3)}{l_1} \\
t_B'(0-)&=(-p_{31},i)\cdot\frac{(l_0,-l_3)}{l_1}
\quad
t_B'(0+)=(p_{21},i)\cdot\frac{(l_0,-l_3)}{l_1} \;.
\end{split}
\ee

From the photon propagator we have
\be
\exp\left\{-il_3[z_B(0)-z_C(0)]-il_0[t_B(0)-t_C(0)]\right\} \;,
\ee
so the saddle-point equation for $l_3$ implies
\be\label{zt0}
z_B(0)-z_C(0)=-\frac{l_3}{l_0}[t_B(0)-t_C(0)] \;.
\ee

After imposing~\eqref{dz0}, \eqref{dt0} and~\eqref{zt0}, we have 12 real unknowns: $z_C(0)$, $t_C(0)$, $t_B(0)$ and $l_3$ (all of which can be complex), and $L_C^\LCm$, $L_C^\LCp$, $L_B^\LCm$ and $L_B^\LCp$, where $L_C^\LCm$ is $L$ for the $u<0$ part of the $C$ instanton etc. We have four real conditions in~\eqref{Imdz}, which means we need 8 additional real conditions. These are the physical-einbein conditions,
\be\label{Imzt}
\text{Im }z_{C,B}(u_{0,1})=\text{Im }t_{C,B}(u_{0,1})=0 \;.
\ee

Thus, we vary $z_C(0)$, $t_C(0)$, $t_B(0)$, $k_3$, $L_C^\LCm$, $L_C^\LCp$, $L_B^\LCm$ and $L_B^\LCp$ until we find instantons satisfying~\eqref{Imdz} and~\eqref{Imzt}. We want to find the dependence on $\gamma_z$, so we need to find instantons for each value between $\gamma_z=0$ and some maximum $\gamma_z$. To speed up the process, we use a numerical continuation similar to~\cite{Schneider:2018huk}, so we first find the instantons at $\gamma_z=0$, use that as starting point for finding the instantons at $\gamma_z=\delta\gamma_z$, where $\delta\gamma_z$ is small, and so on. 

Fortunately, for $\gamma_z=0$ we can actually find the instantons analytically for a Sauter pulse. We find~\cite{DegliEsposti:2021its}
\be\label{tzCBtime}
\begin{split}
t_C&=\theta(-u)t(u,p_1)+\theta(u)t(u,p_{11}) \\
z_C&=\theta(-u)z(u,p_1)+\theta(u)z(u,p_{11}) \\
t_B&=\theta(-u)t(-u,p_{31})+\theta(u)t(u,p_{21}) \\
z_B&=-\theta(-u)z(-u,p_{31})+\theta(u)z(u,p_{21}) \;.
\end{split}
\ee
where $\theta(.)$ is a step function for $u$, but $u$ follows a complex contour, so by this we mean $\theta(-u)=1$ before the photon is emitted (for $C$) or absorbed (for $B$) and $\theta(-u)=0$ afterwards, and vice versa for $\theta(u)$, and
\be\label{tSauterTime}
\begin{split}
t=&\frac{1}{\gamma}\text{arcsinh}\bigg\{\frac{\gamma m_\LCperp}{\sqrt{1+(\gamma m_\LCperp)^2}}\sinh\bigg[\sqrt{1+(\gamma m_\LCperp)^2}u\\
&+i\text{arcsin}\left(\frac{1}{m_\LCperp}\sqrt{\frac{1+(\gamma m_\LCperp)^2}{1+\gamma^2}}\right)\bigg]\bigg\}
\end{split}
\ee
and
\be\label{zSauterTime}
\begin{split}
z=&\frac{1}{\gamma\sqrt{1+(\gamma m_\LCperp)^2}}\text{arcsinh}\bigg\{\gamma m_\LCperp\cosh\bigg[\\
&\sqrt{1+(\gamma m_\LCperp)^2}u+i\text{arcsin}\left(\frac{1}{m_\LCperp}\sqrt{\frac{1+(\gamma m_\LCperp)^2}{1+\gamma^2}}\right)\bigg]\bigg\} \;.
\end{split}
\ee
To compare with~\cite{DegliEsposti:2021its}, note that we have here rescaled $q^\mu\to q^\mu/E$ and $u\to u/E$. 

\begin{figure*}
\includegraphics[width=.45\linewidth]{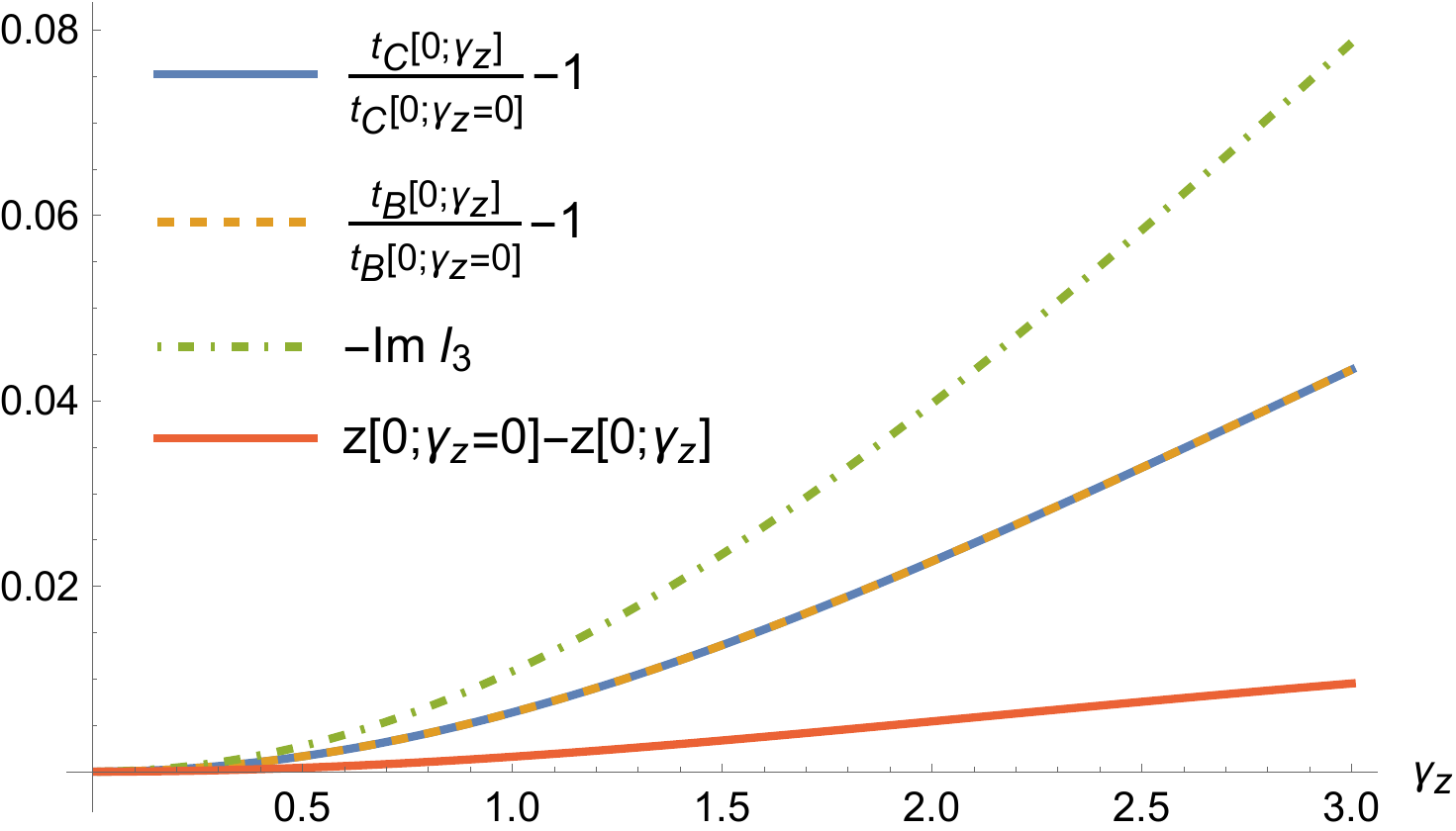}
\includegraphics[width=.45\linewidth]{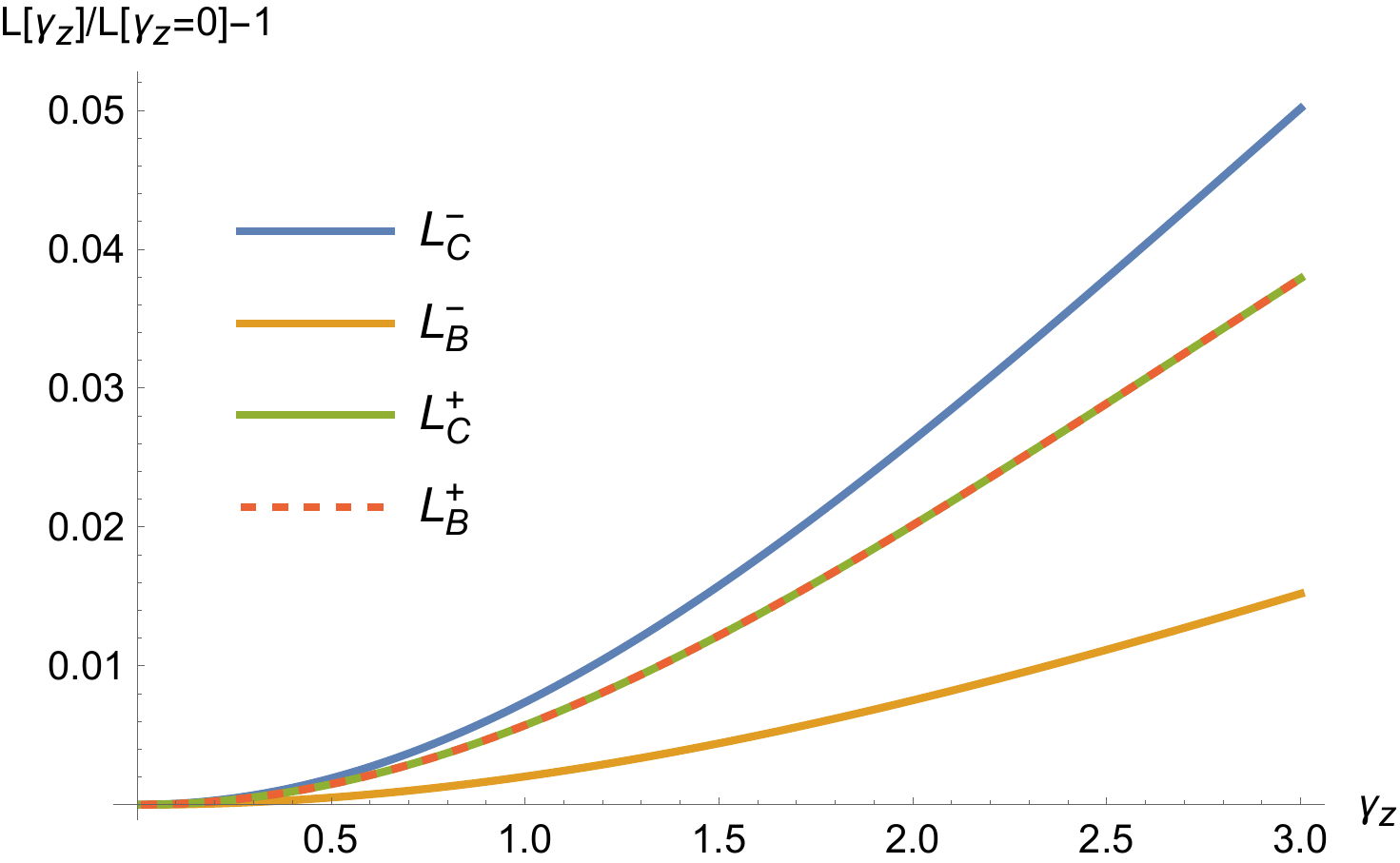}\\
\includegraphics[width=.45\linewidth]{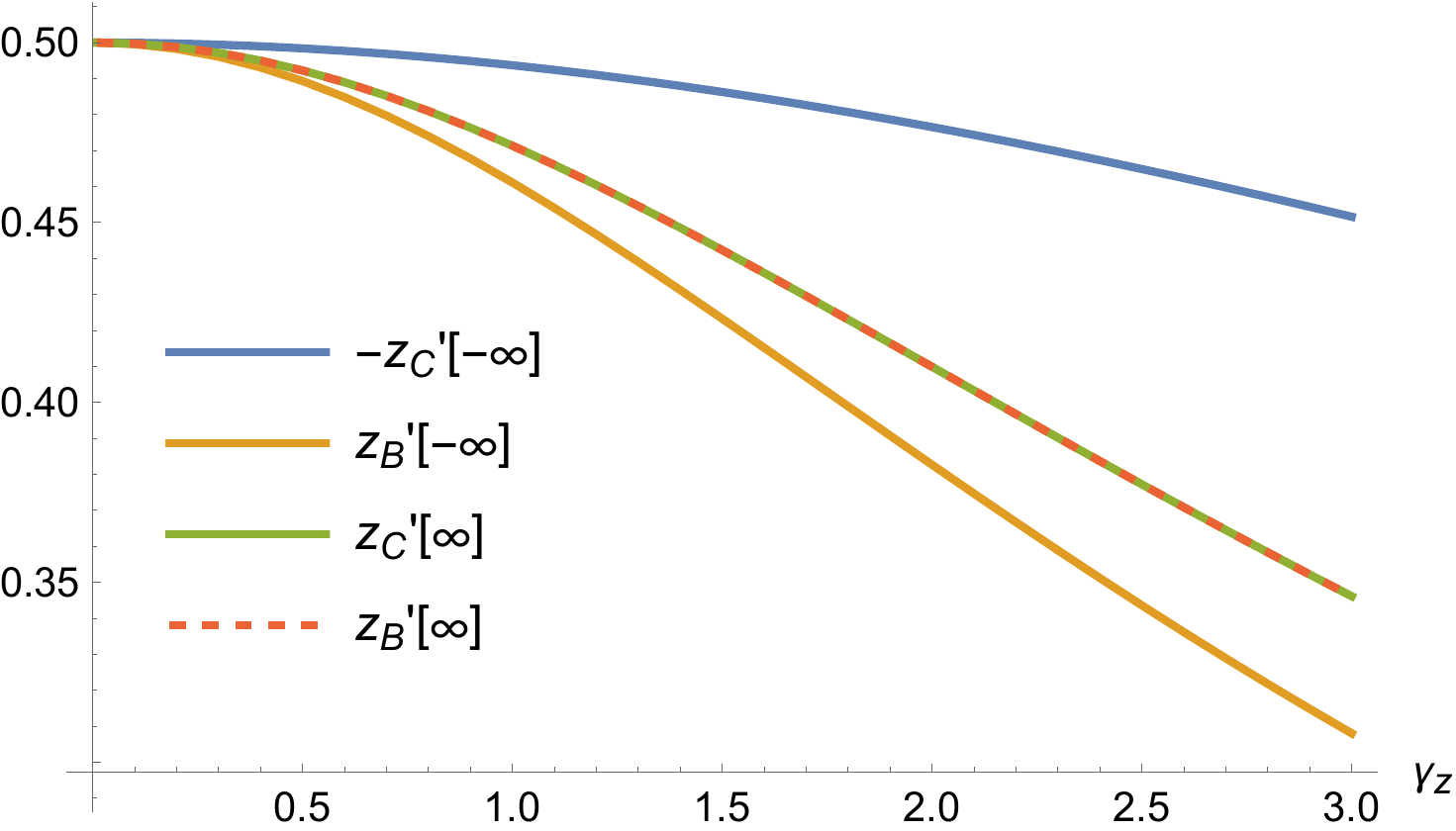}
\includegraphics[width=.45\linewidth]{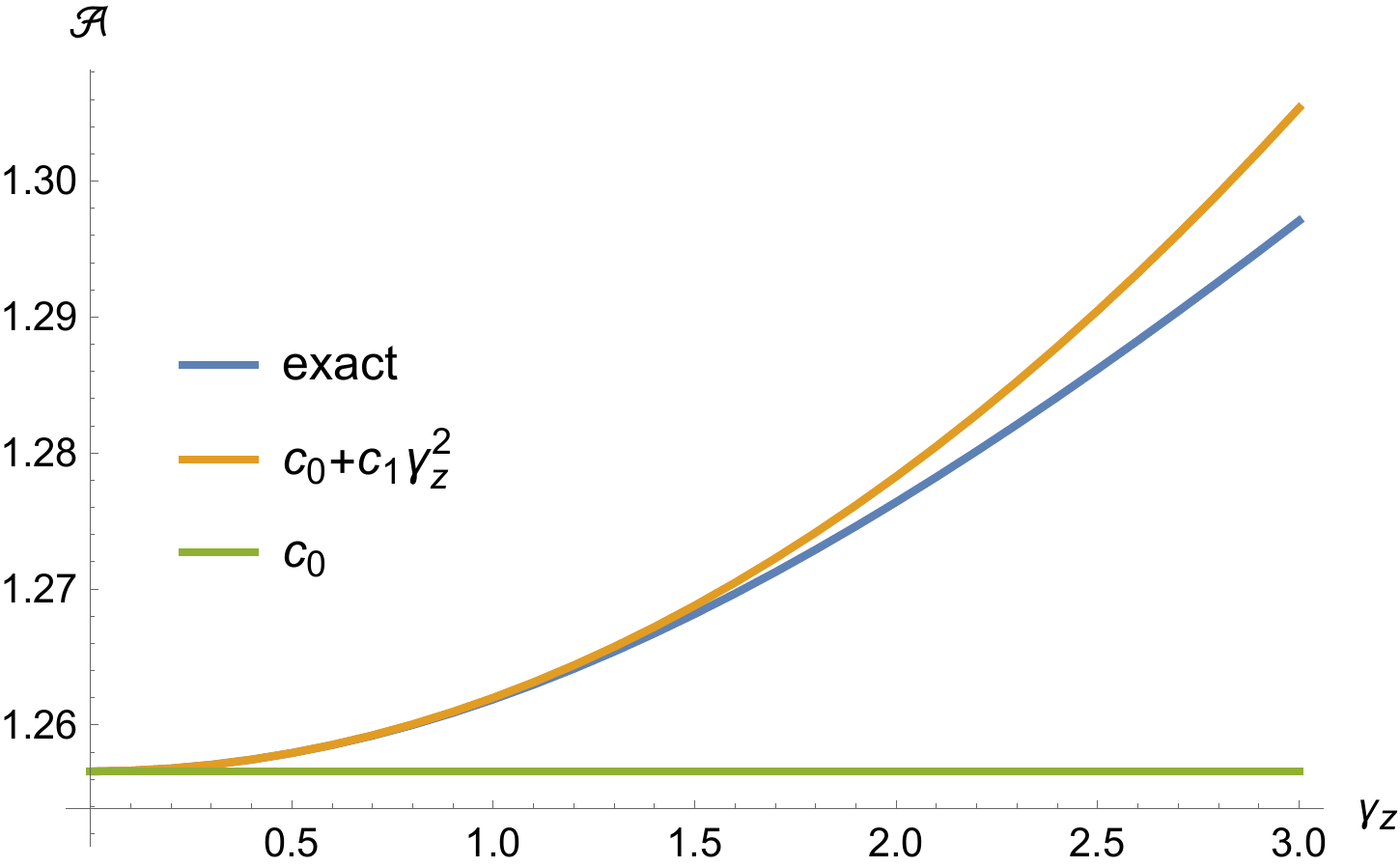}
\caption{Instanton quantities for $\gamma_t=2$ and $p_1=2$ as a function of the spatial inhomogeneity $\gamma_z$. $t_C[0;\gamma_z=0]=t_B[0;\gamma_z=0]$ is given by~\eqref{t0Sauter}, so $t_C(u=0)=t_B(u=0)$ is purely imaginary. The momentum of the intermediate photon is also purely imaginary, $\text{Re }l_3=0$. $z[0;\gamma_z=0]$ is the WKB result~\eqref{zsaddle} ($\Sigma=0$). In this particular case we have $a\approx0.30$ and $b\approx0.014$, so $z_s\approx-0.023$, so $z_C[u=0;\gamma_z]$ is purely real. From~\eqref{einL} and~\eqref{otherL} we find $L_C^\LCm(\gamma_z=0)\approx0.25$ and $L_C^\LCp(\gamma_z=0)=L_B^\LCpm(\gamma_z=0)\approx0.51$. $\mathcal{A}$ is the exponential part of the probability, $P\propto e^{-\mathcal{A}/E}$, obtained from~\eqref{expAC}. $c_1$ is given by the WKB result~\eqref{expgammaz2}.}
\label{fig22}
\end{figure*}

\begin{figure*}
\includegraphics[width=0.45\linewidth]{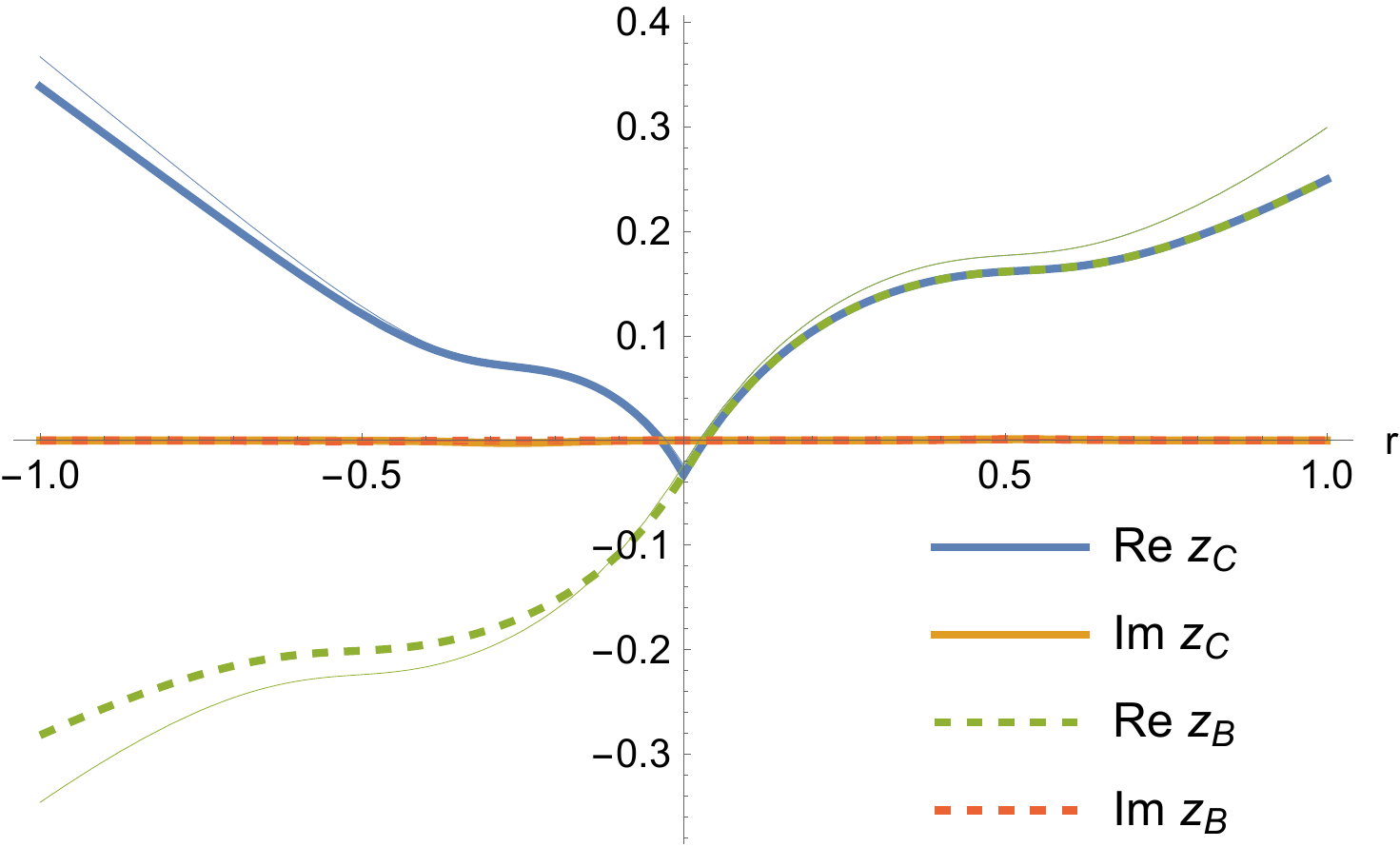}
\includegraphics[width=0.45\linewidth]{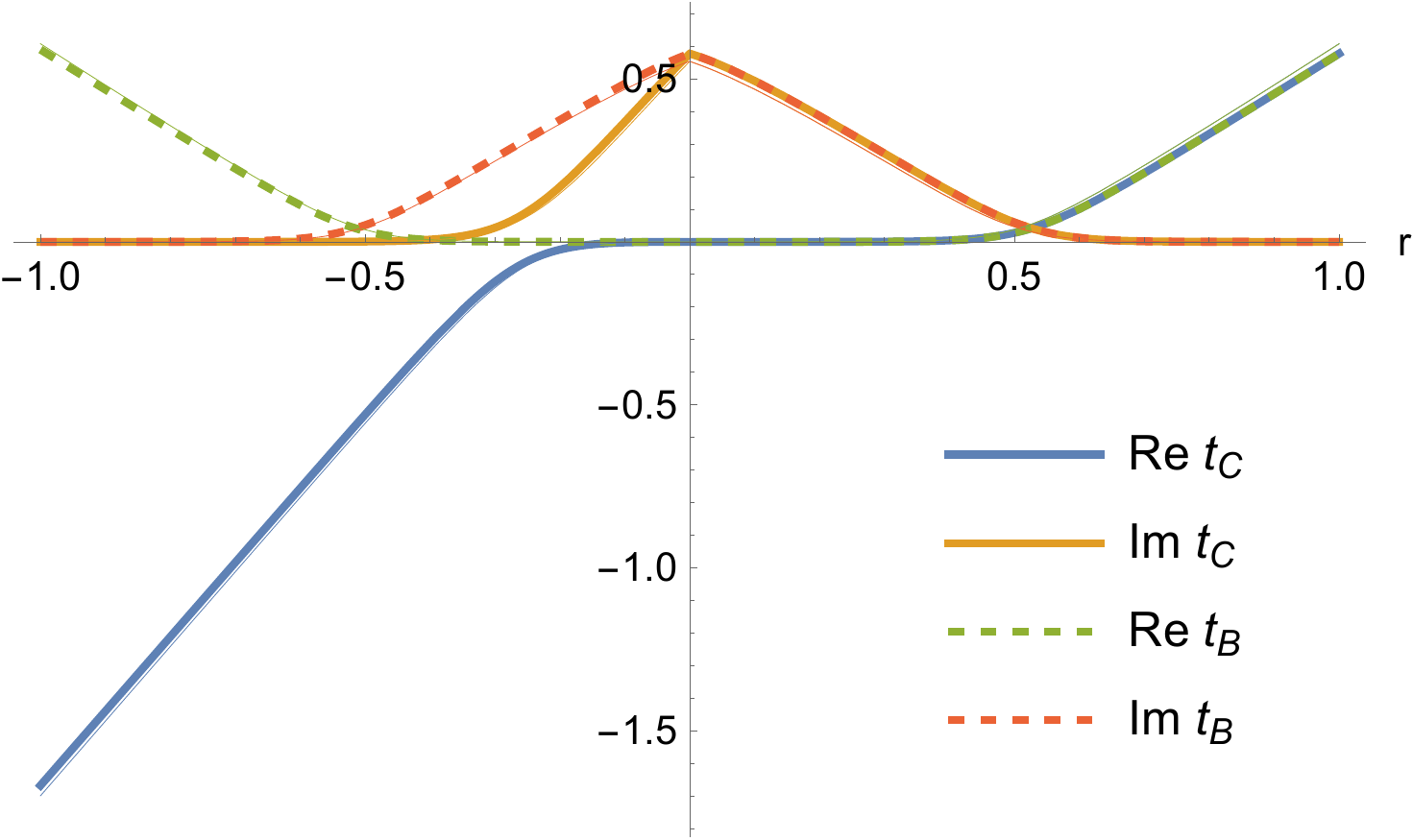}\\
\includegraphics[width=0.45\linewidth]{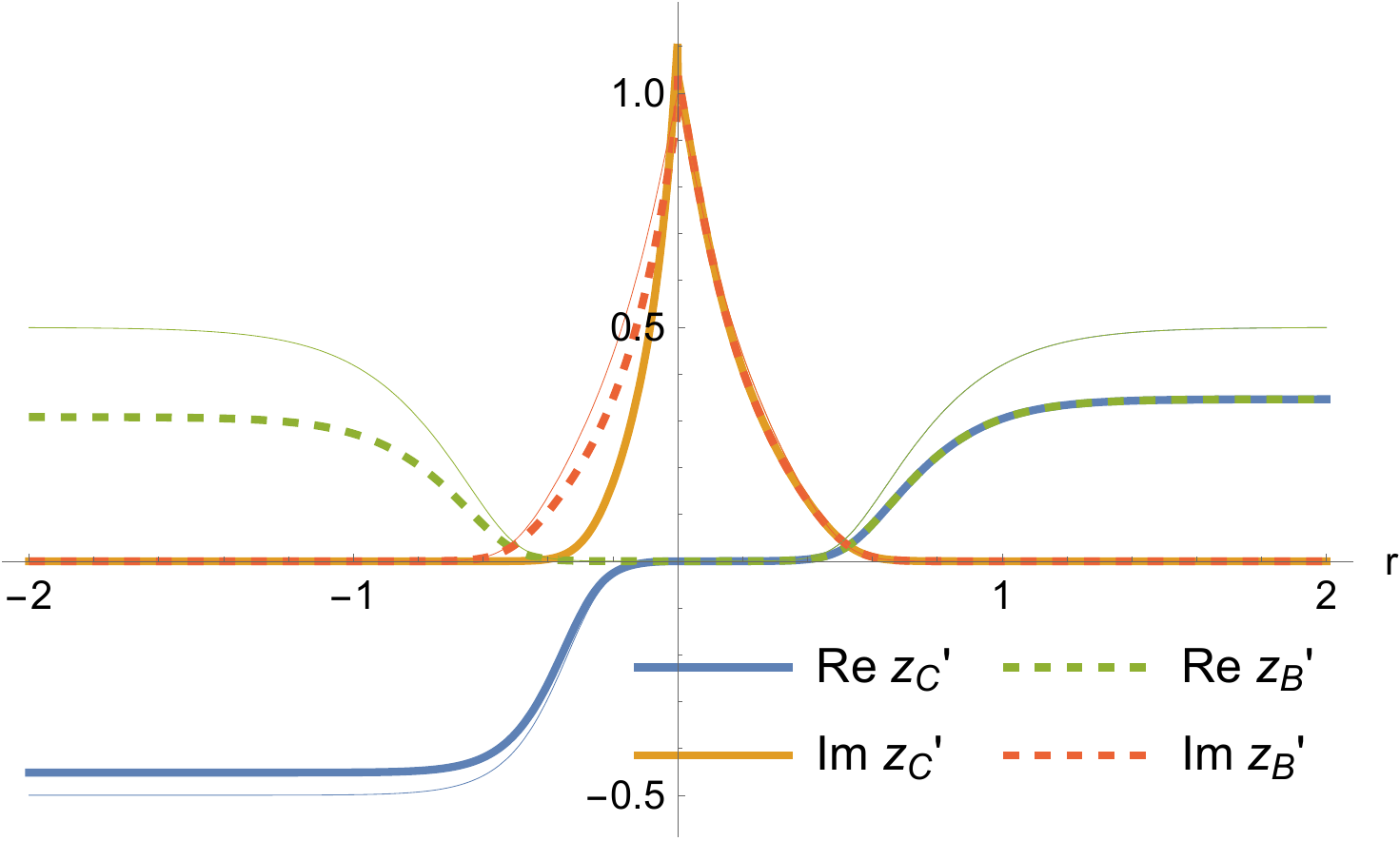}
\includegraphics[width=0.45\linewidth]{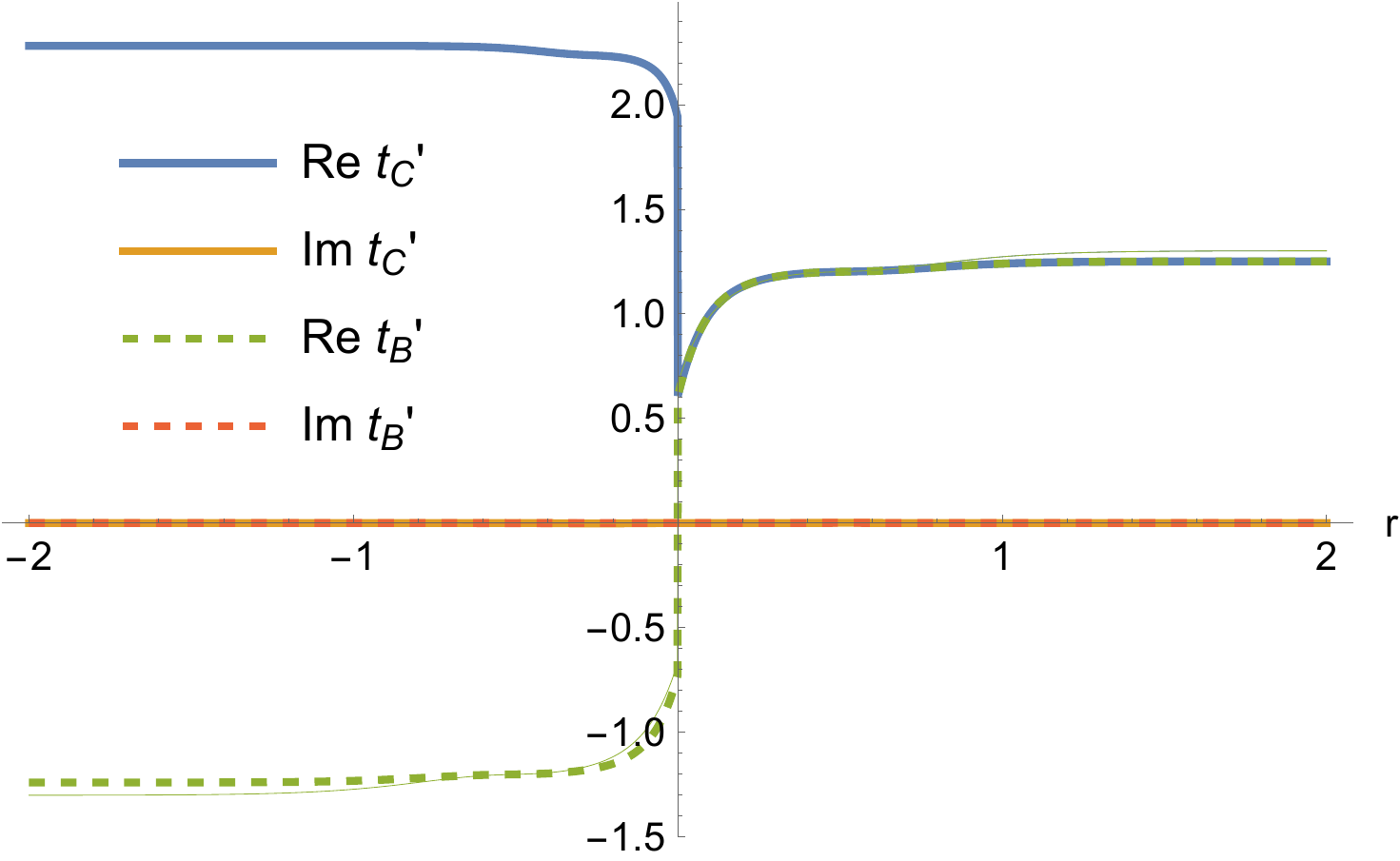}
\caption{Instantons for $\gamma_t=2$, $p_1=2$, $\gamma_z=0.1$ for the thin lines and $\gamma_z=3$ for the other lines. The velocities are $q_\mu'=\ud q_\mu/\ud u$, not $\ud q_\mu/\ud r$. The imaginary part of $z'$ is discontinuous at $r=0$, but, as we can see from~\eqref{dqDiff} and the first plot in Fig.~\ref{fig22}, the discontinuity is relatively small.}
\label{inst22}
\end{figure*}

In~\cite{DegliEsposti:2021its} we chose an integration constant such that $z(0)=0$. This constant is irrelevant for a purely time-dependent field. We found in the previous section that in the WKB approach we have a saddle point at a nonzero $z$, so we actually expect that $z(0)$ will be nonzero for any nonzero value of $\gamma_z$. So starting with exactly $\gamma_z=0$ does not give any idea of what $z(0)$ will be for $\gamma_z>0$. This is actually not a problem, because the above analytical results do give us good starting points for the other constants, and root-finding with $z(0)=0$ arbitrarily chosen as starting point does in fact give the correct result.  

On the other hand, \eqref{tSauterTime} does give us a starting point for $t(0)$,
\be\label{t0Sauter}
t(0)=\frac{i}{\gamma}\text{arctan}(\gamma) \;,
\ee
which will be close to $t(0)$ for $\gamma_z>0$ if $\gamma_z$ is small enough. 

The solutions~\eqref{tSauterTime} and~\eqref{zSauterTime} are analytic functions of $u$ and can be evaluated at different contours in the complex $u$ plane. As mentioned, here we are especially interested in the physical contour, for which the second half of the contour (the part after photon emission or absorption) starts at $u=0$ and first follows the imaginary axis until $u_c$, defined as the point where $t$ becomes real. From~\eqref{tSauterTime} we see that this happens at
\be\label{ucSauter}
u_c=-\frac{i}{\sqrt{1+(\gamma m_\LCperp)^2}}\text{arcsin}\left(\frac{1}{m_\LCperp}\sqrt{\frac{1+(\gamma m_\LCperp)^2}{1+\gamma^2}}\right) \;.
\ee
To find the corresponding length $L$ of the bump of the einbein, we first integrate~\eqref{dudr},
\be
u(r)=r-(\epsilon i+1)\frac{W}{2}\coth\left[\frac{L}{W}\right]\ln\left(\frac{\cosh\left[\frac{L+r}{W}\right]}{\cosh\left[\frac{L-r}{W}\right]}\right) \;,
\ee
where $r$ is real.
For moderately small $W$, the corner of the $u$ contour at $\sim u_c$ is smooth and $u$ does not actually become exactly equal to $u_c$ there. But $\text{Im }u$ does converge to $\text{Im }u_c$ for asymptotic $r$. For $|r|\gg1$ we have
\be\label{ImFromL}
\text{Im }u(r)=-\text{sign}(\epsilon r)L\coth\left[\frac{L}{W}\right] \;.
\ee
For the initial electron, $L$ is determined by
\be\label{einL}
L\coth\left[\frac{L}{W}\right]=-\text{Im }u_c(p_1) \;,
\ee
where $u_c(p_1)$ is given by~\eqref{ucSauter},
and, considering ${\bf p}_1={\bf p}_2={\bf p}_3$, $L$ for the other particles is determined by
\be\label{otherL}
L\coth\left[\frac{L}{W}\right]=-\text{Im }u_c(p_{11}) \;.
\ee
For sufficiently small $W$, the transcendental equation reduces to $L=-\text{Im }u_c$. But for $W=0.1$ and for the values of $\gamma$ and $p_1$ that we have considered, it is better to use the exact solutions of~\eqref{einL} and~\eqref{otherL}. 

Thus, to find the integration constants at the first nonzero value of $\gamma_z$, say $\gamma_z=0.1$, we use as starting points: $z(0)=0$, $t(0)$ given by~\eqref{t0Sauter}, $l_3=0$, $L_C^\LCm$ given by~\eqref{einL} and $L_C^\LCp=L_B^\LCm=L_B^\LCp$ given by~\eqref{otherL}.

The exponential part of the integrated probability, $P\propto e^{-\mathcal{A}/E}$, is given by a sum of the contributions from the $q_C^\mu$ and $q_B^\mu$ worldlines, $\mathcal{A}=\mathcal{A}_C+\mathcal{A}_B$, where (see~\cite{DegliEsposti:2021its,DegliEsposti:2023fbv})
\be\label{expAC}
\mathcal{A}_C=2\text{Im}\int\ud u\,q_C^\mu\partial_\mu A_\nu(q_C)\frac{\ud q_C^\nu}{\ud u}
\ee
and similarly for $\mathcal{A}_B$.

As an example, we consider $\gamma_t=2$ and $p_1=2$. The results are shown in Fig.~\ref{fig22}. 

The instantons are shown in Fig.~\ref{inst22}. We see that $z_C$ starts at $z_C=+\infty$, moves towards the field, passes through the peak at $z=0$, then turns and passes over the peak again, and goes to $z_C=+\infty$. Although $z_C$ is real over the whole trajectory, the velocity $z_C'(u)$ is real asymptotically but imaginary in a region around the turning point. This is consistent because the propertime $u$ follows a contour in the complex plane. $z_C'(u)$ is discontinuous at $u=0$, but it is difficult to see that in Fig.~\ref{inst22} because the discontinuity is relatively small. $t_C$ starts in the asymptotic past, becomes imaginary around the turning point, and goes to the asymptotic future. The energy $t_C'(u)$ is real over the whole trajectory and drops discontinuously at $u=0$, where some of the electron energy is given to the intermediate photon. This discontinuity is larger because we have chosen the transverse momentum $p_1=2$, so $l_1=(2/3)2$, which is significantly larger than the (imaginary) $l_3$ component, so $l_0=\sqrt{l_1^2+l_3^2}\sim l_1$.

For $r>0$ we have $z_C(u)=z_B(u)$ and $t_C(u)=t_B(u)$, so the two final-state electrons follow the same path, even though one of them is connected via a complex worldline to the initial electron while the other is connected to the positron.  

The $r<0$ part of $q_B^\mu$ describes a positron. If we start at $r=-\infty$ and go to $r=0$, then the positron starts in the future and goes backwards in time. If we instead start at $r=0$ and go to $r=-\infty$, then $z_B$ and $t_B$ describe the actual trajectory of the positron, which goes forward in time. We see that the positron moves towards $z=-\infty$, i.e. in the opposite direction compared to the electrons.

By comparing the results for $\gamma_z=0.1$ and $\gamma_z=3$, we see that the instanton is quite insensitive to $\gamma_z$. Indeed, for the time component it is actually difficult to see the difference. But for the asymptotic velocities, $z'(\pm\infty)$, there is a significant, albeit not huge, difference.

\subsection{Momentum dependence}

For $E(t,z)$ we can take $\Pi_a=\{p_{11}, p_{21}, p_{13}, p_{23}, p_{33} \}$ to be the independent momentum variables. To find their saddle points we need $\partial\psi/\partial\Pi_a$, and for the widths of the spectrum around those saddle points we need $\partial^2\psi/\partial\Pi_a\partial\Pi_b$. 
The easiest way to calculate these derivatives is to first express the exponent on the amplitude level as
\be\label{momentumTermsInPsi}
\begin{split}
\psi=&-ip q_C(u_0)+ip_1 q_C(u_1)+ip_2q_B(u_1) + ip_3q_B(u_0)\\
&-ik[q_B(0)-q_C(0)]+\dots
\end{split}
\ee
where the ellipses represent terms that only depend on the momentum variables implicitly via the saddle-point values of the other integration variables. To simplify notation, we write $q_C(u_1^C)$ as $q_C(u_1)$ etc.  
We will eventually evaluate $k_3$ on its saddle point, which depends on the external momenta, $k_3(\Pi)$. But for now we will treat it as an independent variable. Since the field depends on $z$, there is no delta function for the longitudinal momenta, so $p_{13}$, $p_{23}$, $p_{33}$ and $k_3$ are all treated as independent, and for the first derivative we find 
\be\label{k3saddleEq}
\frac{\partial\psi}{\partial k_3}=-i\left[z_B(0)-z_C(0)+\frac{k_3}{k_0}(t_B(0)-t_C(0))\right] 
\ee
\be\label{p13saddleEq}
\frac{\partial\psi}{\partial p_{13}}=i\left[z_C(u_1)+\frac{p_{13}}{p_{10}}t_C(u_1)\right]
\ee
\be\label{p23saddleEq}
\frac{\partial\psi}{\partial p_{23}}=i\left[z_B(u_1)+\frac{p_{23}}{p_{20}}t_B(u_1)\right]
\ee
\be\label{p33saddleEq}
\frac{\partial\psi}{\partial p_{33}}=i\left[z_B(u_0)+\frac{p_{33}}{p_{30}}t_B(u_0)\right] \;.
\ee
To obtain the transverse derivatives we have to remember that
\be
p_{3\LCperp}=p_\LCperp-p_{1\LCperp}-p_{2\LCperp}
\qquad
k_\LCperp=p_\LCperp-p_{1\LCperp} \;.
\ee
We also use e.g.
\be
x_C(u_1)-x_C(0)=-p_{11}u_1^C \;,
\ee
which can be obtained from the fact that the transverse components of the instantons are straight lines before and after the kink at $u=0$. 
We find
\be\label{p1jsaddleEq}
\begin{split}
    \frac{\partial\psi}{\partial p_{1j}}=&i\bigg[p_{1j}\left(\frac{t_C(u_1)}{p_{10}}-u_1^C\right)-p_{3j}\left(\frac{t_B(u_0)}{p_{30}}+u_0^B\right)\\
    &+\frac{k_j}{k_0}\left(t_B(0)-t_C(0)\right)\bigg]
\end{split}
\ee
and
\be\label{p2jsaddleEq}
    \frac{\partial\psi}{\partial p_{2j}}=i\left[p_{2j}\left(\frac{t_B(u_1)}{p_{20}}-u_1^B\right)-p_{3j}\left(\frac{t_B(u_0)}{p_{30}}+u_0^B\right)\right] \;,
\ee
where $j=1,2$.

The above first-order derivatives determine the momentum saddle points. Setting~\eqref{k3saddleEq} to zero gives~\eqref{zt0}. For the external momenta, squaring the amplitude gives $\Re \pa\psi/\pa\Pi=0$ as the condition for the saddle points of $\Pi$. 
The real part of~\eqref{p13saddleEq}, \eqref{p23saddleEq} and~\eqref{p33saddleEq} already vanish for the physical einbein~\eqref{Imzt}. From~\eqref{p1jsaddleEq} and~\eqref{p2jsaddleEq} we find
\be\label{transverseSaddleEq}
\begin{split}
    \Re \frac{\pa \psi}{\pa p_{1j}} &=-\Im \left(p_{1j} u^C_1 + p_{3j}u^B_0  \right) \\
    \Re \frac{\pa \psi}{\pa p_{2j}} &=-\Im \left(p_{2j} u^B_1 + p_{3j}u^B_0  \right) \;.
\end{split}
\ee
Since $p_2=0$, \eqref{transverseSaddleEq} vanishes trivially for $p_{12}=p_{22}=p_{32}=0$. So for the $y$ and $z$ components we find nothing new.  
But for the $x$ component we have an additional, nontrivial condition on the instanton,
\be\label{p1121fromIm}
\begin{split}
p_{11}&=\frac{\Im u^B_0\Im u^B_1}{\Im u^B_0\Im u^B_1+\Im u^B_0\Im u^C_1-\Im u^B_1\Im u^C_1}p_1\\
p_{21}&=\frac{\Im u^B_0\Im u^C_1}{\Im u^B_0\Im u^B_1+\Im u^B_0\Im u^C_1-\Im u^B_1\Im u^C_1}p_1 \;.
\end{split}
\ee
From~\eqref{ImFromL} we have, $\Im(u_1^B) \approx -L_B^+$, $\Im(u_0^B) \approx L_B^-$, and $\Im(u_1^C) \approx -L_C^+$, so this relation can also be expressed as
\be\label{eq:MomentaAndL}
p_{11} L_C^+ = p_{31} L_B^-, \qquad p_{21} L_B^+ = p_{31} L_B^- \; .
\ee
For a time-dependent field, where the saddle point momenta are all equal~\eqref{eq:equalOutMomenta}, $p_{11}=p_{21}=p_{31}$, \eqref{eq:MomentaAndL} simplifies to $L_C^+ = L_B^+ = L_B^-$. However, for a spacetime field, the momenta and the $L$'s are in general different.
Either~\eqref{p1121fromIm} or~\eqref{eq:MomentaAndL} reduce the number of unknown parameters to determine from the six parameters $p_{11}$, $p_{21}$, $L_C^{\pm}$ and $L_B^{\pm}$ to four of them.

By calculating the second-order derivatives, i.e. the Hessian matrix, and evaluating it at the saddle point, we obtain the widths of the momentum spectrum.
This is trivial for all the $y$ components of the Hessian, because, for $j=2$, all the terms in~\eqref{p1jsaddleEq} and~\eqref{p2jsaddleEq} are proportional to $p_{12}$ or $p_{22}$, which both vanish at the saddle point, so we find (the $-2$ superscript is just part of the name)
\be
\begin{split}
d_{12,12}^{-2} &:= 
-\Re \frac{\pa^2 \psi}{\pa p_{12}^2} = \Im \left[\frac{t_C(u_1)}{p_{10}} + \frac{t_B(u_0)}{p_{30}} -u_1^C + u_0^B  \right] \\
d_{12,22}^{-2} &:= 
-\Re \frac{\pa^2 \psi}{\pa p_{12} \pa p_{22}} = \Im \left[\frac{t_B(u_0)}{p_{30}} + u_0^B  \right] \\
d_{22,22}^{-2} &:= 
-\Re \frac{\pa^2 \psi}{\pa p_{22}^2} = \Im \left[\frac{t_B(u_1)}{p_{20}} + \frac{t_B(u_0)}{p_{30}} -u_1^B + u_0^B  \right] \;.
\end{split}
\ee
Note that there is no mixing with the $x$ and $z$ momentum components. For the physical einbein this simplifies further,
\be
d_{12,12}^{-2}=d_{22,22}^{-2}= \Im(u_0^B-u_1^C)
\qquad
d_{12,22}^{-2}=\Im u_0^B \;.
\ee

To obtain the other second-order derivatives, we have to calculate the derivative of the instanton with respect to $\Pi$ and $k_3$. We do this as in~\cite{DegliEsposti:2023fbv}. We first note that we do not need the full $\Pi$ dependence of $\pa_\Pi q_{C,B}(\Pi)$, only $\pa_\Pi q_{C,B}(\Pi_s)$ evaluated at the saddle point $\Pi=\Pi_s$. This can be obtained from $q(\Pi)=q(\Pi_s)+\delta q\delta \Pi$, where $\delta\Pi=\Pi-\Pi_s$ and $\delta q=\partial_\Pi q(\Pi_s)$. In particular,  
\be
\begin{split}
  q^\mu &\to q^\mu + \delta k_3 \, \delta_k q^\mu(u)  \\ 
  q^\mu &\to q^\mu + \delta \Pi_\alpha \, \delta_\alpha q^\mu(u)  
\end{split}
\ee
(no sum over $\alpha$) with $\Pi_\alpha$ denoting any of the momentum variables $\{p_{11}, p_{21}, p_{13}, p_{23}, p_{33} \}$, and similarly for $k_3$. In other words, $\delta_\alpha q$ denotes the derivative of the instanton with respect to the variable $\Pi_\alpha$ evaluated at all the saddle points $\{\Pi_{\alpha, s}, k_{3,s}\}$. 

All variations are solutions to the following equation, which is obtained by making a linear perturbation of the Lorentz-force equation,
\be\label{dtdzeq}
\begin{split}
\delta t''&=E\delta z'+\nabla E\cdot\{\delta t,\delta z\}z' \\
\delta z''&=E\delta t'+\nabla E\cdot\{\delta t,\delta z\}t' \;,
\end{split}
\ee
where $\nabla E=\{\partial_t E,\partial_z E\}$. Some of the conditions which distinguish the different $\delta q$'s are asymptotic conditions at $u=\pm\infty$. But that does not mean that we have to use a shooting method to find the $\delta q$'s, because~\eqref{dtdzeq} is linear, so we can first solve~\eqref{dtdzeq} with four different initial conditions, where the only nonzero conditions are
\be
\delta t_{[1]}(0)=
\delta z_{[2]}(0)=
\delta t_{[3]}'(0)=
\delta z_{[4]}'(0)=1 \;.
\ee
and then an arbitrary solution can be expressed as a superposition of these four basis solutions,
\be\label{qjsum}
\delta q(u)=\sum_{j=1}^4 a_j\delta q_{[j]}(u) \;,
\ee
where the coefficients $a_j$ are determined algebraically.

$a_3$ and $a_4$ in~\eqref{qjsum} are determined directly by simply matching $\delta t'(0\pm)$ and $\delta z'(0\pm)$ with the derivative of~\eqref{dz0} and~\eqref{dt0} with respect to $k_3$ for $\delta_k q$ or $\Pi_\alpha$ for $\delta_\alpha q$. 
Note that, due to the discontinuity of $q_\mu'$ at $u=0$ in~\eqref{dz0} and~\eqref{dt0}, we have $a_3=a_3^\LCm$ and $a_4=a_4^\LCm$ for $u<0$ and $a_3=a_3^\LCp\ne a_3^\LCm$ and $a_4=a_4^\LCp\ne a_4^\LCm$ for $u>0$.
$a_1$ and $a_2$, on the other hand, must be the same for both $u<0$ and $u>0$, because $q^\mu$ is continuous at $u$. We therefore only need two conditions to determine $a_1$ and $a_2$, which we have by differentiating the asymptotic conditions~\eqref{asymptoticConditions} for $z'(\pm\infty)$.

Having obtained the variations, $\delta_k q$ and $\delta_\alpha q$, we can now obtain the Hessian and hence the momentum widths from
\be\label{eq:X}
X_{\alpha \beta} = \frac{\pa^2 \psi}{\pa \Pi_\alpha \pa \Pi_\beta} \qquad X_{k \alpha} = \frac{\pa^2 \psi}{\pa k_3 \pa \Pi_\alpha} \qquad X_{kk} = \frac{\pa^2 \psi}{\pa k_3^2} \; .
\ee
Differentiating~\eqref{k3saddleEq} with respect to $k_3$ gives
\be
X_{kk} = -i\left[\delta_k z_B(0) - \delta_k z_C(0) +\frac{k_3}{k_0} \bigr(\delta_k t_B(0) - \delta_k t_C(0) \bigr)\right] \;.
\ee

We can obtain $X_{kp_{13}}$ by either differentiating~\eqref{p13saddleEq} with respect to $k_3$, or by differentiating~\eqref{k3saddleEq} with respect to $p_{13}$. In the latter case, the result simplifies since $\delta_{13}q_B=0$, which follows from the fact that none of the conditions for $q_B$ in \eqref{dz0}, \eqref{dt0} or~\eqref{asymptoticConditions} depends on $p_{13}$. We find
\be
\begin{split}
X_{k p_{13}} =& i\biggr[ \delta_k z_C(u_1) + \frac{p_{13}}{p_{10}} \delta_k t_C(u_1) \biggr] \\
=&  -i\biggr[ - \delta_{13} z_C(0) -\frac{k_3}{k_0} \delta_{13} t_C(0) \biggr] \;,
\end{split}
\ee
The fact that partial derivatives commute has given us two different representations of $X_{k p_{13}}$ and hence a highly nontrivial relation between $\delta_k q$ and $\delta_{13}q$. Computing both representations and comparing them allows one to estimate the precision of the numerical results.    

Similarly, using $\delta_{23}q_C = \delta_{33} q_C = \delta_{21} q_C = 0$ and $t_C(0)=t_B(0)$ we find
\be
\begin{split}
X_{k p_{23}} =& i\biggr[ \delta_k z_B(u_1) + \frac{p_{23}}{p_{20}} \delta_k t_B(u_1) \biggr] \\
=& -i\biggr[ \delta_{23} z_B(0) +\frac{k_3}{k_0} \delta_{23} t_B(0) \biggr]
\end{split}
\ee
\be
\begin{split}
X_{k p_{33}} =& i\biggr[ \delta_k z_B(u_0) + \frac{p_{33}}{p_{30}} \delta_k t_B(u_0) \biggr] \\
=& -i\biggr[ \delta_{33} z_B(0) +\frac{k_3}{k_0} \delta_{33} t_B(0) \biggr]
\end{split}
\ee

\be
\begin{split}
X_{k p_{11}} =& i\biggr[ \frac{p_{11}}{p_{10}} \delta_k t_C(u_1) - \frac{p_{31}}{p_{30}}\delta_k t_B(u_0) \\
&+ \frac{k_1}{k_0}\bigr( \delta_k t_B(0) - \delta_k t_C(0) \bigr)\biggr] \\
=& -i\biggr[ \delta_{11} z_B(0) - \delta_{11} z_C(0) \\
&+\frac{k_3}{k_0} \bigr(\delta_{11} t_B(0) - \delta_{11} t_C(0) \bigr) \biggr]
\end{split}
\ee
\be
\begin{split}
X_{k p_{21}} =& i\biggr[ \frac{p_{21}}{p_{20}} \delta_k t_B(u_1) - \frac{p_{31}}{p_{30}}\delta_k t_B(u_0) \biggr] \\
=& -i\biggr[ \delta_{21} z_B(0) +\frac{k_3}{k_0}\delta_{21} t_B(0) \biggr] \;.
\end{split}
\ee
As mentioned, there is no mixing with the $y$ component, so $X_{kp_{12}}=X_{kp_{22}}=0$.

For the $x$ components we find
\be
\begin{split}
X_{p_{11}p_{11}} =& i\biggr[ \frac{t_C(u_1)}{p_{10}} - u_1^C - \frac{p_{11}^2}{p_{10}^3} t_C(u_1) + \frac{p_{11}}{p_{10}} \delta_{11} t_C(u_1)  \\
&+ \frac{t_B(u_0)}{p_{30}} + u_0^B -\frac{p_{31}^2}{p_{30}^3} t_B(u_0) - \frac{p_{31}}{p_{30}} \delta_{11} t_B(u_0) \\
&+\frac{\delta_{11}t_B(0) - \delta_{11} t_C(0)}{k_0}k_1\biggr]
\end{split}
\ee
\be
\begin{split}
X_{p_{21}p_{21}} =& i\biggr[ \frac{t_B(u_1)}{p_{20}} - u_1^B - \frac{p_{21}^2}{p_{20}^3} t_B(u_1) + \frac{p_{21}}{p_{20}} \delta_{21} t_B(u_1)  \\
&+
\frac{t_B(u_0)}{p_{30}} + u_0^B -\frac{p_{31}^2}{p_{30}^3} t_B(u_0) - \frac{p_{31}}{p_{30}} \delta_{21} t_B(u_0)\biggr]
\end{split}
\ee
\be
\begin{split}
X_{p_{11}p_{21}} =& i\biggr[ \frac{t_B(u_0)}{p_{30}} +u_0^B - \frac{p_{31}^2}{p_{30}^3}t_B(u_0) \\
&-\frac{p_{31}}{p_{30}}\delta_{21}t_B(u_0) +\frac{k_1}{k_0} \delta_{21} t_B(0) \biggr] \\
=& i \biggr[ \frac{t_B(u_0)}{p_{30}} +u_0^B - \frac{p_{31}^2}{p_{30}^3}t_B(u_0) \\
& -\frac{p_{31}}{p_{30}}\delta_{11}t_B(u_0) +\frac{p_{21}}{p_{20}}\delta_{11} t_B(u_1)\biggr] \;.
\end{split}
\ee

For the cross terms between the $x$ and $z$ components we find
\be
\begin{split}
    X_{p_{11}p_{13}} 
    =& i\left[ \frac{p_{11}}{p_{10}}\delta_{13} t_C(u_1) -\frac{k_1}{k_0}\delta_{13}t_C(0) - \frac{p_{11}p_{13}}{p_{10}^3}t_C(u_1) \right] \\
    =&i\left[ \delta_{11}z_C(u_1) +\frac{p_{13}}{p_{10}}\delta_{11}t_C(u_1) - \frac{p_{11}p_{13}}{p_{10}^3}t_C(u_1) \right]
\end{split}
\ee
\be
\begin{split}
   X_{p_{11}p_{23}} 
   =& i\left[ \delta_{11}z_B(u_1) +\frac{p_{23}}{p_{20}} \delta_{11}t_B(u_1) \right] \\
   =& i\left[ \frac{k_1}{k_0} \delta_{23} t_B(0)-\frac{p_{31}}{p_{30}} \delta_{23} t_B(u_0) \right]
\end{split}
\ee
\be
\begin{split}
   X_{p_{11}p_{33}} 
   =& i\left[ \delta_{11}z_B(u_0) +\frac{p_{33}}{p_{30}} \delta_{11}t_B(u_0) +\frac{p_{33}p_{31}}{p_{30}^3}t_B(u_0) \right] \\
   =& i\left[ \frac{k_1}{k_0} \delta_{33} t_B(0) -\frac{p_{31}}{p_{30}} \delta_{33} t_B(u_0) +\frac{p_{33}p_{31}}{p_{30}^3}t_B(u_0) \right]
\end{split}
\ee
\be
\begin{split}
    &X_{p_{21}p_{23}} \\
    &= i\left[ \delta_{21}z_B(u_1) + \frac{p_{23}}{p_{20}}\delta_{21} t_B(u_1) -\frac{p_{21}p_{23}}{p_{20}^3} t_B(u_1) \right] \\
    &= i\left[ \frac{p_{21}}{p_{20}} \delta_{23}t_B(u_1) - \frac{p_{31}}{p_{30}}\delta_{23}t_B(u_0) -\frac{p_{21}p_{23}}{p_{20}^3}t_B(u_1) \right]
\end{split}
\ee
\be
\begin{split}
    X_{p_{21}p_{33}}
    =& i\left[ \frac{p_{21}}{p_{20}} \delta_{33} t_B(u_1) -\frac{p_{31}}{p_{30}} \delta_{33} t_B(u_0) + \frac{p_{31}p_{33}}{p_{30}^3} t_B(u_0)\right] \\
    =& i\left[ \delta_{21}z_B(u_0) + \frac{p_{33}}{p_{30}} \delta_{21} t_B(u_0) + \frac{p_{31}p_{33}}{p_{30}^3} t_B(u_0)\right]
\end{split}
\ee
while $X_{p_{21}p_{13}}=0$.

For the $z$ components we find
\be
\begin{split}
    X_{p_{13}p_{13}} 
    =&i\biggr[ \delta_{13}z_C(u_1)+ \frac{p_{13}}{p_{10}}\delta_{13}t_C(u_1) \\
    &+\frac{t_C(u_1)}{p_{10}} -\frac{p_{13}^2}{p_{10}^3}t_C(u_1)\biggr]
\end{split}
\ee
\be
\begin{split}
   X_{p_{23}p_{23}} 
   &= i\biggr[ \delta_{23}z_B(u_1)+ \frac{p_{23}}{p_{20}}\delta_{23}t_B(u_1) \\
    &+\frac{t_B(u_1)}{p_{20}} -\frac{p_{23}^2}{p_{20}^3}t_B(u_1) \biggr] 
\end{split}
\ee
\be
\begin{split}
   X_{p_{33}p_{33}} 
   &= i\biggr[ \delta_{33}z_B(u_0)+ \frac{p_{33}}{p_{30}}\delta_{33}t_B(u_0) \\
    &+\frac{t_B(u_0)}{p_{30}} -\frac{p_{33}^2}{p_{30}^3}t_B(u_0) \biggr] 
\end{split}
\ee
\be
\begin{split}
    X_{p_{23}p_{33}} 
    &= i\left[ \delta_{33}z_B(u_1)+ \frac{p_{23}}{p_{20}}\delta_{33}t_B(u_1) \right] \\
    &= i\left[ \delta_{23}z_B(u_0)+ \frac{p_{33}}{p_{30}}\delta_{23}t_B(u_0) \right]
\end{split}
\ee
while $X_{p_{13}p_{23}} = X_{p_{13}p_{33}} = 0$ due to the vanishing variations $\delta_{13} q_B = 0$ etc. mentioned above. 

In matrix form we have
\be
\textbf{X} =
\left[
\begin{array} {cc|ccc}
    X_{p_{11} p_{11}} & X_{p_{11} p_{21}} & X_{p_{11} p_{13}} & X_{p_{11} p_{23}} & X_{p_{11} p_{33}} \\
    X_{p_{11} p_{21}} & X_{p_{21} p_{21}} & 0 & X_{p_{21} p_{23}} & X_{p_{21}p_{33}} \\ \hline
    X_{p_{11} p_{13}} & 0 & X_{p_{13} p_{13}} & 0 & 0 \\
    X_{p_{11} p_{23}} & X_{p_{21} p_{23}} & 0 & X_{p_{23} p_{23}} & X_{p_{23} p_{33}} \\
    X_{p_{11} p_{33}} & X_{p_{21}p_{33}} & 0 & X_{p_{23} p_{33}} & X_{p_{33} p_{33}}
\end{array}
\right]
\ee
where we have highlighted a transverse block with elements $X_{p_{11} p_{11}}$, etc., a longitudinal block with elements $X_{p_{13} p_{13}}$, etc., and mixed blocks. 

To compute the widths themselves, we need to take into account the nontrivial $k_3$ integral. As in~\cite{DegliEsposti:2023fbv}, since the $k_3$ saddle point depends on the momentum, $k_s(\Pi)$, we obtain
\be\label{symmetricAfterkInt}
\psi\big|_\text{quad. part}=\delta\Pi_\alpha\left[X_{\alpha \beta} - \frac{X_{k \alpha} X_{k \beta}}{X_{kk}}\right]\delta\Pi_\beta \;.
\ee
While the intermediate steps have not been symmetric under the exchange of the momenta of the two final-state electrons, ${\bf p}_1\leftrightarrow{\bf p}_2$, \eqref{symmetricAfterkInt} is actually symmetric. This makes the next step much simpler, because the two terms in the trident amplitude, $M=M_{12}-M_{21}$, both have the same exponential, so $|M_{12}|^2$, $|M_{21}|^2$ and $2\Re M_{12}\bar{M}_{21}$ all have the same exponential. Thus, only the real part of~\eqref{symmetricAfterkInt} contributes, so the width matrix is given by    
\be\label{dalphabeta}
d_{\alpha \beta}^{-2} = -\Re \left[ X_{\alpha \beta} - \frac{X_{k \alpha} X_{k \beta}}{X_{kk}}\right]
\ee
where $\alpha, \beta \in \{p_{11}, p_{21}, p_{13}, p_{23}, p_{33} \}$. The spectrum is proportional to
\be
P(\Pi)\propto\exp\left(-[\Pi-\Pi_s]_\alpha d_{\alpha \beta}^{-2}[\Pi-\Pi_s]_\beta\right) \;.
\ee

\begin{figure*}
    \centering
    \includegraphics[width=.49\linewidth]{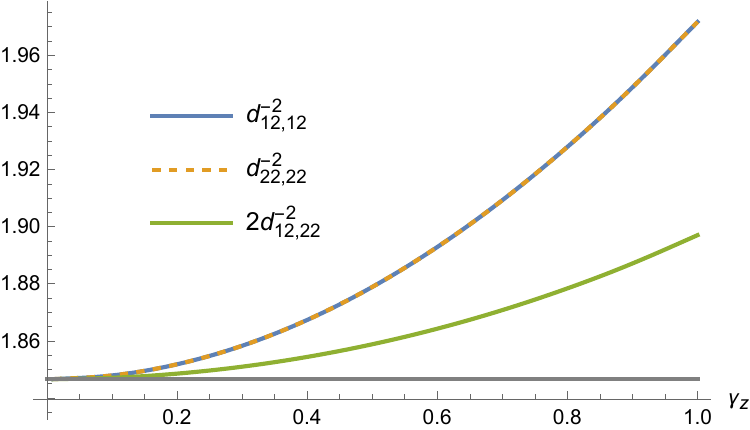}
    \includegraphics[width=.49\linewidth]{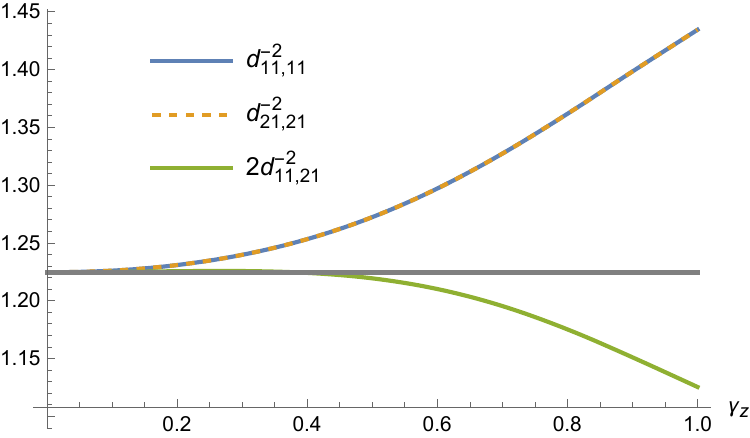}\\
    \includegraphics[width=.49\linewidth]{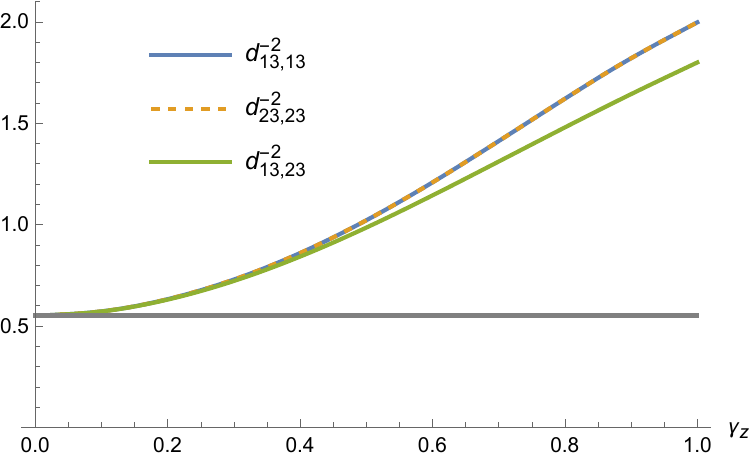} 
    \includegraphics[width=0.49\linewidth]{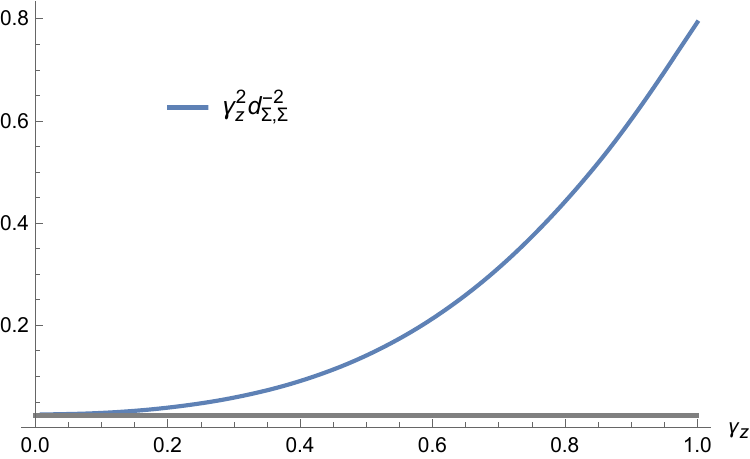}\\
    \includegraphics[width=0.49\linewidth]{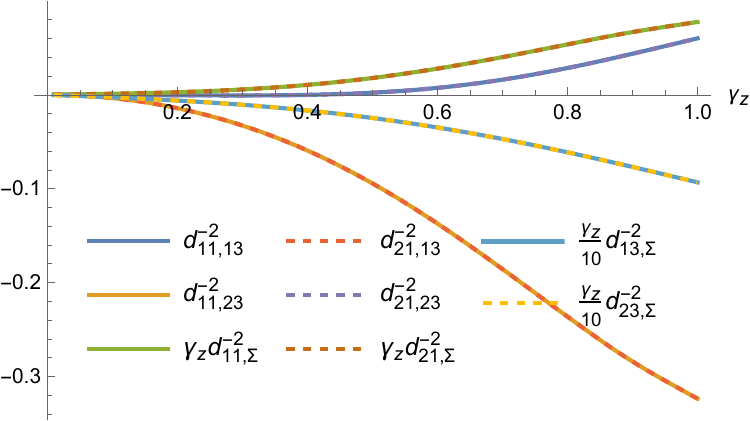}
    \includegraphics[width=0.49\linewidth]{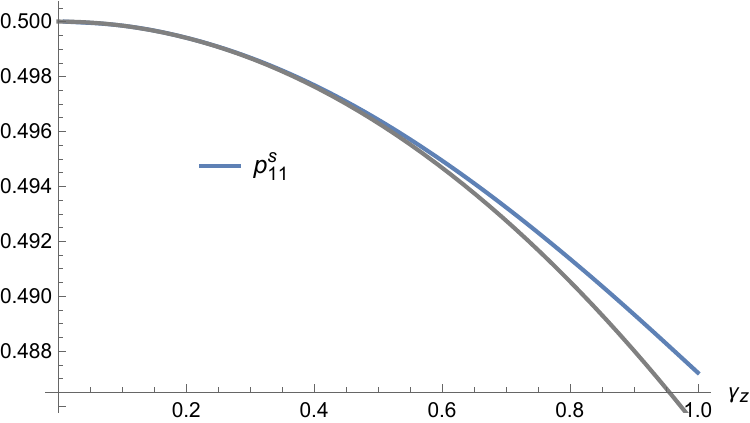}
    \caption{The widths~\eqref{dalphabeta}, or rather $d^{-2}_{(\Sigma)}$ in~\eqref{d2SigmaCoord} and without the trivial factor of $1/E$, for $\gamma_t=3/4$ and $p_1=3/2$, with $\delta\Sigma=\delta p_{13}+\delta p_{23}+\delta p_{33}$ instead of $\delta p_{33}$. The gray, straight lines in the first three plots show the $\gamma_z=0$ results from~\eqref{Qdef}. The gray, straight line in the fourth plot shows the $\Sigma^2$ coefficient in~\eqref{expgammaz2}, and the gray (lower) line in the last plot shows~\eqref{deltaP1saddle}.}
    \label{fig:widthPlots}
\end{figure*}

Fig.~\ref{fig:widthPlots} shows all the components of $d^2$ for $\gamma_t=3/4$ and $p_1=3/2$. For $\gamma_z=1$ we have
\be\label{d2Ex}
d_{\alpha\beta}^{-2}=\begin{pmatrix}
    1.43 & 0.563 & 0.138 & -0.246 & 0.0775 \\
 0.563 & 1.43 & -0.246 & 0.138 & 0.0775 \\
 0.138 & -0.246 & 0.923 & -0.174 & -0.140 \\
 -0.246 & 0.138 & -0.174 & 0.923 & -0.140 \\
 0.0775 & 0.0775 & -0.140 & -0.140 & 0.793 
\end{pmatrix}
\;.
\ee
We see that it is symmetric, $(d^{-2})^T=d^{-2}$, which is a nontrivial check of the implementation of the above formulas, since we have used both versions for the mixed derivatives. It is less immediately obvious that it is symmetric under exchange of the final-state electrons, but with
\be
S=\begin{pmatrix}
    0 & 1 & 0 & 0 & 0 \\
 1 & 0 & 0 & 0 & 0 \\
 0 & 0 & 0 & 1 & 0 \\
 0 & 0 & 1 & 0 & 0 \\
 0 & 0 & 0 & 0 & 1
\end{pmatrix} 
\ee
as the matrix that exchanges the electrons, i.e.
\be
\begin{split}
&(\delta p_{21},\delta p_{11},\delta p_{23},\delta p_{13},\delta p_{33})\\
&=S(\delta p_{11},\delta p_{21},\delta p_{13},\delta p_{23},\delta p_{33}) \;,
\end{split}
\ee
we can check that~\eqref{d2Ex} is indeed symmetric,
\be
S^T d^{-2}S=d^{-2} \;.
\ee

For $\gamma_z=0.01$ we find
\be
\begin{split}
&d^{-2}=\\
&\begin{pmatrix}
    1.224 & 0.6122 & 0.009202 & 0.009168 & 0.009204
   \\
 0.6122 & 1.224 & 0.009168 & 0.009202 & 0.009204
   \\
 0.009201 & 0.009168 & 249.0 & 248.7 & 248.7 \\
 0.009169 & 0.009201 & 248.7 & 249.0 & 248.7 \\
 0.009204 & 0.009203 & 248.7 & 248.7 & 249.0
\end{pmatrix} \;.
\end{split}
\ee
If we were to truncate the elements of this matrix to two digits, then it would be singular, $\det d^{-2}\to0$. As this suggests, it can become harder to maintain a high precision as $\gamma_z$ becomes smaller. However, we can use our $\gamma_z\ll1$ approximations to check the precision. To do this we first change variable from $\delta p_{33}$ to $\delta\Sigma=\delta p_{13}+\delta p_{23}+\delta p_{33}$, which we can do with
\be
T=\begin{pmatrix}
     1 & 0 & 0 & 0 & 0 \\
 0 & 1 & 0 & 0 & 0 \\
 0 & 0 & 1 & 0 & 0 \\
 0 & 0 & 0 & 1 & 0 \\
 0 & 0 & -1 & -1 & \gamma_z
\end{pmatrix} \;,
\ee
so that
\be
\begin{split}
&(\delta p_{11},\delta p_{21},\delta p_{13},\delta p_{23},\delta p_{33})\\
&=T(\delta p_{11},\delta p_{21},\delta p_{13},\delta p_{23},\delta \Sigma/\gamma_z) \;,
\end{split}
\ee
where we have included a factor of $\gamma_z$ since $\delta\Sigma\sim\gamma_z$ is a natural scale for $\delta\Sigma$. What we plot in Fig.~\ref{fig:widthPlots} is
\be\label{d2SigmaCoord}
d^{-2}_{(\Sigma)}=T^T d^{-2}T \;.
\ee
The plots show convergence towards the WKB results in~\eqref{Qdef} (for the two blocks) and~\eqref{expgammaz2} (for the $\Sigma\Sigma$ component),
\be
d^{-2}_{(\Sigma)}=
\begin{pmatrix}
     1.2 & 0.61 & 0 & 0 & 0 \\
 0.61 & 1.2 & 0 & 0 & 0 \\
 0 & 0 & 0.55 & 0.28 & 0 \\
 0 & 0 & 0.28 & 0.55 & 0 \\
 0 & 0 & 0 & 0 & 0.023
\end{pmatrix} \;.
\ee

\subsection{$\gamma_z\ll1$}

In Sec.~\ref{space time WKB sec}, we obtained $\mathcal{O}(\gamma_z^2)$ corrections to the purely time-dependent case using WKB. We can obtain the same results with much less effort using the worldline approach. The trick~\cite{DegliEsposti:2023qqu,DegliEsposti:2023fbv} is similar to how we obtained the first-order momentum derivatives in the previous section, i.e. we express the exponential part of the amplitude by its original form, but with all the integration variables replaced by their saddle points. Then when we differentiate with respect to $\gamma_z$, we have
\be
\frac{\ud}{\ud\gamma_z^2}f(x_s[\gamma_z],\gamma_z)=x_s'(\gamma_z)\partial_x f+\partial_{\gamma_z^2}f=\partial_{\gamma_z^2}f \;,
\ee
so we only have to consider the explicit dependence on $\gamma_z^2$, which comes from
\be
\int\ud u\, A_3(t,z)z'\approx\int\ud u\, A_3(t)[1-(\gamma_z z)^2] \;.
\ee
Since $A_3$ does not vanish for large $u$, and $|z|\to\infty$, we should actually make a partial integration before we make this expansion, so the $\mathcal{O}(\gamma_z^2)$ correction to the instanton action is given by
\be\label{instantonAgammaz2}
-i\frac{\gamma_z^2}{3}\int\ud u\, z^3E(t)t' \;,
\ee
where $z$ and $t$ are now the instantons for the time-dependent case~\eqref{tzCBtime}; we have one~\eqref{instantonAgammaz2} term for $q_C$ and one for $q_B$. Recall that $z(0)$ is arbitrary (a zero mode) for $\gamma_z=0$. But~\eqref{instantonAgammaz2} fixes this zero mode. To make this dependence explicit, we redefine $z(u)\to z_0+z(u)$, where now $z(0)=0$ and $z_0$ is the constant we will determine. For $\gamma_z=0$, we have $\ud z/\ud t=-\pi_3/\pi_0$, so we can change integration variable in~\eqref{instantonAgammaz2} from propertime $u$ to $t$, and then we do not actually have to find the instanton, i.e. we do not need to use~\eqref{tSauterTime} and~\eqref{zSauterTime}. We immediately recover the WKB expression~\eqref{deltaSfinz}. The rest of the calculation is the same as in the WKB approach.

\section{Gamow-Sommerfeld suppression}\label{SommerfeldSection}

We saw in the previous section that, at the saddle-point values of the momenta, the two final-state electrons follow the same worldline. It is therefore natural to ask whether the Coulomb repulsion between these particles is important. In the absence of a background field, it is well known that the Coulomb repulsion is important when the relative velocity of two charged particles is small, which means that one can treat the Coulomb interaction non-relativistically~\cite{Sommerfeld:1931qaf,Sakharov:1948plh}. The result, which can be found in e.g. the textbooks~\cite{Merzbacher,LLnonRel,LLQED,Schiff}, is that the scattering cross section/probability with and without the Coulomb interaction are related by a simple multiplicative factor,
\be
P(\text{with Coulomb})=C^2P(\text{without Coulomb}) 
\ee
where $C^2$ is the Gamow-Sommerfeld factor,
\be\label{SommerfeldC2}
C^2=\frac{2\pi\mu}{e^{2\pi\mu}-1} 
\qquad
\mu=\frac{q_1q_2}{4\pi v}\;,
\ee  
where $q_1$ and $q_2$ are the charges of the two particles, and $v$ is their relative velocity.

Our focus in this section will be on the repulsive case, $q_1q_1>0$, but we first mention that the attractive case, $q_1q_2<0$, has an interesting $v\to0$ limit. Consider the annihilation of an $e^\LCp e^\LCm$ pair into $\mu^\LCp \mu^\LCm$, $\tau^\LCp \tau^\LCm$, $p\bar{p}$ or some other pair of particles. At the threshold, $v\to0$ in the final state and
\be
P(\text{without Coulomb})\propto v\to0 \;,
\ee  
but the Sommerfeld factor diverges,
\be
C^2\to 2\pi|\mu|\propto\frac{1}{v} \;,
\ee
so $P(\text{with Coulomb})$ remains nonzero as $v\to0$, i.e. as a function of the initial energy the cross section has a discontinuous step at threshold~\cite{Voloshin:1989pe,Smith:1993vp,Brodsky:2009gx,Baldini:2007qg}. Such steps have been observed in experiments~\cite{BaBar:2005pon,BaBar:2007fsu,BESIII:2023ioy} (see also references in~\cite{BESIII:2023ioy}, e.g. for discussions about the existence of steps even for neutral baryon pairs). Sommerfeld enhancement has also been important in studies of dark matter, see~\cite{Arkani-Hamed:2008hhe} for a pedagogical explanation and for further references.

In the repulsive case, the Sommerfeld factor gives, for $v\ll1$, an exponential suppression rather than an enhancement,
\be\label{SommerfeldSuppression}
C^2\approx C_{\rm exp}^2\left(\frac{2\pi\alpha}{v}\right)
\qquad
C_{\rm exp}^2(x)=x e^{-x} \;,
\ee
where we have assumed that the two particles are electrons. This approximation is valid for $v\ll 2\pi\alpha\approx0.046$. 

If one forgets for a moment where the function $C_{\rm exp}^2$ comes from, one sees a function which can be expanded in a Taylor series in $\alpha$, $x e^{-x}=x-x^2+\dots$ for $x\ll1$. However, that expansion is not relevant since $C_{\rm exp}^2$ is only valid as a $x\gg1$ approximation of $C^2$, so there is no reason to expect that the Taylor expansion of $C_{\rm exp}^2$ should agree with the Taylor expansion of $C^2$. In this case it is easy to see that the true expansion is indeed different, $C^2(x)=1-(x/2)+\dots$. 

The Sommerfeld factor~\eqref{SommerfeldC2} is usually derived by solving the nonrelativistic Schr\"odinger equation. Here we will derive it using worldline instantons.
After integrating out the (quantum) photon field, one finds a nonlocal term in the action
\be\label{Sint}
S_{\rm int}=-\frac{ie^2}{8\pi^2}\int_0^1\ud\tau_1\ud\tau_2\frac{\dot{q}_1(\tau_1)\dot{q}_2(\tau_2)}{(q_1(\tau_1)-q_2(\tau_2))^2-i\epsilon} \;.
\ee 
For $q_1^\mu\ne q_2^\mu$, $S_{\rm int}$ describes interaction between two different worldlines, e.g. the two final-state electrons in trident. 

As an aside, if one instead considers a closed loop and $q_1^\mu=q_2^\mu$, then $S_{\rm int}$ gives after renormalization the AAMLR result~\cite{Ritus:1975pcc,Affleck:1981bma,Lebedev:1984mei,Huet:2010nt,Huet:2018ksz} for the imaginary part of the effective action in a constant electric field
\be\label{AAMLR}
\text{Im }\Gamma\sim E^2\exp\left(-\frac{\pi m^2}{E}+\alpha\pi\right) \;.
\ee  
We present a derivation of~\eqref{AAMLR} in Appendix~\ref{AAMLRsection}.
For electrons and positrons, the $\alpha$ term in~\eqref{AAMLR} is of course negligible, but it is important for magnetic monopoles~\cite{Gould:2017fve}. In~\cite{Ritus:1975pcc,Lebedev:1984mei}, the result~\eqref{AAMLR} was obtained by first calculating the $\mathcal{O}(\alpha)$ correction to $\mathcal{O}(\alpha^0)$ and then guessing that higher orders in $\alpha$ would sum to $e^{\pi\alpha}$. In~\cite{Affleck:1981bma}, \eqref{AAMLR} was instead obtained by including~\eqref{Sint} in a saddle-point approximation of the worldline path integral, i.e. without making a perturbative expansion in $\alpha$.  

We will now derive the Sommerfeld factor from~\eqref{Sint}. We assume that two electrons are produced at $r\sim0$ in some process whose formation length is short compared to the subsequent Coulomb repulsion dynamics. This is the standard assumption for the Sommerfeld factor, and we will explain below why this is also relevant in the trident case. Given that this effect is only significant when the relative velocity of the particles is small, we can go to the center-of-mass frame where the dynamics is nonrelativistic. We begin by changing integration variables in~\eqref{Sint} from proper time $\tau_j$ to coordinate time $t_j$, and then to $t=(t_1+t_2)/2$ and $\Delta t=t_2-t_1$. In the numerator we have $1-{\bf v}_1\cdot{\bf v}_2\approx1$ for nonrelativistic velocities. In the denominator we have $(q_2-q_1)^2\approx\Delta t^2-[{\bf r}_2(t)-{\bf r}_1(t)]^2$. Performing the integral over $\Delta t$ gives (including an overall factor of $2$ since $q_1\leftrightarrow q_2$ both describe the same process)
\be
S_{\rm int}\approx\int\ud t\, V(|{\bf r}_2-{\bf r}_1|)
\qquad
V=\frac{\alpha}{|{\bf r}_2-{\bf r}_1|} \;,
\ee  
which is the instantaneous Coulomb potential.

The non-interacting parts of the action, $e^{-i(S_1+S_2)}$, are given by
\be\label{worldlineAction}
S_j=\frac{T_j}{2}m+\int_0^1\ud\tau m\frac{\dot{q}_j^2}{2T_j} \;.
\ee  
Substituting the saddle points for $T$ and $t$ gives
\be
S_j=\int\ud t\, m\sqrt{1-v_j^2}\approx\int\ud t\left(-\frac{m v_j^2}{2}\right)+\text{const.}
\ee
With ${\bf r}_1={\bf R}-(1/2){\bf r}$ and ${\bf r}_2={\bf R}+(1/2){\bf r}$ we have
\be\label{S3}
S_1+S_2+S_{\rm int}=\int\ud t\left(-\frac{\mu\dot{\bf r}^2}{2}-\frac{M\dot{R}^2}{2}+V(r)\right) \;,
\ee
where $\mu=m/2$ is the reduced mass and $M=2m$. Since we are outside any background field, the center-of-mass dynamics is trivial, $\dot{\bf R}=0$. The saddle-point equation for the relative motion is
\be
\mu\ddot{\bf r}=-\nabla V \;.
\ee   
Multiplying by $\dot{\bf r}$ and integrating gives
\be\label{dr2}
\dot{\bf r}^2=\frac{2}{\mu}(\mathcal{E}-V) \;,
\ee
where $\mathcal{E}=\mu v^2/2$ is the asymptotic energy. Eq.~\eqref{dr2} is an implicit solution, since $V$ depends on $r$, but we do not need to find an explicit solution. Substituting into~\eqref{S3} gives
\be
S=\mathcal{E}(t_{\rm out}-t_{\rm in})+2\int\ud t(V-\mathcal{E}) \;.
\ee  
The terms outside the integral cancel against similar terms from the asymptotic states. Changing integration variable from $t$ to $r$ and squaring the amplitude gives,
\be
C^2\propto \exp\left(-2\sqrt{2\mu}\int_0^{r_1}\ud r\sqrt{V-\mathcal{E}}\right) \;,
\ee
which is the well-known exponent for nonrelativistic tunneling through a barrier. The integration limit for a Coulomb potential is 
\be\label{r1Coulomb}
r_1=\frac{\alpha}{\mathcal{E}} \;.
\ee 
Performing the integral gives
\be
C^2\propto \exp\left(-\frac{2\pi\alpha}{v}\right) \;,
\ee
in agreement with~\eqref{SommerfeldSuppression}.

From~\eqref{r1Coulomb} we see that the Coulomb repulsion works on a scale of $r_C\sim\alpha/v^2\gtrsim1/\alpha\gg1$, cf.~\cite{LLQED}, which also means $t_C\sim\alpha/v^3\gtrsim1/\alpha^2$. Compare this with the scale of the background field. In the lab frame we have $t_E\sim1/\omega$, which means $t_E\sim1/(p_0\omega)$ in the center-of-mass frame, where $p_0$ is the asymptotic energy of the particles in the lab frame. Comparing these two scales,
\be
\frac{t_C}{t_E}\sim\frac{\alpha}{v}\gamma\frac{p_0}{m}\frac{E}{v^2}\gg1 \;,
\ee   
which is large because $v\sim\alpha$ so $E/v^2\gg1$ for any relevant value of $E$ (we need $E\ll1$ to use the saddle-point method, but for $E\sim\alpha^2$ the probability would be extremely small due to the exponential suppression, so there is no motivation to study such regimes). For a wide field, $\gamma\ll1$, but then the typical energy of the electrons will be $p_0/m\sim1/\gamma$ ($p_0$ can also be large due to large momentum of the initial particle), so $\gamma p_0/m\sim1$ and the ratio is still large.  
Thus, the Coulomb repulsion happens on a much longer time scale compared to the formation process in the background field. One finds a similar conclusion for $z_C/z_E$, though somewhat less favorable.  

\section{Conclusions}

We have shown how to use WKB and worldline-instanton methods to study nonlinear trident in fields which are not plane waves. We studied both the exponential and the pre-exponential parts of the spectrum in detail for time-dependent electric fields $E(t)$. At first this might not seem like a step beyond plane waves, but the results for $E(t)$ have a richer analytical dependence. Indeed, it is possible to obtain the plane-wave results by taking the high-energy limit of the results for $E(t)$, but it is not possible do the reverse. Studying $E(t)$ therefore allows us to explicitly see how and under what conditions the spectrum converges to the plane-wave results in the high-energy limit. As a rule of thumb, one expects that plane-wave results should be applicable for sufficiently high energies, but, in studying these WKB approximations, one can see explicitly that the plane-wave approximation may break down if the particles have high energy but along the electric field direction or if the background field is simultaneously very strong, $a_0\gg1$. Both of these cases can be relevant when particles have high energies as a result of being accelerated by a strong field. 

We have shown how to generalize the incoherent-product approximation, where the trident probability is approximated by the incoherent product of the probabilities of nonlinear Compton scattering and Breit-Wheeler pair production. The incoherent-product approximation is very useful and often the only method available in cases where higher orders in $\alpha$ are important. This provides motivation for further studies of how to construct an incoherent-product approximation for non-plane-wave fields.  

WKB methods are often convenient for 1D fields, but for fields which depend on more than one coordinate, we have shown that a worldline-instanton approach based on open worldlines allows one to study in principle arbitrary multidimensional fields, and also to take into account higher orders in $\alpha$, e.g. for the Coulomb interaction of the two final-state electrons in trident.

\acknowledgements

G. T. is supported by the Swedish Research Council, Contract No. 2020-04327.

\appendix
\section{AAMLR}\label{AAMLRsection}

In Sec.~\ref{SommerfeldSection} we studied the Coulomb repulsion between two final-state electrons. The resulting Sommerfeld factor is all-orders in $\alpha$, and the reason that all orders are important is because the effective parameter is $\alpha/v$, which is not small if the relative velocity is nonrelativistic. That Sommerfeld factor is an interaction between two different worldlines.
In this section we will consider the self-interaction of a single worldline with itself. So in this section we consider Schwinger~\cite{Sauter:1931zz,Schwinger:1951nm} rather than trident pair production. The goal is to rederive~\eqref{AAMLR}. For electron-positron pairs, $\alpha\ll1$ and one should for consistency not include that term, because it is smaller than other corrections, e.g. NLO terms in the $E\ll1$ expansion. Indeed, for the NLO correction in the $E$ expansion to be negligible compared to the $\alpha$ terms, one would need $E\ll\alpha$, but the probability is so extremely small for such weak fields that there is no point in considering it. One would therefore consider the regime $\alpha<E\ll1$, where the $\alpha$ term should be neglected. So instead of electrons and positrons, one can consider magnetic monopoles~\cite{Affleck:1981bma,Gould:2017fve}, for which the corresponding $\alpha$ is large. Dirac's quantization condition says that the charge $q_M$ of a monopole is related to the electron/positron charge by $q_M=2\pi n/e$, where $n$ is an integer (cf.~\cite{Schwinger:1975ww}), so 
\be
\alpha_M=\frac{q_M^2}{4\pi}=\frac{n^2}{4\alpha}\approx34 n^2 \;,
\ee
and $\pi\alpha_M\approx108n^2$.
In the following we will not assume that we are dealing specifically with monopoles, so we will keep the same notation as in previous sections.

The self-interaction leads to a UV divergence, which will be absorbed into a renormalization of the mass. In the quenched approximation (only one fermion line/loop), the mass renormalization is the only renormalization needed. We use $m_0$ to denote the bare mass. The exponential part of the path integral is given by $e^{-iS-iS_{\rm int}}$, where
\be
S=\frac{m_0T}{2}+\int_0^1\ud\tau\left(\frac{m_0\dot{q}^2}{2T}+A\dot{q}\right)
\ee 
and
\be
S_{\rm int}=-\frac{i\alpha}{2\pi}\int_0^1\ud\tau_1\ud\tau_2\dot{q}(\tau_1)\dot{q}(\tau_1)R([q(\tau_1)-q(\tau_2)]^2) \;,
\ee
where $R(x)$ is a UV-regularized version of $1/(x^2-i\epsilon)$, for example
\be\label{areg}
R_a=\frac{1}{[q(\tau_1)-q(\tau_2)]^2-a^2-i\epsilon} \;,
\ee 
where $a\to0$ after renormalization of the mass. 

For a constant field, the instanton is a circle in the $\text{Im}(t)-z$ plane for $\alpha=0$. Because of symmetry, one expects~\cite{Affleck:1981bma} that the instanton will still be a circle for $\alpha>0$. We therefore take the following as an ansatz,
\be
t=i\rho\cos(2\pi\tau)
\qquad
z=\rho\sin(2\pi\tau) \;,
\ee  
where $\rho$ is a constant to be determined. $T$ does not appear in $S_{\rm int}$. Inserting the saddle-point value $T=-2\pi i\rho$ into $S$ gives
\be
-iS=\frac{\pi}{E}(-2m_0\rho+\rho^2) \;.
\ee   
For the regularized interaction term we find
\be
\begin{split}
-iS_{\rm int}&=\pi\alpha\left(1-\frac{2\rho^2+(Ea)^2}{Ea\sqrt{4\rho^2+(Ea)^2}}\right)\\
&=\pi\alpha-\frac{\pi}{E}\frac{\alpha\rho}{a}+\mathcal{O}(a) \;.
\end{split}
\ee
The saddle-point value of $\rho$ is
\be
\rho=m_0+\frac{\alpha}{2a}
\ee
and the exponent becomes
\be
\exp\left(-\frac{\pi}{E}\left[m_0+\frac{\alpha}{2a}\right]^2+\pi\alpha\right) \;.
\ee 
By writing this in terms of the physical, renormalized mass,
\be\label{m0m}
m=m_0+\frac{\alpha}{2a} \;,
\ee
we recover~\eqref{AAMLR}.

By reinstating $c$ and $\hbar$ in~\eqref{m0m},
\be
m_0=m-\frac{1}{2}\frac{e^2}{4\pi\hbar c}\frac{\hbar}{ca}=m\left(1-\frac{r_e}{2a}\right) \;,
\ee 
we see that the mass renormalization does not depend on $\hbar$. In fact, it is exactly the same as the mass renormalization in the derivation of the Lorentz-Abraham-Dirac (LAD) equation, see~\cite{Nakhleh:2012ji}. In the case of LAD, \eqref{m0m} is exact, i.e. it is not just the first term in some expansion in $\alpha$.

While our final result~\eqref{AAMLR} agrees with Eq.~(2.29) in~\cite{Affleck:1981bma}, we cannot make a concrete comparison with their implicit regularization or their indirect discussion about renormalization. However, from the perturbative treatment in the appendix of~\cite{Affleck:1981bma}, it seems that they might have had a more traditional perturbative renormalization in mind.

\end{document}